\documentclass[A4paper,floatfix,5p,10pt,twocolumn,preprint]{elsarticle}

\makeatletter
\let\c@author\relax
\makeatother

\expandafter\let\csname ver@natbib.sty\endcsname\relax

\makeatletter
\def\ps@pprintTitle{%
    \let\@oddhead\@empty
    \let\@evenhead\@empty
    \def\@oddfoot{}%
    \let\@evenfoot\@oddfoot}
\makeatother

\usepackage{amsmath}
\usepackage{amssymb}
\usepackage{graphicx}%
\usepackage{dcolumn}
\usepackage{hyperref}
\usepackage{subcaption}
\usepackage[normalem,normalbf]{ulem}
\usepackage{enumerate}
\usepackage{tikz}
\usepackage{algpseudocode}
\usepackage{algorithm}
\usepackage{multirow}
\usepackage{geometry}
\numberwithin{equation}{section}

\colorlet{mdtRed}{red!50!black}

\hypersetup{pdfpagemode={UseOutlines},
bookmarksopen=true,
bookmarksopenlevel=0,
hypertexnames=false,
colorlinks=true,
citecolor=magenta,
linkcolor=red,
urlcolor=mdtRed,
pdfstartview={FitV},
unicode,
breaklinks=true,
}

\usepackage[backend=biber, citestyle=numeric-comp, bibstyle=ieee, mcite=true, subentry, sorting=none, citetracker]{biblatex}
\defbibentryset{set1}{PCRGTG2019REPORT, PCRBTG2019, PCRBTG2019DATA}
\defbibentryset{set2}{DHTGDW2017,TGDWCMDH2005}
\journal{{}}
\usetikzlibrary{arrows,chains,matrix,positioning,scopes}
\makeatletter
\tikzset{join/.code=\tikzset{after node path={%
\ifx\tikzchainprevious\pgfutil@empty\else(\tikzchainprevious)%
edge[every join]#1(\tikzchaincurrent)\fi}}}
\makeatother
\tikzset{>=stealth',every on chain/.append style={join},
         every join/.style={->}}
\tikzstyle{labeled}=[execute at begin node=$\scriptstyle,
   execute at end node=$]
 
\newcommand{\be}{\begin{equation}}
\newcommand{\ee}{\end{equation}}

\begin{document}
\title{Revisiting the Epps effect using volume time averaging: An exercise in R}
\author[uct-sta]{Patrick Chang}
\ead{CHNPAT005@myuct.ac.za}
\author[uct-sta]{Roger Bukuru}
\ead{BKRROG001@myuct.ac.za}
\author[uct-sta]{Tim Gebbie}
\ead{tim.gebbie@uct.ac.za}
\address[uct-sta]{Department of Statistical Science, University of Cape Town, Rondebosch 7701, South Africa}

\begin{abstract}
We revisit and demonstrate the Epps effect using two well-known non-parametric covariance estimators; that of Malliavin and Mancino (MM) \cite{MM2002, MM2009, MR2017} and the estimator suggested by Hayashi and Yoshida (HY) \cite{HY2005}. We show the existence of the Epps effect \cite{EPPS1979} in the top 10 stocks from the Johannesburg Stock Exchange (JSE) by various methods of aggregating Trade and Quote (TAQ) data. Concretely, we compare calendar time sampling with two volume time sampling methods: asset intrinsic volume time averaging, and volume time averaging synchronised in volume time across assets. We reaffirm the argument made in much of the literature that the MM estimator is more representative of trade time reality because it does not over-estimate short-term correlations in an asynchronous event driven world. We confirm well know market phenomenology with the aim of providing some standardised R based simulation tools that can ease replication while providing a volume time view of the loss of correlations at short term sampling rates.    
\end{abstract}


\maketitle

\tableofcontents

\section{Introduction} \label{intro}

Covariation is a key parameter in quantitative finance with traditional applications including portfolio optimisation and risk management \cite{MEF2005}, but also including more recent applications in unsupervised state discovery to discern changes in the financial markets in the high-frequency domain \cite{DHPRL2017,DHTGDW2016,DHPHD}. The availability of high-frequency financial data has seemly ameliorated many of the apparent estimation problems relating to large portfolio correlation measures by moderating the problem of data scarcity \cite{MMZ2011}. This has in turn provided new challenges for co-movement estimation both due to the variety of different price generating processes, but also due to the very nature of trading itself. 

First, the price discovery process generating financial market price data is often different on different time-scales {\it e.g.} daily-sampled closing prices are the result of a closing auction at the end of each days trading. The price resulting from a closing auction and can be well approximated by some equilibrium price derived from balancing supply with demand. While transaction prices resulting from intraday tick-by-tick data are generated by price discovery in a continuous double auction market. The resulting prices are dynamically achieved as a function of order-flow and are not the result of an equilibrium between supply and demand.

Even within the intraday regime different traders and market participants trade and interact in a variety of different ways with different strategic objectives. Because different types of agents can trade in different ways on different times scales, and use different market structures, even for the same asset, the correlations between assets can differ due to the structure of markets rather than being intrinsic to some set of stochastic processes, or properties specific to a particular asset. Second, having high-frequency financial data comes with additional caveats because of the asynchronous nature of the generating processes that create tick-by-tick trade and quote data and the potential for these processes to generate non-linear dependency of correlations relative to temporal averaging scales.

It then becomes unclear which particular type of measure of co-movement can best suit a specific decision time-scale, and then whether or not it can be meaningfully extracted using non-averaged high-frequency data in event form to proxy co-movements on the required decision time-scale. The key point: Decision making in finance takes place on a particular time-scale, with a particular application, within a particular theoretical framework, and with a business model and regulatory framework. 

These types of model constraints inform the modeller about what averaging scales could be appropriate; such as seconds, minutes, and weeks in calendar time, as well as the possible measure of co-variation that could be effective to make the required decision {\it e.g.} using covariances, correlations, copulas, relative-entropy and so on \cite{MEF2005}. Particularly when derived from actual trade events or aggregated event data \cite{DGMOP2001,Z2010,BLPP2019} these types of decisions can then lead directly into the fundamental issue of whether or not to smooth, down-sample data, or average the data so that it can conform to the theoretical assumptions and data-requirements for a particular measure and use case. The choice of decision tools and theoretical framework can cloud decision making because of the necessity of implicit averaging whatever the form of the resulting asymptotic biases are.  

In the context of realised volatility estimators, many popular approaches remain problematic when used with high-frequency asynchronous data because they require a choice of synchronisation, data interpolation, and averaging which leads to biases induced in the estimates \cite{HY2005, BR2002, Z2010, BLPP2019}. Normative decisions taken to frame high-frequency finance problems themselves can lead to ineffective modeling decisions in support of the quest for robust asymptotic results. 
Here we take a simplified introductory approach from the perspective of intraday trading on South African equity markets and present two well-known non-parametric estimators designed specifically to deal with asynchrony. The first estimator proposed by Malliavin and Mancino \cite{MM2002, MM2009, MR2017} adopts a Fourier approach, and the second estimator proposed by Hayashi and Yoshida \cite{HY2005} using the contributions from overlapping intervals to overcome the problem faced by the traditional realised volatility. There are a rich variety of estimators grounded in the market micro-structure noise perspective of high-frequency financial markets (See \cite{BLPP2019} and references therein). 

To investigated the challenges related to the averaging scales and the selection of an appropriate clock for high-frequency markets, we consider these two different covariance estimators and their relative perspectives on the relationship between average correlations and the temporal averaging, or measurement scale: the ``Epps effect'', first generically and then explicitly through the lens of synchronised volume-time averaging. 

The Epps effect is often contrasted with the observation that in calendar time volatility is a decreasing function of increasing estimation time scales \cite{SS2014} where this is then considered to be the  ``market micro-structure noise effect". In this view the Epps effect is then just the average statistical overlap of the momentum of overlapping asset prices in the presence of long-memory \cite{SS2014} and the market micro-structure noise effect is the result of negative correlations from the bid-ask bounce. 

The problem we have with this view is that order-flow is the key top-down driver of much of the dynamics of markets and the ebb and flow of order-flow can change the nature of time-scales at which trading events occur, all while trading events themselves are fundamentally discrete. The resulting smearing out of time-scales as covariance measures are aligned in calendar time can then dilute the correlations. The Epps effect can then plausibly be due to averaging and convolution effects of averaging over order-flow as it is realised through the mechanics and structure of the market. It is not clear whether the market micro-structure noise explanations are the result of normative models being forced onto the data, or are in-fact reflecting fundamental properties of the underlying stochastic processes in the presence of order-flow dynamics and market structure.  

Although much of work here does not provide anything fundamentally new, we hope that it provides a useful set of transparent R code patterns that can ease reproducibility \cite{JICD2013} while re-enforcing the need to be careful with regards to merely consider the impact of: i.) lead-lag and ii.) asynchrony when trying to explain the Epps effect. 

And more generally when deciding how to estimate co-movements between assets it remains important to be mindful of, iii.) the correlation structure dependence's on the sampling intervals \cite{MMZ2011}, iv.) the definition of time itself in the context of fundamentally discrete nature high-frequency financial markets \cite{ELO2012A} and its impact on an averaged observable, and v.) the consequence of the choice of the implicit theoretical frameworks that one may choose to place ones tools inside of \cite{Z2010,BLPP2019}.  

For this later reason, we cautiously avoid the more popular market micro-structure noise frameworks and explanations \cite{MS2008, GO2008, SFX2010,BLPP2019} and try retain a pragmatic event driven paradigm \cite{ELO2012A} aimed a solving individual investment and trading problems use-case by use-case. Our approach is perhaps best seen in the context of exploratory data analysis in the spirit of Tukey \cite{T1977}. The Epps effect \cite{EPPS1979} then relates to a collection of empirical phenomena concerned with the decay of correlations associated with smaller and smaller sampling intervals. 

To better understand the estimators we have chosen we explicitly reconstruct an Epps type effect from asynchrony by exploring the relationship between correlation estimation, the level of asynchrony, and the sampling interval under consideration \cite{RENO2001, PI2007, MMZ2011}. We will then use these insights in later sections to inform our visualisation of the decay in correlations under averaging where we briefly consider the Epps effect through the lens of different time-aggregations using Trade and Quote (TAQ) data.

We will consider three time averaging methods based on an event-time perspective: calendar time averaging, intrinsic time averaging using stock specific volume clocks \cite{DERMAN2002}, and a synchronised (across assets) volume clock \cite{ELO2012A,ELO2012B}. 

The paper is structured as follows: in Section \ref{sec:estimators} we will discuss the implementation algorithms for the MM and HY estimators. Section \ref{sec:MC} we conduct simple Monte Carlo experiments to identify how, when and where the two estimators differ in the context of the asynchrony perspective. First, we begin the comparison by assessing how the two estimators perform under asynchrony - specifically asynchrony induced in a missing-data framework. Second, we then compare the two estimators with various stochastic processes. Third, we replicate the results from \cite{RENO2001} and adapt his experiment to highlight the differences in our data-informed approach, compared to the market micro-structure noise approach. Section \ref{sec:description} briefly outlines the data used.

In calendar time (See Section \ref{ssec:calendar}) we specifically we consider Closing Prices, from decimating transaction to the end of the calendar time intervals, and then Volume Weighted Average Price (VWAP) bars using volume averaging across the same uniform calendar time intervals. Finally, we consider an intrinsic time approach in Section \ref{ssec:intrinsic} where we employ a framework first suggested by Derman \cite{DERMAN2002}, and then introduce our synchronised volume time aggregation in Section \ref{ssec:event}. Section \ref{sec:averaging} combines the various aggregation techniques from Section \ref{sec:MC} along with the estimators to study the Epps effect in the Johannesburg Stock Exchange (JSE) data. The work aims to motivate the importance of considering the interplay between averaging scales and event time with regards to the loss of correlations on shorter calendar time-intervals. Finally, we end off with Section \ref{sec:conclude} with a brief conclusion. 


\section{Estimators} \label{sec:estimators}
\subsection{Malliavin-Mancino estimator}

Malliavin and Mancino \cite{MM2002, MM2009} proposed an estimator that is constructed in the frequency domain. It expresses the Fourier coefficients of the volatility process using the Fourier coefficients of the price process $p_i(t) = \ln(S_i(t))$ where $S_i(t)$ is a generic asset price at time $t$. By re-scaling the trading times from $[0, T]$ to $[0, 2\pi]$ (See Algorithm \ref{algo:rescale}) and using Fourier convolution \cite{MM2009} (See Theorem 2.1 \cite{MM2009}) we have that for all $k \in \mathbb{Z}$ and $N$ samples:
\begin{equation} \label{eq:Der:16}
    \begin{aligned}
      \mathcal{F}(\Sigma^{ij})(k) = \lim_{N \rightarrow \infty} \frac{2 \pi}{2N+1} \sum_{|s| \leq N} \mathcal{F}(dp_i)(s) \mathcal{F}(dp_j)(k-s).
    \end{aligned}
\end{equation}
Here $\mathcal{F}(\ast)(\star)$ is the $\star^{\text{th}}$ Fourier coefficient of the $\ast$ process.
Now using previous tick interpolation to avoid a downward bias in the estimator \cite{BR2002} and a simple function approximation for the Fourier coefficients (See \cite{MM2009, MALHERBE2007, PCRGTG2019REPORT}), we obtain a representation for the integrated volatility:
\begin{equation} \label{eq:Der:21}
  \int_0^{2\pi} \Sigma^{ij}(t) dt = \frac{1}{2N+1}  \sum_{\substack{|s|\leq N \\ j=1,i=1}}^{n-1,n-1} e^{is(t_i - t_j)} \delta_{1}(I_i) \delta_{2}(I_j),
\end{equation}
where the trade intervals are $I_i := [t_i, t_{i+1})$ and $I_j := [t_j, t_{j+1})$ while the price fluctuation are $\delta_{1}(I_i) := p_1(t_{i+1}) - p_1(t_i)$ and $\delta_{2}(I_j) := p_2(t_{j+1}) - p_2(t_j)$ for the $i^{\text{th}}$ and $j^{\text{th}}$ asset respectively.

The R code implementation can be found in Algorithm \ref{algo:ComplexFT} at our GitHub site \cite{PCRBTG2019}. For further theoretical details we refer the reader to the original papers by Malliavin and Mancino \cite{MM2002, MM2009} and our supporting materials and code on GitHub \cite{set1}.

\subsection{Hayashi-Yoshida estimator}

Hayashi and Yoshida \cite{HY2005} proposed a cumulative covariance estimator:
\begin{equation} \label{eq:Der:23}
  U_n := \sum_{i=1}^{n} \sum_{j=1}^{n} \Delta P_1(J_i) \Delta P_2(J_j) 1_{\{ J_i \cap J_j \neq \emptyset \}}.
\end{equation}
This is free of the need to synchronize the data prior to analysis. Using the Kanatani weight matrix and a simple integral approximation we find the integrated volatility:
\begin{equation} \label{eq:Der:34}
\begin{aligned}
  &\int_0^T \Sigma^{ij}(t) dt = \sum_{i=1}^{n} \sum_{j=1}^{n} w_{ij} \delta_1(J_i) \delta_2(J_j),
\end{aligned}
\end{equation}
where the trade intervals are $J_i := (t_{i-1}, t_i]$ and $J_j := (t_{j-1}, t_j]$ while the price fluctuations are $\delta_1(J_i) := p_1(t_{i}) - p_1(t_{i-1})$ and $\delta_2(J_j) := p_2(t_{j}) - p_2(t_{j-1})$ for the $i^{\text{th}}$ and $j^{\text{th}}$ asset respectively. The Kanatani weights are defined as 

\begin{equation} \label{eq:Der:26}
  w_{ij} =
  \begin{cases} 
      1 & \mathrm{if} \ J_i \cap J_j \neq \emptyset \\
      0 & \mathrm{otherwise}.
   \end{cases}
\end{equation}

The R code implementation can be found in Algorithm \ref{algo:HY} at our GitHub site \cite{PCRBTG2019}. For further theoretical details we refer the reader to the original papers by Hayashi and Yoshida \cite{HY2005} and our supporting materials and code on GitHub \cite{set1} \footnote{Kanatani's weighted realized volatility can also be used for Malliavin and Mancino's Fourier estimator (See \cite{HNKN2008} or \cite{KANATANI2004}).}.

\section{Simple Monte Carlo Experiments} \label{sec:MC}

Monte Carlo experiments are conducted to identify the differences between the two estimators to highlight which estimator may be better suited for what kinds of contexts from the perspective of data asynchrony. We first represent asynchrony as missing data observations. Here data is decimated from underlying uniformly sampled bivariate stochastic processes to investigate the impact of asynchrony levels as inspired by \cite{PI2007}.  The missing data approach can be informative because it is conceptually simple and the impact of missing data on covariance estimation is generally well understood in the context of Gaussian random matrix theory \cite{WG2007}. Concretely, we consider the efficacy of the respective estimators if the world was manifestly discrete, asynchronous and not the result of sampling some underlying continuous Ito process. Second, we then consider a more sophisticated representation of asynchrony using random arrival times to re-sample the underlying processes to directly expose the relationship between asynchrony and the sampling intervals \cite{RENO2001}. 

\begin{figure*}[t!]
\begin{subfigure}{0.5\textwidth}
\includegraphics[width=\textwidth]{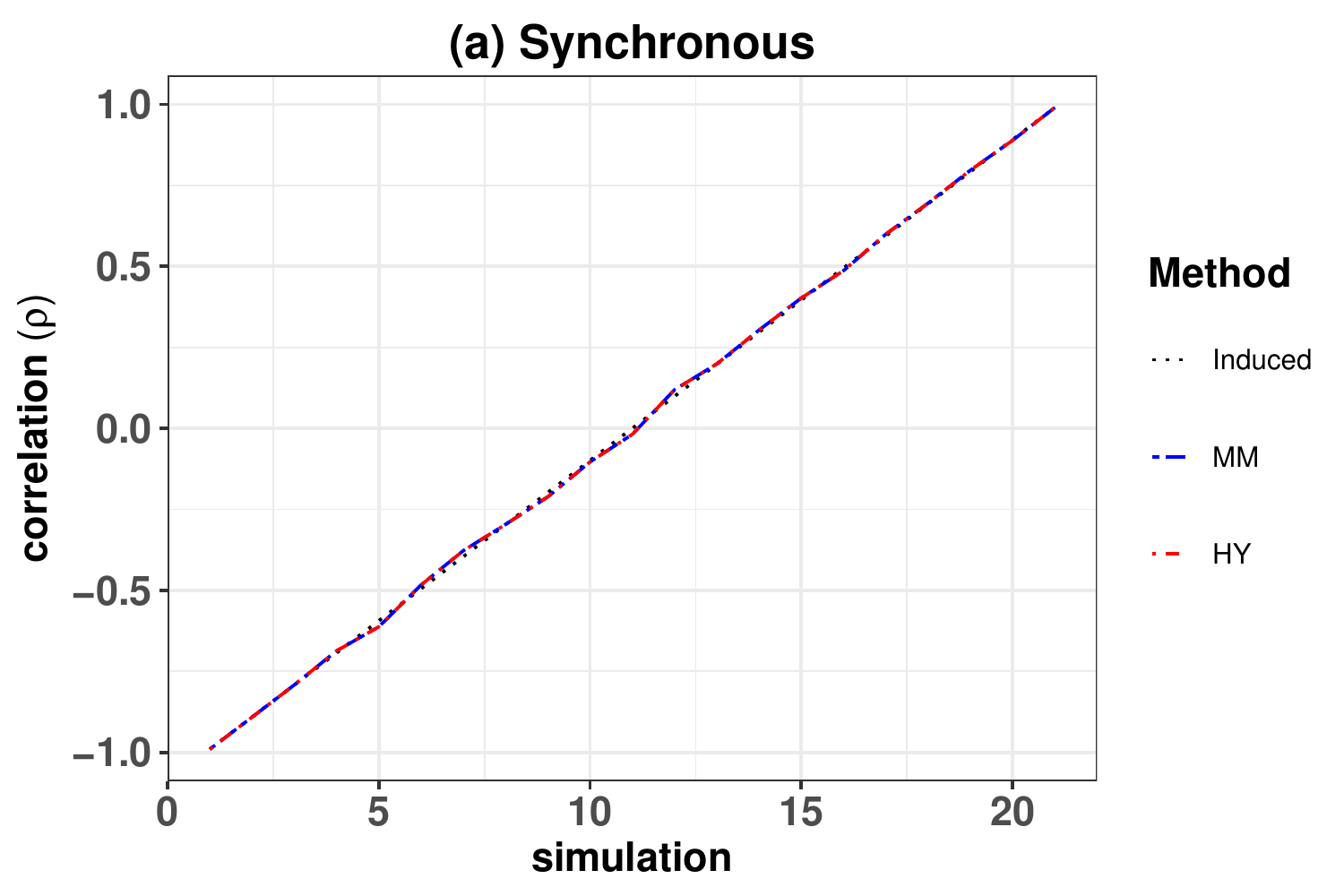}
\end{subfigure}
\hfill
\begin{subfigure}{0.5\textwidth}
\includegraphics[width=\textwidth]{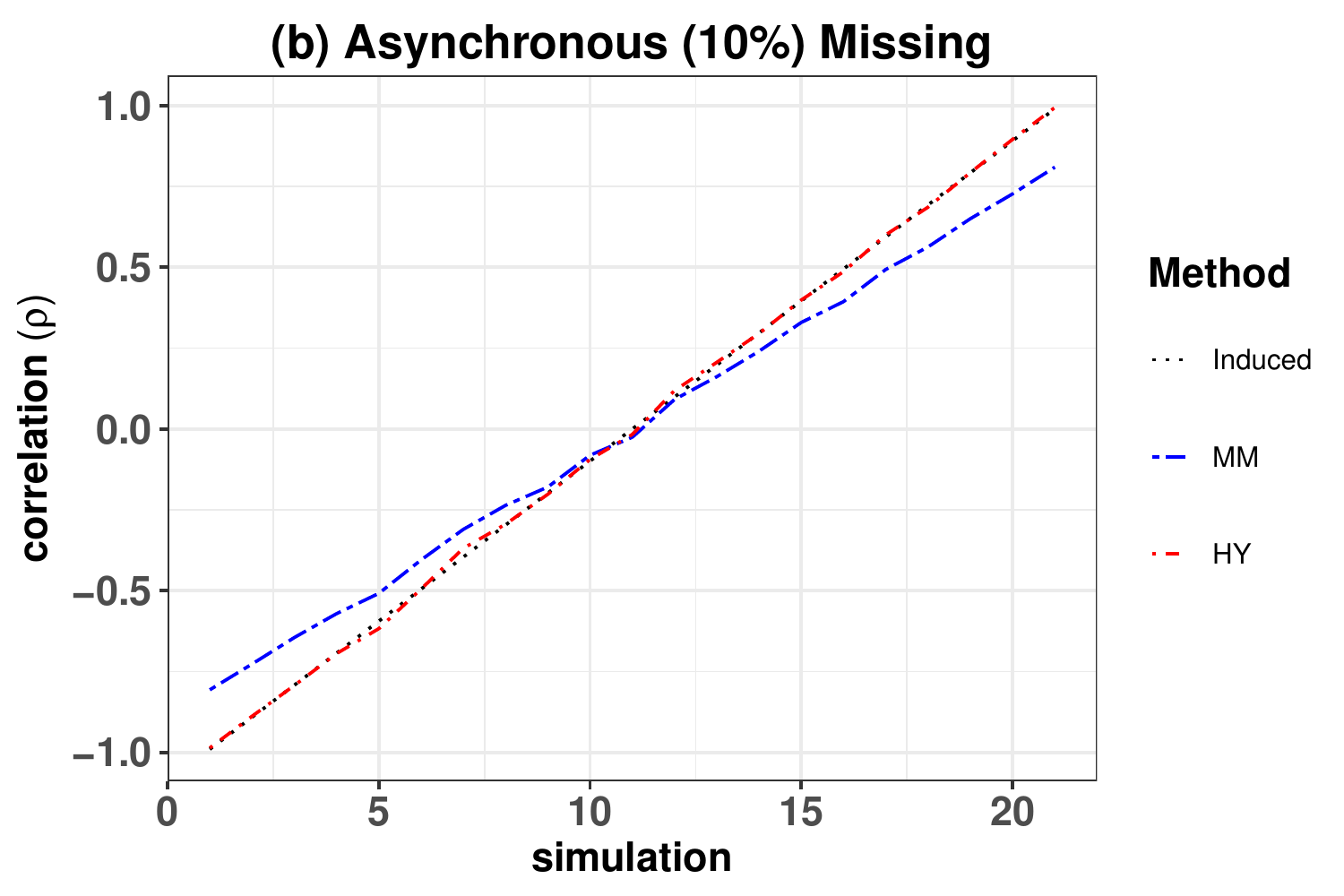}
\end{subfigure}

\begin{subfigure}{0.5\textwidth}
\includegraphics[width=\textwidth]{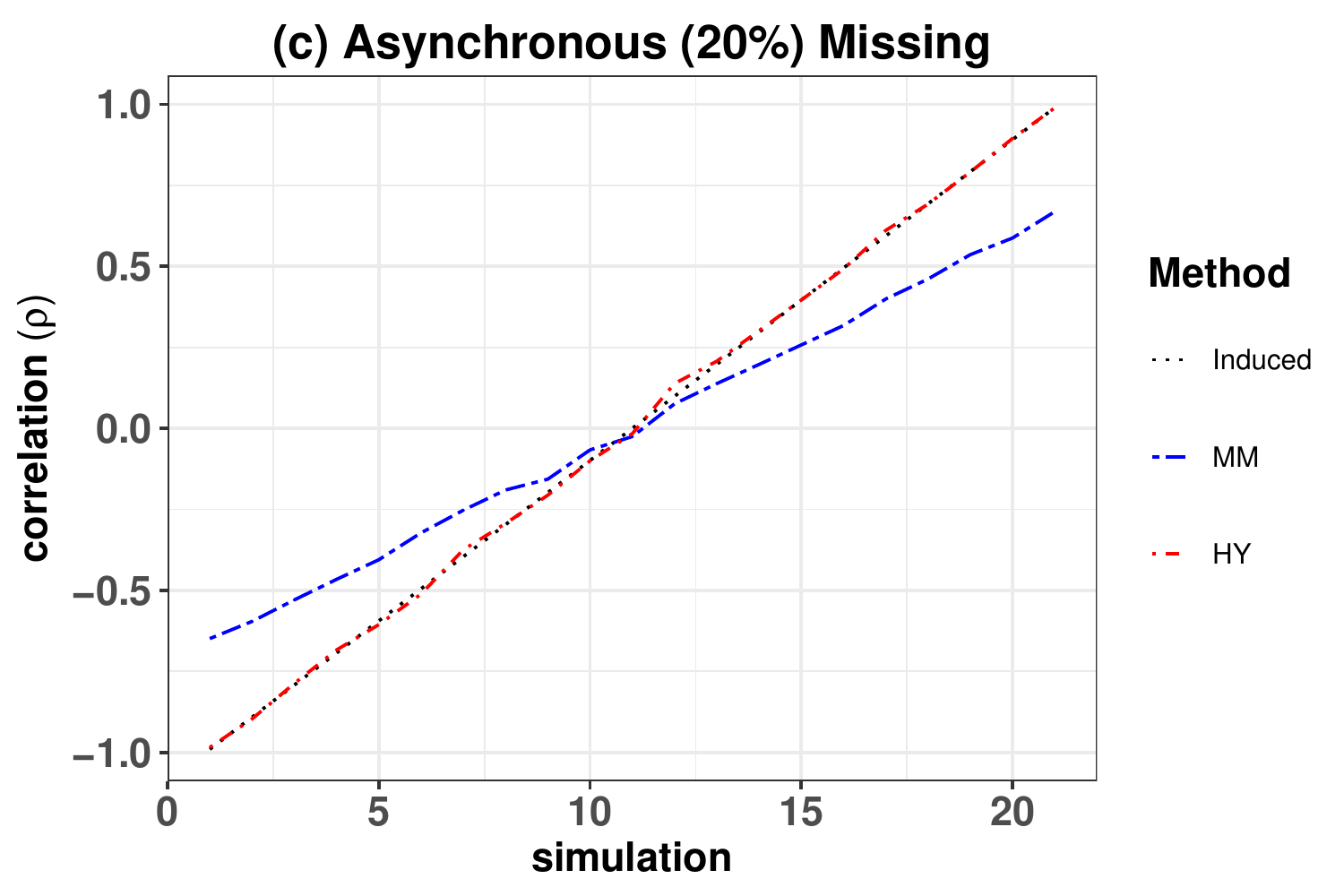}
\end{subfigure}
\hfill
\begin{subfigure}{0.5\textwidth}
\includegraphics[width=\textwidth]{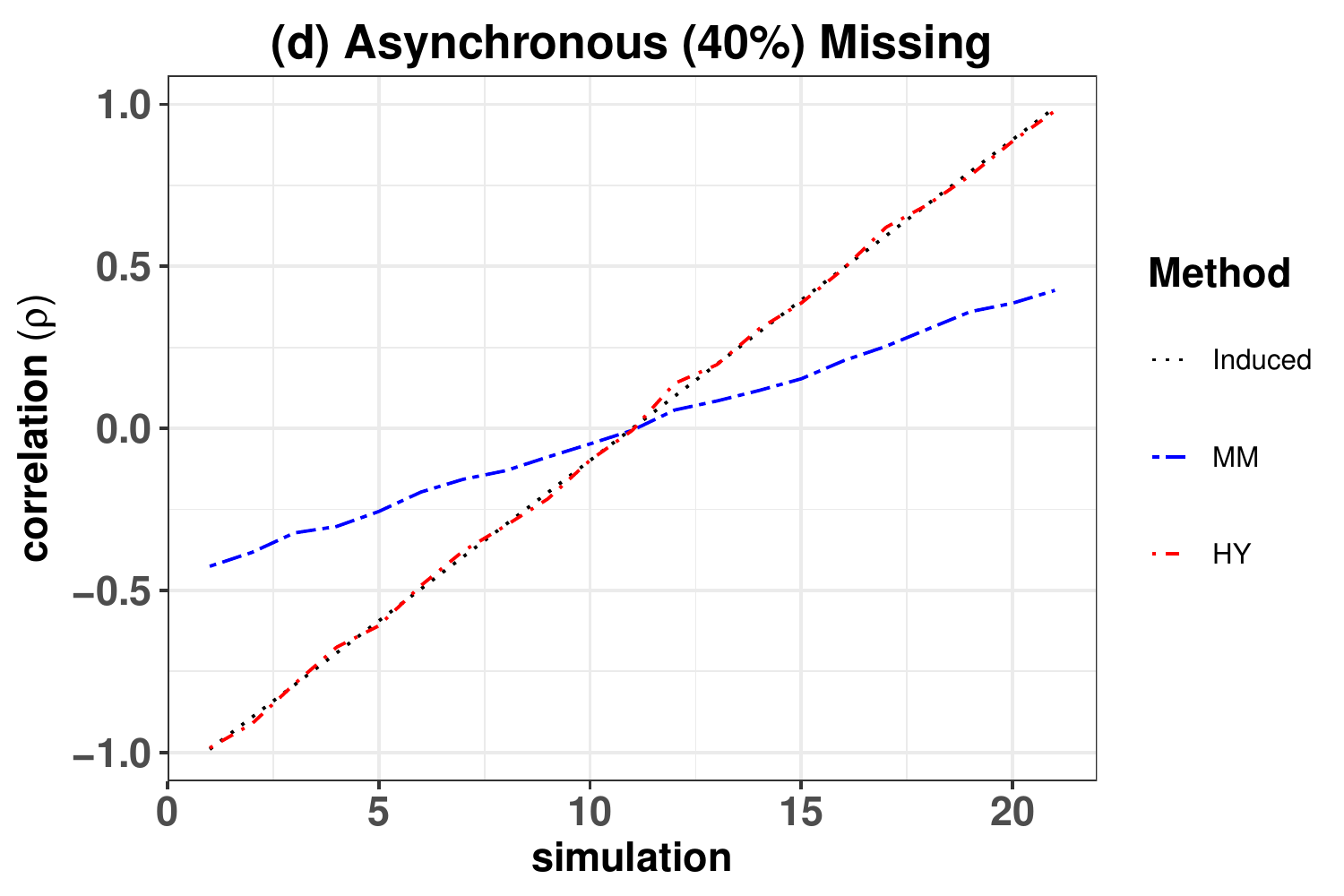}
\end{subfigure}
\caption[The effect of missing data.]{Comparing different levels of missing data to demonstrate the bias between the MM and HY estimators. Concretely, (a) through to (d) show 0\%, 10\%, 20\% and 40\% of each sample path replaced by missing data to replicate asynchrony. As per the figure legend, the blue dotted line is the MM estimator, estimated using Algorithm \ref{algo:ComplexFT}; the red dotted line is the HY estimator, estimated using Algorithm \ref{algo:HY} and the black dotted line is the induced correlation of the GBM. Each simulation with varying correlation is done by simulating 10,000 seconds from a bivariate GBM satisfying \eqref{eq:BM:1} and numerically simulated using Algorithm \ref{algo:GBM}. The figure shows that if the data is discrete and asynchronous then the MM estimator is most appropriate, however, if the data is sampled discretely from an underlying continuous-time GBM then the HY estimator is more appropriate. We argue that the Epps effect from asynchrony is real and as such due to the discrete nature of financial market trades one should use the MM estimator rather than the HY estimator. The figures can be recovered using the R script file \href{https://github.com/rogerbukuru/Exploring-The-Epps-Effect-R/blob/master/Monte\%20Carlo\%20Plots/MissingData.R}{MissingData.R} on the GitHub resource \cite{PCRBTG2019}.}
\label{fig:GBMcor}
\end{figure*}

\subsection{Missing data representation} \label{ssec:missing}

The missing data experiments focus on how the two estimators differ when asynchrony is induced by randomly decimating the price path data points. The benchmark experiment is conducted by simulating 10,000 realisations, each perhaps representing a sample 1 second apart, from a bivariate geometric Brownian motion with daily parameters $\mu_1 = 0.01$, $\mu_2 = 0.01$, $\sigma^2_1 = 0.1$, $\sigma^2_2 = 0.2$, $\rho_{12}$ ranging from $(-1, 1)$ and a starting price of R100. The non-synchronicity, or asynchrony, is achieved by randomly sampling a percentage of the observations from each sample path and removing them. The first missing data experiment found that the two estimators differ under asynchrony as represented by missing data observations where the level of asynchrony determined how different the estimators behave. 

To further gain insight into the two estimators, various stochastic processes are studied to identify alternative situations where these two estimators differ, by again representing asynchrony as missing data points. The additional experiments focused on how the alternative stochastic processes will cause the two estimators to differ with commensurate levels of missing data. 

Here we first consider the benchmark geometric Brownian motion (See Section \ref{sssec:GBM}), and then a collection of alternative random bivariate processes: the Merton model (Section \ref{sssec:MM}), Variance Gamma (Section \ref{sssec:VG}), GARCH (1,1) (Section \ref{sssec:Garch}) and a mean-reverting Ornstein Uhlenbeck process (Section \ref{sssec:OU}).

\subsubsection{Geometric Brownian motion} \label{sssec:GBM}

The bivariate Geometric Brownian Motion (GBM) satisfies the following system of SDEs
\begin{equation} \label{eq:BM:1}
  \frac{dS_i(t)}{S_i(t)} = \mu_i dt + \sigma_i dW_i(t), \ \ \ \ i = 1, 2.
\end{equation}
Figure \ref{fig:GBMcor} (a), we see that both MM (blue dotted line) and HY (red dotted line) perfectly recover the induced correlation (black dotted line) for the synchronous case. From Figure \ref{fig:GBMcor} (b) through to (d), it is clear that as the level of asynchrony increases, MM appears to have a downward bias towards zero which \cite{RENO2001} attributes to the Epps effect \cite{EPPS1979} while HY recovers the induced correlation regardless of the level of asynchrony.

Hayashi and Yoshida claimed that the Epps effect is a bias that arises from the estimator for which their estimator is immune to \cite{HY2005}. Looking at Figure \ref{fig:GBMcor}, this can seem to be the case. However, this goes against the findings of \cite{RENO2001, TK2009, MMZ2011, PI2007}. The current literature has identified the main sources for the Epps effect to be: smaller sampling intervals \cite{EPPS1979, TK2009}, lead-lag \cite{RENO2001, MMZ2011} and asynchrony \cite{RENO2001, PI2007}. Closed-form expressions recovering the Epps effect can be found in \cite{TK2009, MMZ2011} - indicating that the Epps effect is not a simple bias from the estimator. Rather the HY is upward biased even though it recovers the correct induced correlation.

Although the experiment recovers the Epps effect arising from this type of asynchrony it is not the result of a truly asynchronous process. It the result of missing observation in uniformly sampled data. Therefore, the argument is that the HY estimator is the better estimator of the two if one believes that the observed prices in the market are discrete samples of an underlying continuous stochastic process and that asynchrony is a missing data problem. Then the HY estimator will be able to reproduce the true underlying correlation between the assets by allowing multiple contributions to the estimator. This would imply that the MM estimator has a downward bias attached to it. On the other hand, the MM estimator is the better estimator if one is of the view that the financial market data is not a missing data problem relative to an underlying synchronous continuous stochastic process this is only partially revealed; but is rather disconnected, discrete and asynchronous where the events and their relationships to each other are the fundamental entities and measurables of the finance world. Then the MM estimator will produce the true correlation in the system as it is lossless interpolation between the events. This then implies that the HY estimator will have an upward bias that is caused by the multiple contributions \cite{DHTGDW2017}.

TAQ data is truly discontinuous \cite{MW2014}, discrete and made up of in-homogeneous collections of asynchronous events; the MM estimator is a better tool of the two for studying co-movements between asset attributes with this type of data representation. 

Additional issues regarding the HY estimator are pointed out in \cite{SFX2010, GO2008}. First, when the processes are highly asynchronous the HY estimator deletes observations through its multiple contributions ({\it e.g.} Fig. 1 in \cite{SFX2010}) and therefore it does not utilise all available observations. Second, a critical assumption underlying the HY estimator is that the correlation between two assets does not extend beyond the intervals where returns fully or partially overlap. This implies that information regarding the correlation is fully accounted for when a price update arrives. This assumption does not hold in practice and causes the HY estimator to be biased as shown by \cite{GO2008}.

\subsubsection{Merton} \label{sssec:MM}

The bivariate Merton model satisfies the following system of SDEs:
\begin{equation} \label{eq:Mert:1}
  \frac{dS_i(t)}{S_i(t^{-})} = \mu_i dt + \sigma_i dW_i(t) + dJ_i(t), \ \ \ \ i = 1, 2.
\end{equation}
Here the correlation is $\mathrm{Corr}(dW_1, dW_2) = \rho_{12}$ and the intervals $J_i$ are independent of the $W_i$ with piece-wise constant paths \cite{GLASSERMAN2004}. J is defined as: \begin{equation} \label{eq:Mert:2}
  J_i(t) = \sum_{j=1}^{N(t)} (Y_j - 1),
\end{equation}
where $N(t)$ is a Poisson process with $Y_j \sim LN(a,b)$ i.i.d and also independent of $N(t)$ \footnote{The $t^{-}$ on the LHS of \eqref{eq:Mert:1} is used to indicate the C\`{a}dl\`{a}g nature of the process near jumps.}.

\begin{figure*}[t!]
\begin{subfigure}{0.5\textwidth}
\includegraphics[width=\textwidth]{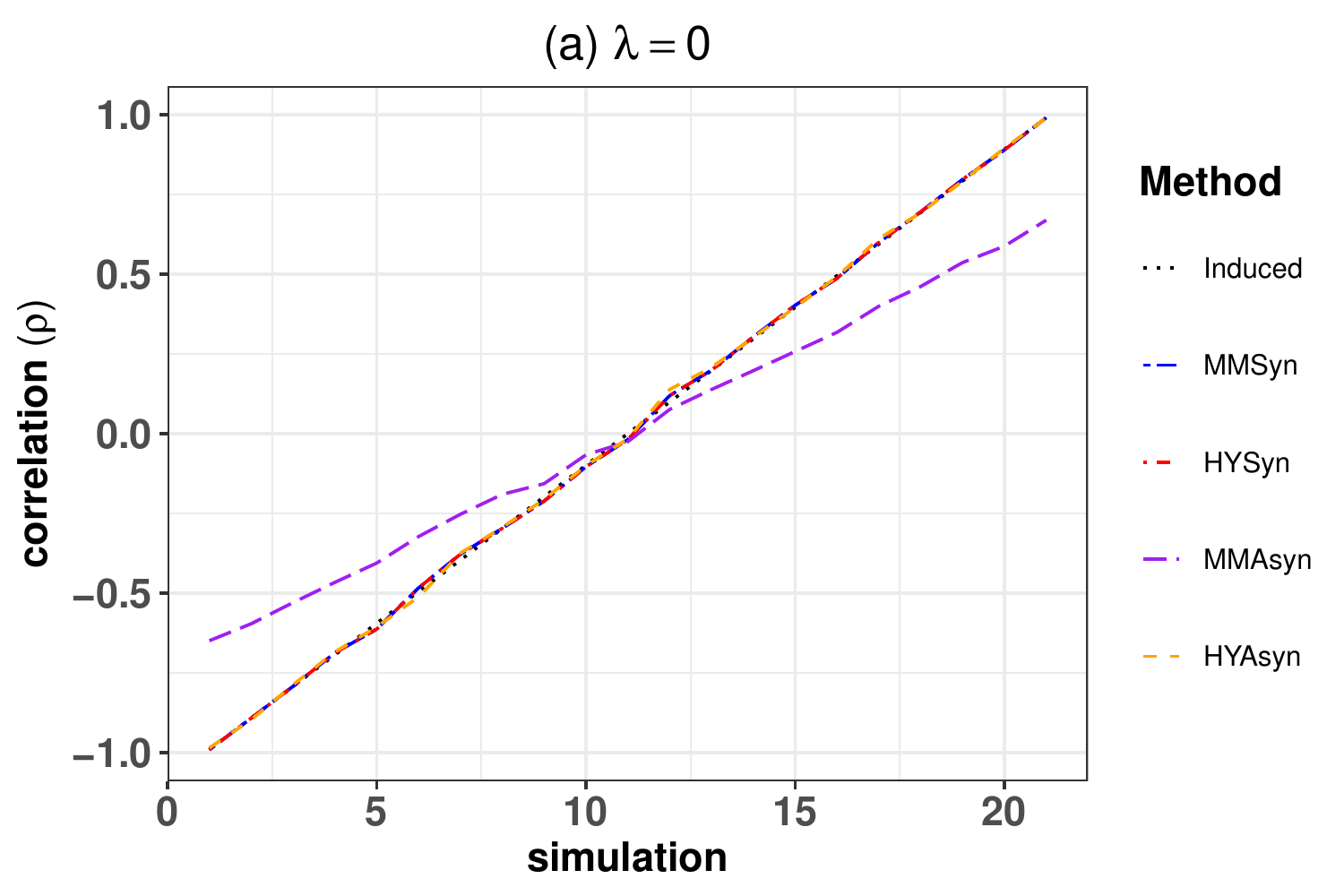}
\end{subfigure}
\hfill
\begin{subfigure}{0.5\textwidth}
\includegraphics[width=\textwidth]{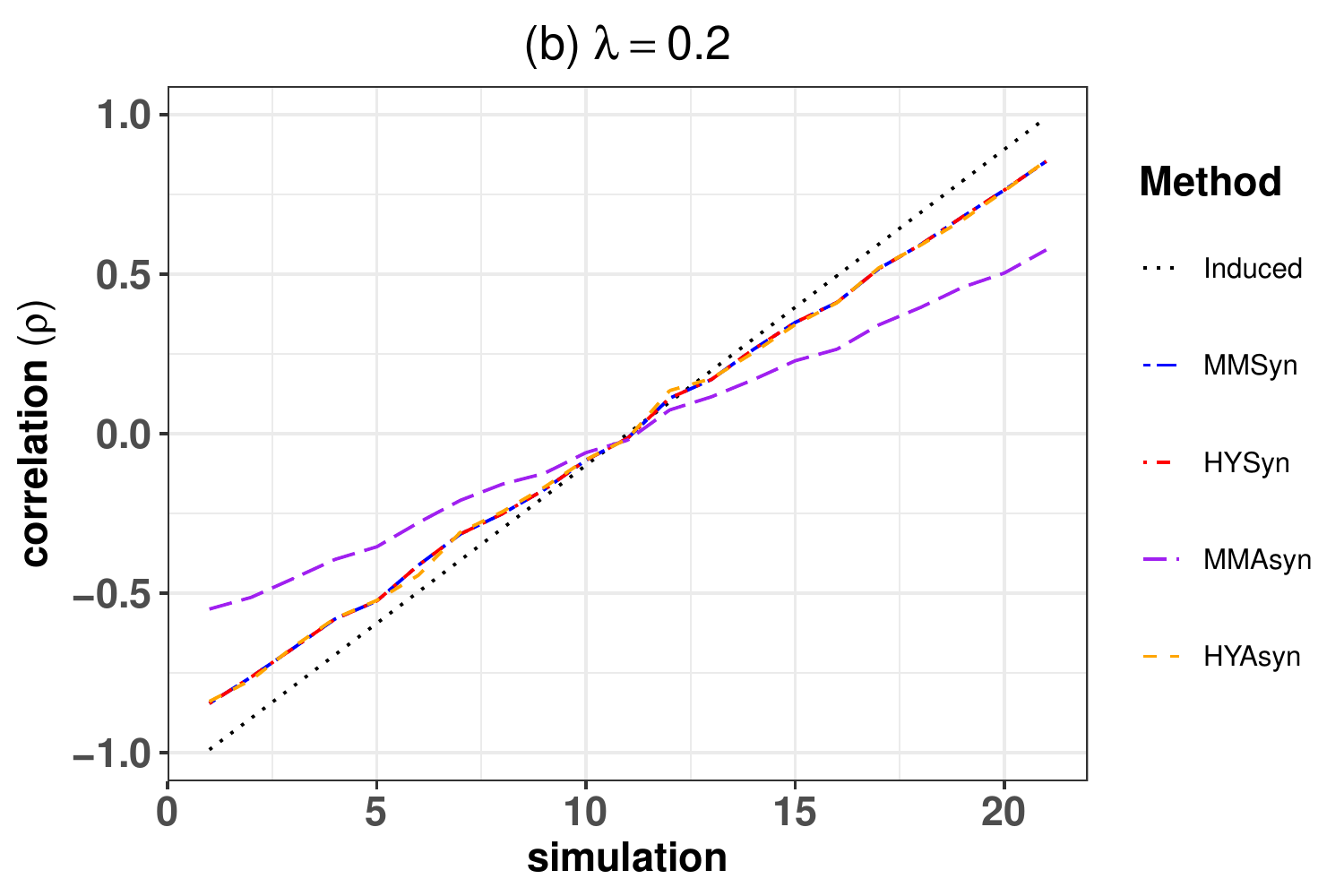}
\end{subfigure}
\begin{subfigure}{0.5\textwidth}
\includegraphics[width=\textwidth]{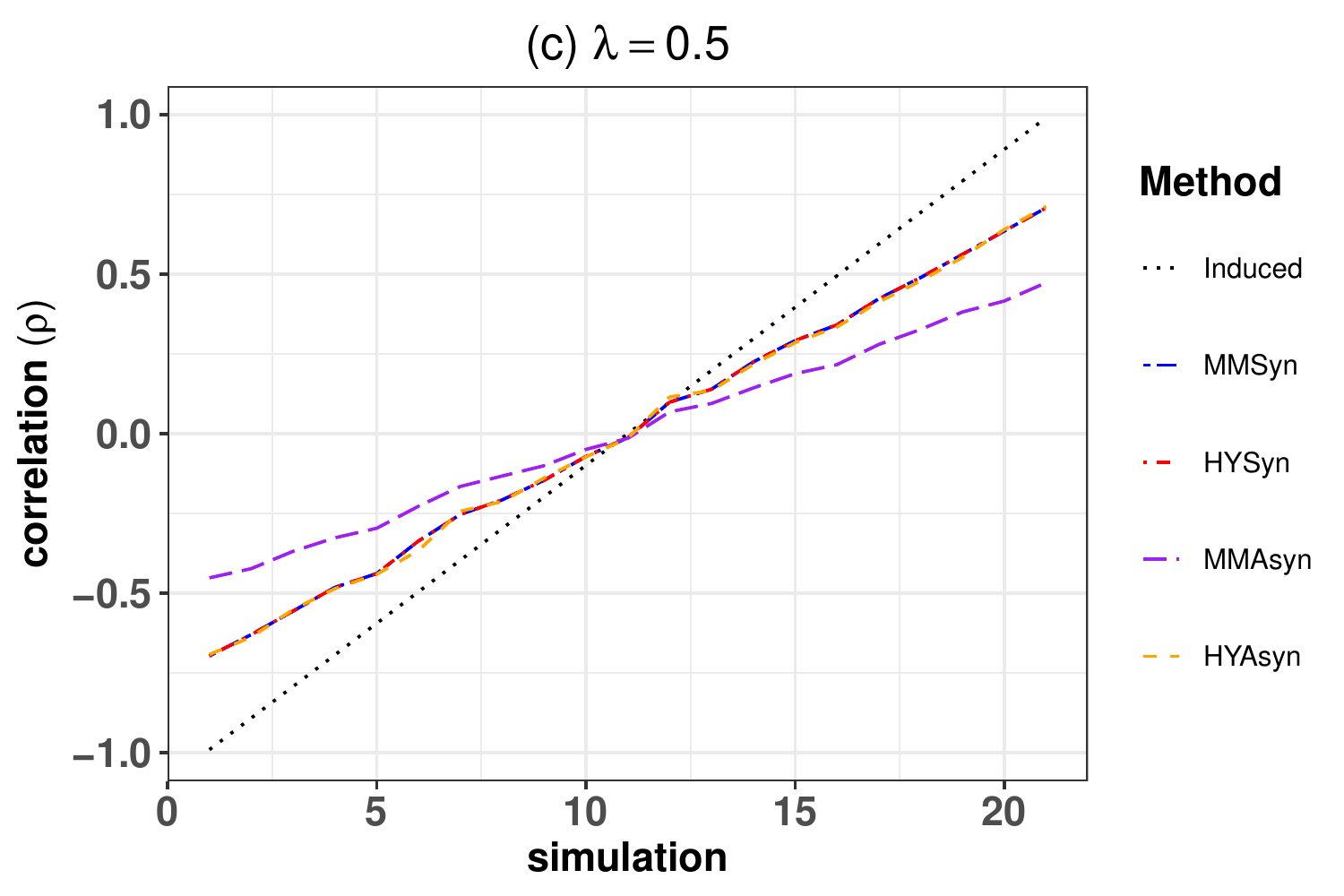}
\end{subfigure}
\hfill
\begin{subfigure}{0.5\textwidth}
\includegraphics[width=\textwidth]{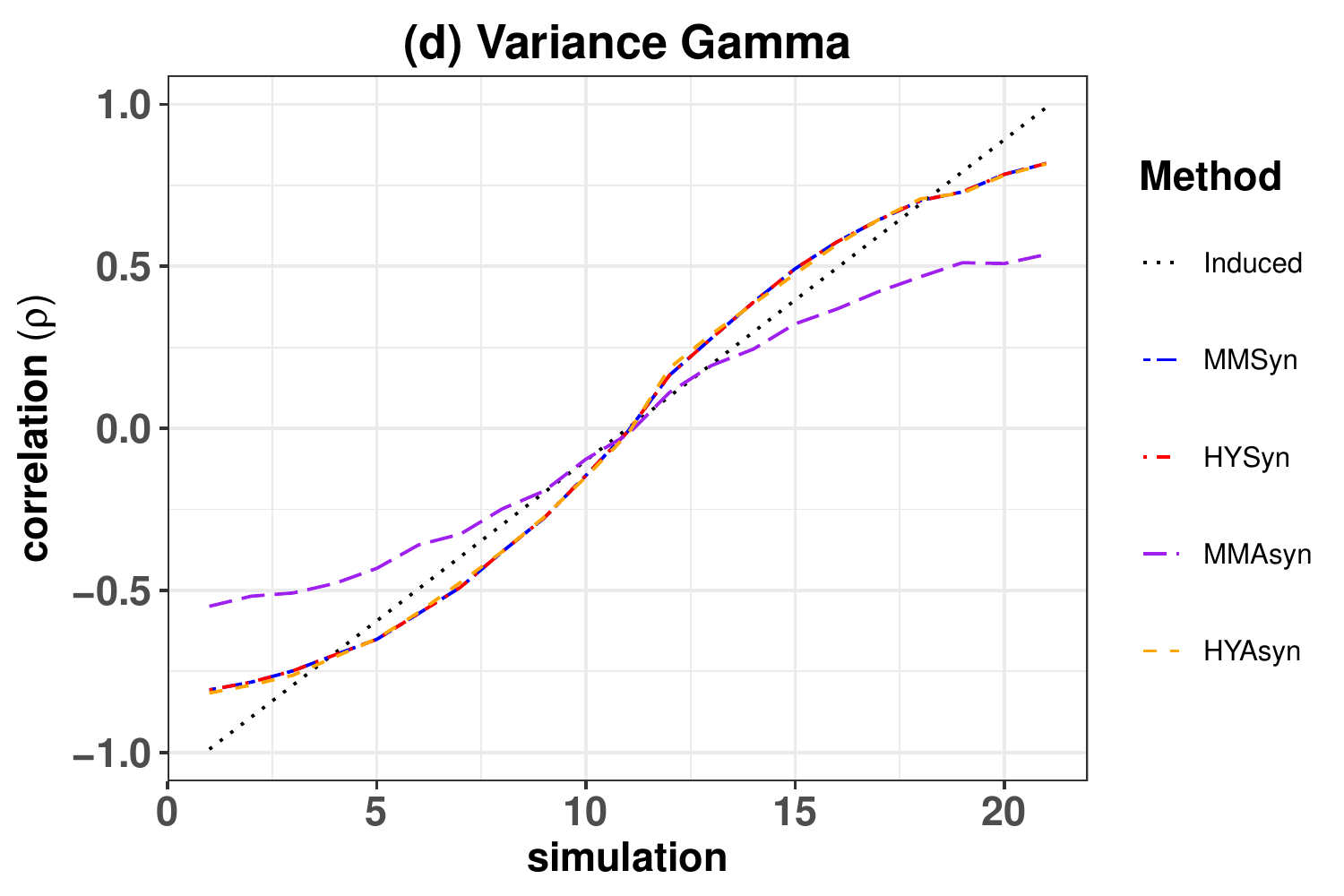}
\end{subfigure}
\caption[The effect of various diffusion processes.]{Comparing various stochastic processes to demonstrate the effect on the MM and HY estimators. Concretely, (a) shows a pure diffusion process, (b) and (c) show a jump-diffusion process and (d) shows a pure jump process. As per the figure legend, the blue and purple dotted lines are the MM estimator estimated using Algorithm \ref{algo:ComplexFT} under synchronous and asynchronous observations respectively. The red and orange dotted line is the HY estimator estimated using Algorithm \ref{algo:HY} under synchronous and asynchronous observations respectively. The black dotted line is the induced correlation from the various stochastic processes. (a) through to (c) is 10,000 seconds simulated from a bivariate Merton model satisfying \eqref{eq:Mert:1} and numerically simulated using Algorithm \ref{algo:Mert}. (d) is from a bivariate Variance Gamma model satisfying \eqref{eq:VG:1} and numerically simulated using Algorithm \ref{algo:VG}. The asynchrony is induced by down-sampling 20\% of the observations from each sample path. The figure shows that both MM and HY produce the same estimates regardless of the underlying process and that it seems the difference between them arises due to asynchrony. Under asynchrony, HY recovers the synchronous estimates while the correlation for MM drops. The figures can be recovered using the R script file \href{https://github.com/rogerbukuru/Exploring-The-Epps-Effect-R/blob/master/Monte\%20Carlo\%20Plots/SDE1.R}{SDE1.R} on the GitHub resource \cite{PCRBTG2019}.}
\label{fig:Mertcor}
\end{figure*}

The daily parameters used for the Merton model are $\mu_1 = 0.01$, $\mu_2 = 0.01$, $\sigma^2_1 = 0.1$, $\sigma^2_2 = 0.2$, $\rho_{12}$ ranging from $(-1, 1)$, $a_1 = 0$, $a_2 = 0$, $b_1 = 100$, $b_2 = 100$ and varying $\lambda$ to move from a pure diffusion process to a jump-diffusion process. A sample path of 10,000 seconds is simulated starting at R100. This model will determine if the two estimators differ due to the effect of jumps.

Figure \ref{fig:Mertcor} (a) through to (c), $\lambda_1 = \lambda_2 = 0$, $0.2$ and $0.5$ respectively. The asynchronicity is induced by down-sampling each price path by 20\%. For all the plots in Figure \ref{fig:Mertcor}, the synchronous MM (blue dotted line) is the same as the synchronous HY (red dotted line). The asynchronous HY (orange dotted line) recovers the synchronous estimates while the asynchronous MM (purple dotted line) has a lower correlation estimate than the synchronous case. In Figure \ref{fig:Mertcor} (a) through to (c), both the synchronous MM and HY estimators produce the same estimate which drops towards zero as $\lambda$ increases. 

This drop in correlation is not due to any bias from the estimators but rather due to the fact that the jump process of the Merton model is independent of the underlying diffusion process. Therefore as the intensity of jumps increase, the impact from the independence, seeps through to change the correlation structure of the overall jump-diffusion process. This is the case because when the trades are synchronous, HY becomes the Realized Volatility (RV) \cite{HNKN2008} and the RV is consistent under jumps \cite{AT2009}. Therefore, the two estimators seem to only differ under asynchrony and it is not dependent on the type of diffusion process.

\subsubsection{Variance Gamma} \label{sssec:VG}

The bivariate Variance Gamma (VG) process satisfies the following SDEs:
\begin{equation} \label{eq:VG:1}
  S_i(t) = U(t) - D(t), \ \ \ \ i = 1, 2,
\end{equation}
with U and D being independent gamma processes satisfying
\begin{equation} \label{eq:VG:2}
\begin{aligned}
  &U(t_{i+1}) - U(t_i) \sim \text{Gamma}(\alpha (t_{i+1} - t_i), \beta),   \\
  &D(t_{i+1}) - D(t_i) \sim \text{Gamma}(\alpha (t_{i+1} - t_i), \beta).
\end{aligned}
\end{equation}
U and D are limited to have the same shape and scale parameters allowing an alternative representation $W(G(t))$ where $W$ is a standard Brownian motion, $G$ a Gamma process \cite{GLASSERMAN2004} and $\mathrm{Corr}(dW_1, dW_2) = \rho_{12}$ ranging from (-1, 1). A sample path of 10,000 seconds is simulated starting at R100 with daily parameters $\mu_1 = \mu_2 = 0.01$, $\sigma_1^2 = 0.1$, $\sigma_2^2 = 0.2$ and $\beta_1 = \beta_2 = 1$. This model will determine if a pure jump process will cause the estimators to differ. The Merton model is contrasted with the Variance Gamma model in Figure \ref{fig:Mertcor}. 

\subsubsection{GARCH(1,1)} \label{sssec:Garch}

The bivariate GARCH (1,1) model satisfies the following SDEs:
\begin{equation} \label{eq:Garch:1}
\begin{aligned}
  &dp_i(t) = \sigma_i(t) dW_i(t), \ \ i = 1, 2,
\end{aligned}
\end{equation}
and
\begin{equation} \label{eq:Garch:2}
\begin{aligned}
  &d\sigma_1^2(t) = \theta_1 [w_1 - \sigma_1^2]dt + \sqrt{2 \lambda_1 \theta_1} \sigma_1^2(t) dW_3(t),           \\
  &d\sigma_2^2(t) = \theta_2 [w_2 - \sigma_2^2]dt + \sqrt{2 \lambda_2 \theta_2} \sigma_2^2(t) dW_4(t),            
\end{aligned}
\end{equation}
where $\mathrm{Corr}(dW_1, dW_2) = \rho_{12}$ $\in (-1, 1)$. We simulate a sample path of 10,000 seconds starting at R100 using the parameters from \cite{RENO2001, AB1998} i.e. $\theta_1 = 0.035$, $\theta_2 = 0.054$, $w_1 = 0.636$, $w_2 = 0.476$, $\lambda_1 = 0.296$ and $\lambda_2 = 0.48$ \footnote{We note the SDE specified by \cite{RENO2001} is different to what \cite{AB1998} has, we followed \cite{AB1998}.}. This model will determine if stochastic volatility and volatility clustering will cause the estimators to differ. The model is given for various levels of missing data in Figure \ref{fig:SPCor}.

\subsubsection{Ornstein Uhlenbeck} \label{sssec:OU}

The bivariate Ornstein Uhlenbeck process satisfies the following SDEs:
\begin{equation} \label{eq:OU:1}
  dp_i(t) = \theta_i(\mu_i - p_i(t))dt + \sigma_i dW_i(t), \ \ \ \ i = 1, 2,
\end{equation}
where $\mathrm{Corr}(dW_1, dW_2) = \rho_{12}$ $\in (-1, 1)$. A sample path of 10,000 seconds is simulated starting at R100 with parameters $\mu_1 = \mu_2 = 100$, $\sigma_1^2 = 0.1$, $\sigma_2^2 = 0.2$, $\theta_1 = 0.035$ and $\theta_2 = 0.054$. This model compares the two estimators when there is mean-reversion.
\begin{figure*}[t!]
\begin{subfigure}{0.5\textwidth}
\includegraphics[width=\textwidth]{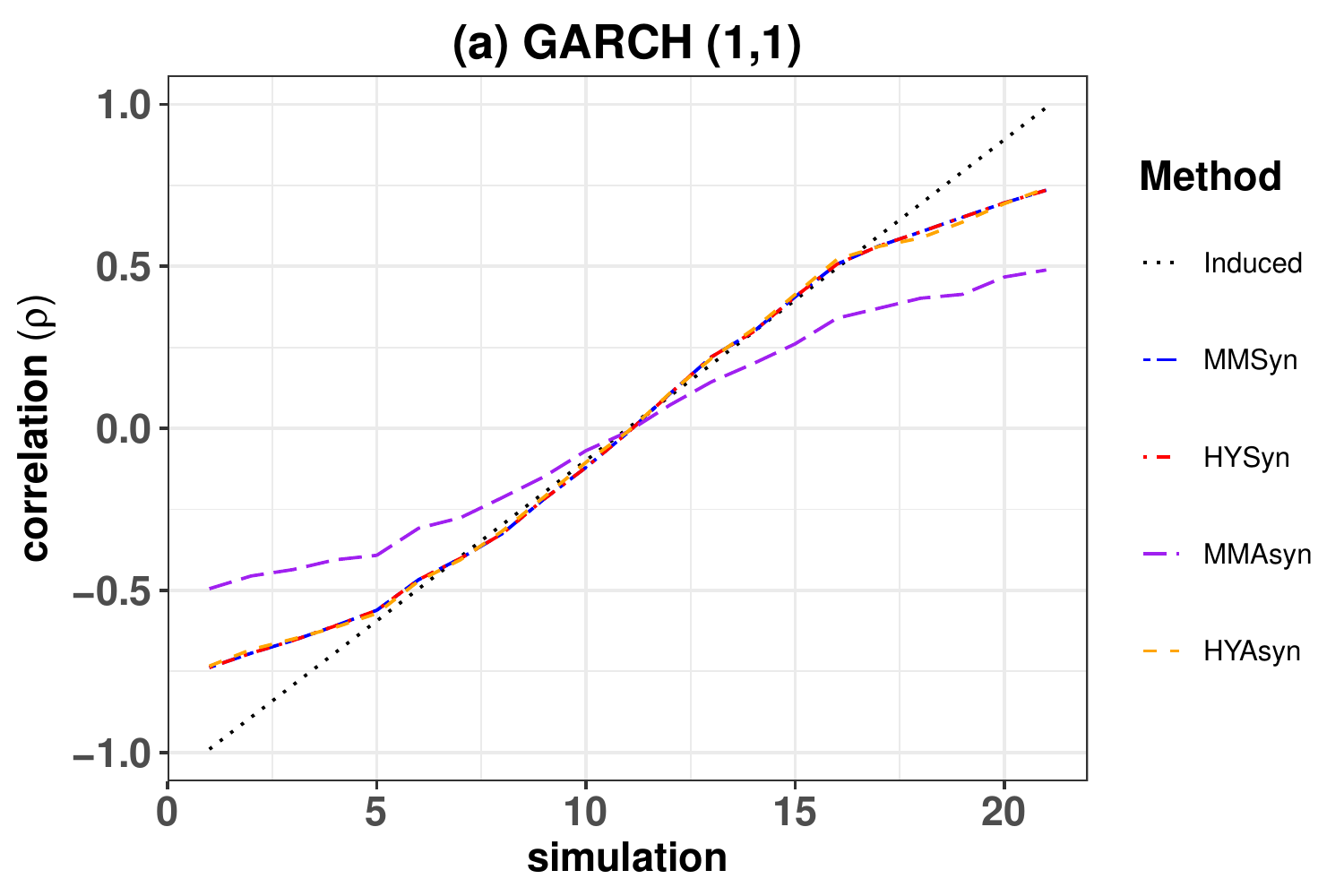}
\end{subfigure}
\hfill
\begin{subfigure}{0.5\textwidth}
\includegraphics[width=\textwidth]{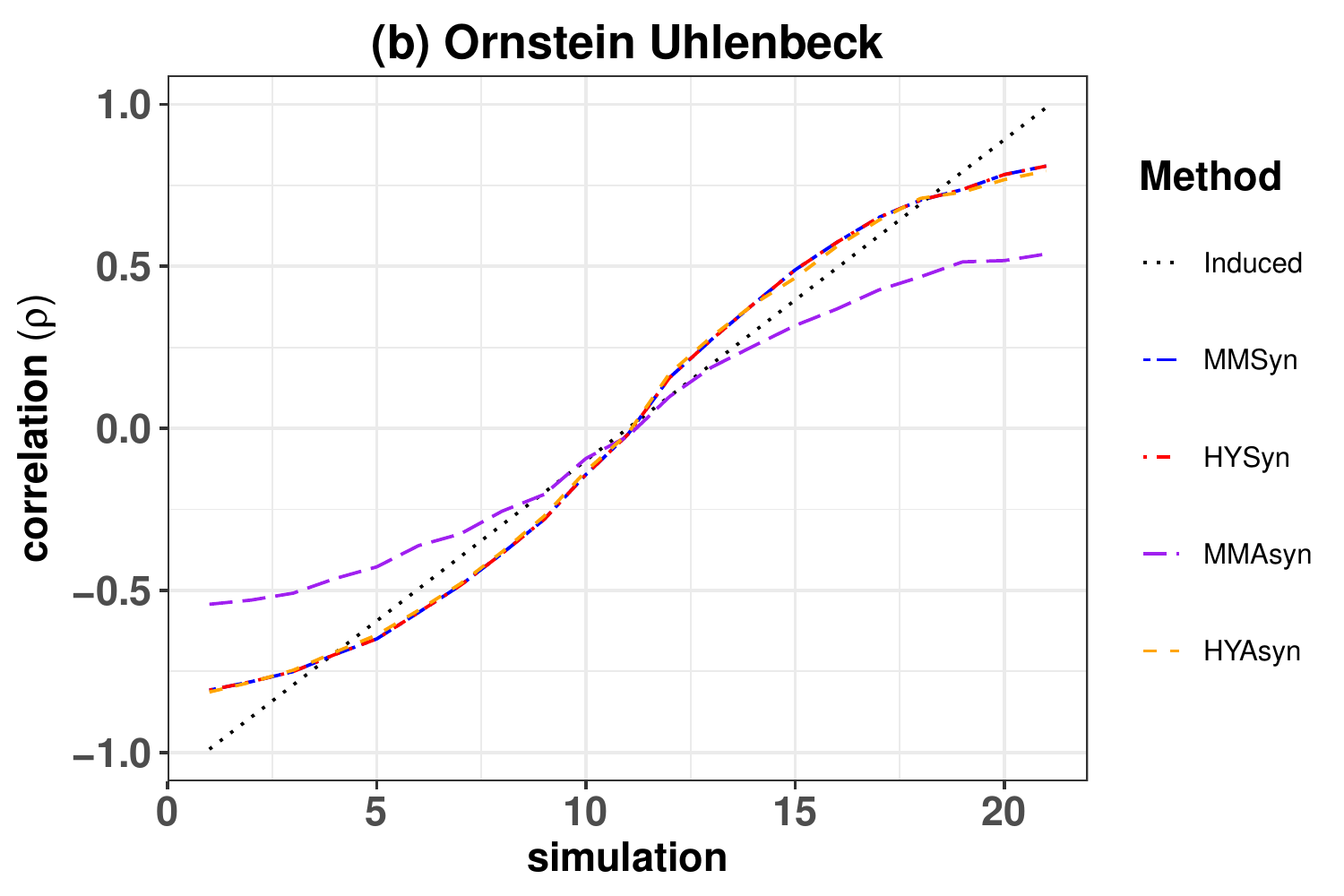}
\end{subfigure}
\caption[The effect of volatility clustering and mean-reversion.]{Comparing the two estimators based on volatility clustering (a) and mean-reversion (b). As per the figure legend, the blue and purple dotted lines are the MM estimator estimated using Algorithm \ref{algo:ComplexFT} under synchronous and asynchronous observations respectively. The red and orange dotted line is the HY estimator estimated using Algorithm \ref{algo:HY} under synchronous and asynchronous observations respectively. The black dotted line is the induced correlation from the various stochastic processes. (a) is 10,000 seconds simulated from a bivariate GARCH(1,1) satisfying \eqref{eq:Garch:1} and numerically simulated using Algorithm \ref{algo:Garch}. (b) is from a bivariate Ornstein Uhlenbeck satisfying \eqref{eq:OU:1} and numerically simulated using Algorithm \ref{algo:OU}. The figure shows that both estimators produce the same estimates in the synchronous case while for the asynchronous case, HY recovers the synchronous estimates but the correlation drops for MM. The figures can be recovered using the R script file \href{https://github.com/rogerbukuru/Exploring-The-Epps-Effect-R/blob/master/Monte\%20Carlo\%20Plots/SDE2.R}{SDE2.R} on the GitHub resource \cite{PCRBTG2019}.}
\label{fig:SPCor}
\end{figure*}

Figure \ref{fig:SPCor}, the asynchrony is induced by down-sampling each price path by 20\%. Once again, the synchronous MM (blue dotted line) is the same as the synchronous HY (red dotted line) and the asynchronous HY (orange dotted line) recovers the synchronous estimates while the asynchronous MM (purple dotted line) has a lower correlation estimate than the synchronous case. Figure \ref{fig:SPCor} provides the insight that neither volatility clustering nor mean-reversion causes the two estimators to differ under synchronicity. The two estimators only seem to produce different estimates under asynchronous conditions.


\subsection{Arrival time representation} \label{ssec:asynchrony}

In the previous experiments, asynchrony is induced by removing observations from the sample paths: the ``missing data" approach. The missing data approach is informative because it can be easily applied to uniformly sampled data to investigate the impact on the bias. It does not reflect the random nature of the time of event arrivals. For this reason we will now follow a similar methodology to that used by \cite{RENO2001} and \cite{PI2007}. The focus will be to achieve asynchronous sample paths that behave more like tick-by-tick TAQ data and investigate the effect the number of Fourier coefficients (N in \eqref{eq:Der:21}) has on the estimates using random event arrival times. 

\begin{figure*}[t!]
\begin{subfigure}{0.5\textwidth}
\includegraphics[width=\textwidth]{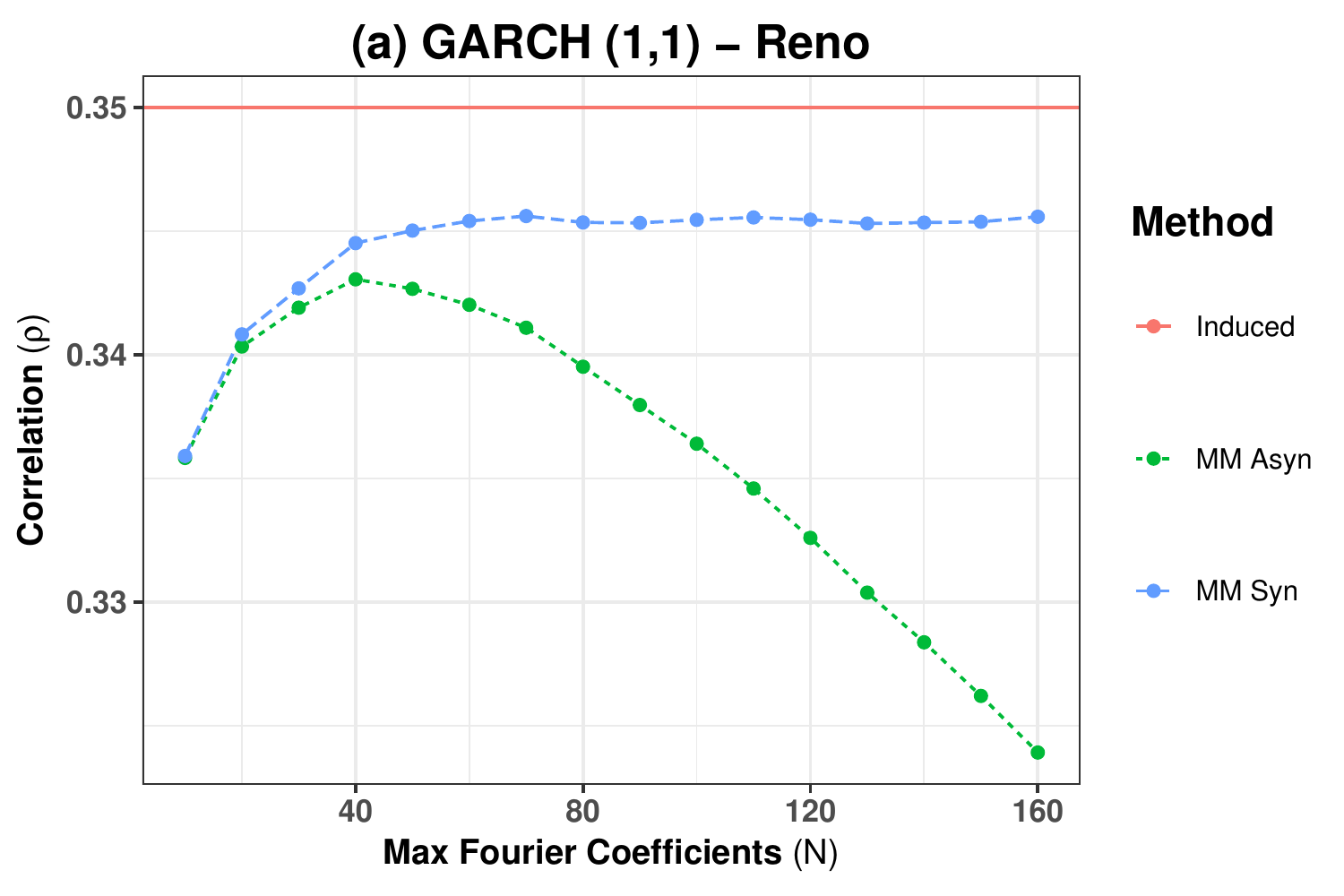}
\end{subfigure}
\hfill
\begin{subfigure}{0.5\textwidth}
\includegraphics[width=\textwidth]{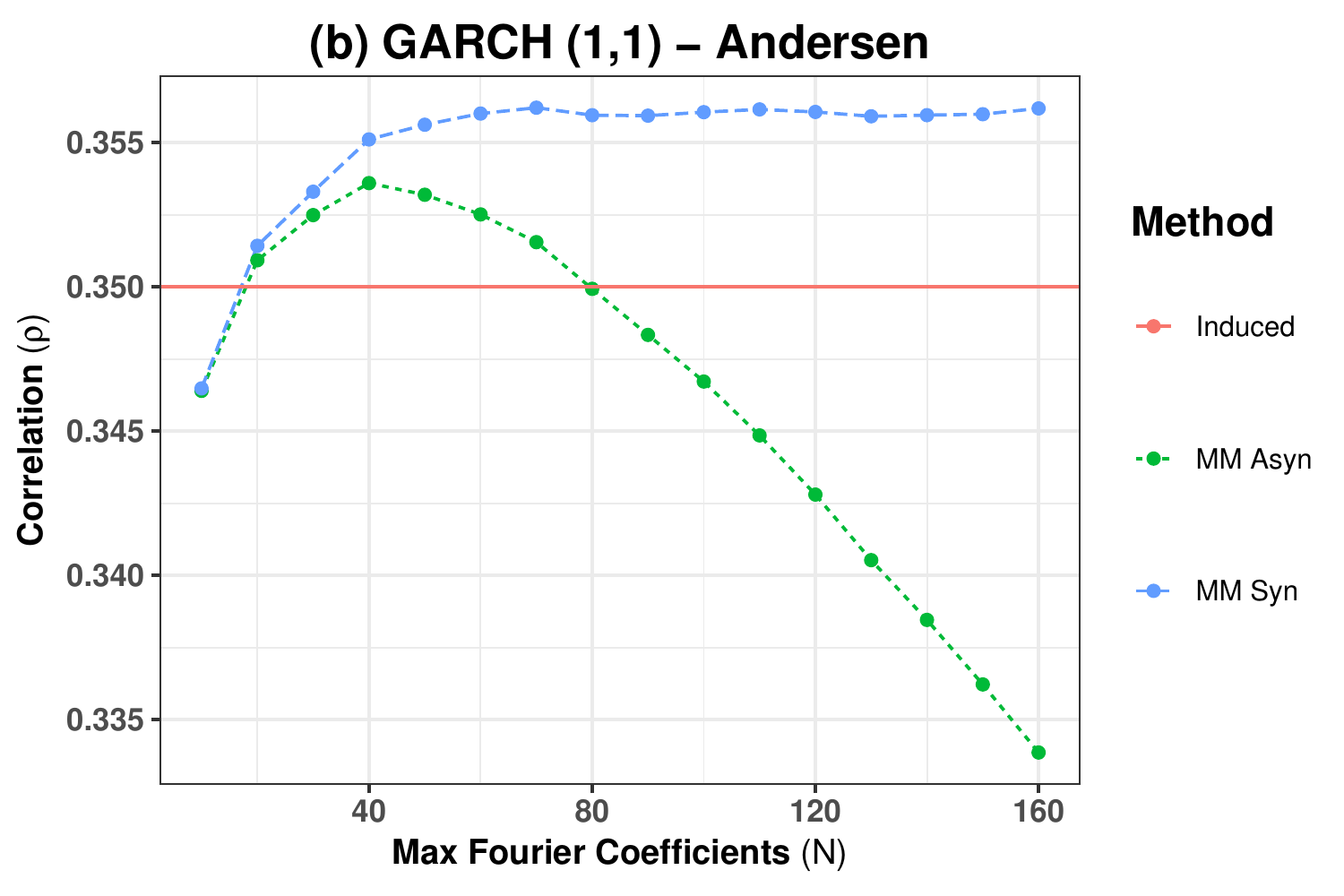}
\end{subfigure}
\caption[Recovering the results from \cite{RENO2001}.]{We recover the result from \cite{RENO2001} using the complex exponential Fourier Transform (See Algorithm \ref{algo:ComplexFT}). The average correlation is plotted as a function of the sampling frequency N in \eqref{eq:Der:21}. Concretely, the asynchronous sample paths for (a) and (b) are exponential inter-arrival time samples from 86,400 seconds of simulated data. The exponential inter-arrival times have a mean of 15 seconds and 45 seconds respectively for asset 1 and asset 2. The synchronous sample paths for (a) and (b) are achieved by forcing the first time series to be observed at the same times as the second time series. (a) is simulated by adjusting \eqref{eq:Garch:1} to how \cite{RENO2001} defined the SDE and implemented by adjusting Algorithm \ref{algo:Garch} accordingly. (b) is simulated from \eqref{eq:Garch:1} using Algorithm \ref{algo:Garch}. As per the figure legend, the green dots and blue dots are the asynchronous and synchronous sample paths estimated using Algorithm \ref{algo:ComplexFT} respectively. The orange line is the induced correlation between \eqref{eq:Garch:1}. The results are obtained through 10,000 replications. The figures can be recovered using the R script file \href{https://github.com/rogerbukuru/Exploring-The-Epps-Effect-R/blob/master/Monte\%20Carlo\%20Plots/Reno\%20Recovery.R}{Reno Recovery.R} on the GitHub resource \cite{PCRBTG2019}.}
\label{fig:RR}
\end{figure*}

We first replicate the results from \cite{RENO2001}. The first interesting replication point is that \cite{RENO2001} used a different specification of the GARCH(1,1) process when compared to the original specification \cite{AB1998} from where the parameters were borrowed. For interest, in Figure \ref{fig:RR} we present both specifications of the GARCH(1,1) model \footnote{Prior to the so-called ``replication crises" many publications did not provide either the data nor the code to generate figures and simulations to ease reproducibility.}. 

The experiment in Figure \ref{fig:RR} is conducted by first simulating price paths of 86,400 realisations (where each realisation can be considered to be 1 second in calendar time) from a bivariate GARCH(1,1) with parameters from above. The asynchrony is induced by sampling the price path with an exponential inter-arrival time with a mean of 15 seconds and 45 seconds from asset 1 and asset 2 respectively. 

The synchronous case is achieved by forcing the first time series to be observed at the same time as the second time series {\it i.e.} the price paths are sampled with the same exponential inter-arrival time with a mean of 45 seconds. For each of the asynchronous and synchronous cases we compute the correlation estimate using Algorithm \ref{algo:ComplexFT} for N ranging from 10 to 160. The effect achieved by studying the range of Fourier coefficients (N) is that it allows various sampling frequencies to be studied through the relationship of the two directly from Fourier analysis itself.

From Figure \ref{fig:RR} an Epps like effect is clearly demonstrated. As N increases, so does the sampling frequency and from the asynchronous case (green dots) it is clear that the correlation drops as the sampling frequency increases. However, for the synchronous case (blue dots), the correlation does not drop as the sampling frequency increases. This is because the stochastic processes we have studied have dimensionless correlation (independent of the sampling intervals). The Epps effect recovered here is due to asynchrony which differs from the original Epps effect presented by Thomas Epps \cite{EPPS1979} where the correlation drops from synchronous observations as the sampling interval decreases. 

Researchers have investigated this - specifically \cite{TK2009} was able to derive an analytical expression for the Epps effect arising from smaller sampling intervals by decomposing the correlation at a time scale $\Delta t$ as a function of lagged auto-correlations and correlations of smaller time scales $\Delta t_0$. While the earlier work of \cite{RENO2001} was extended by \cite{MMZ2011} where they analytically deriving an Epps effect explicitly from asynchrony as a function of $\Delta t$.

\begin{figure*}[p]
\begin{subfigure}{0.5\textwidth}
\includegraphics[width=\textwidth]{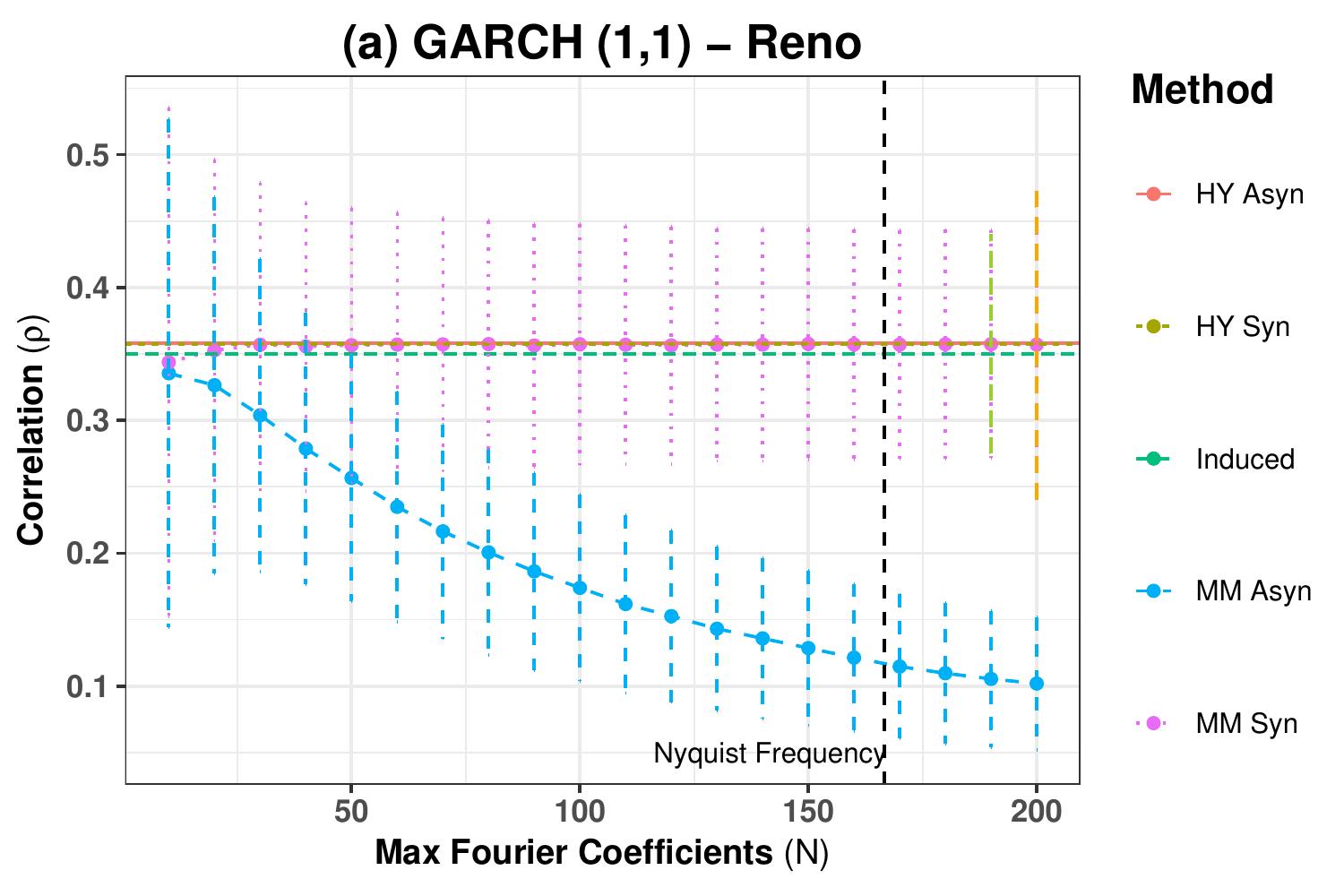}
\end{subfigure}
\hfill
\begin{subfigure}{0.5\textwidth}
\includegraphics[width=\textwidth]{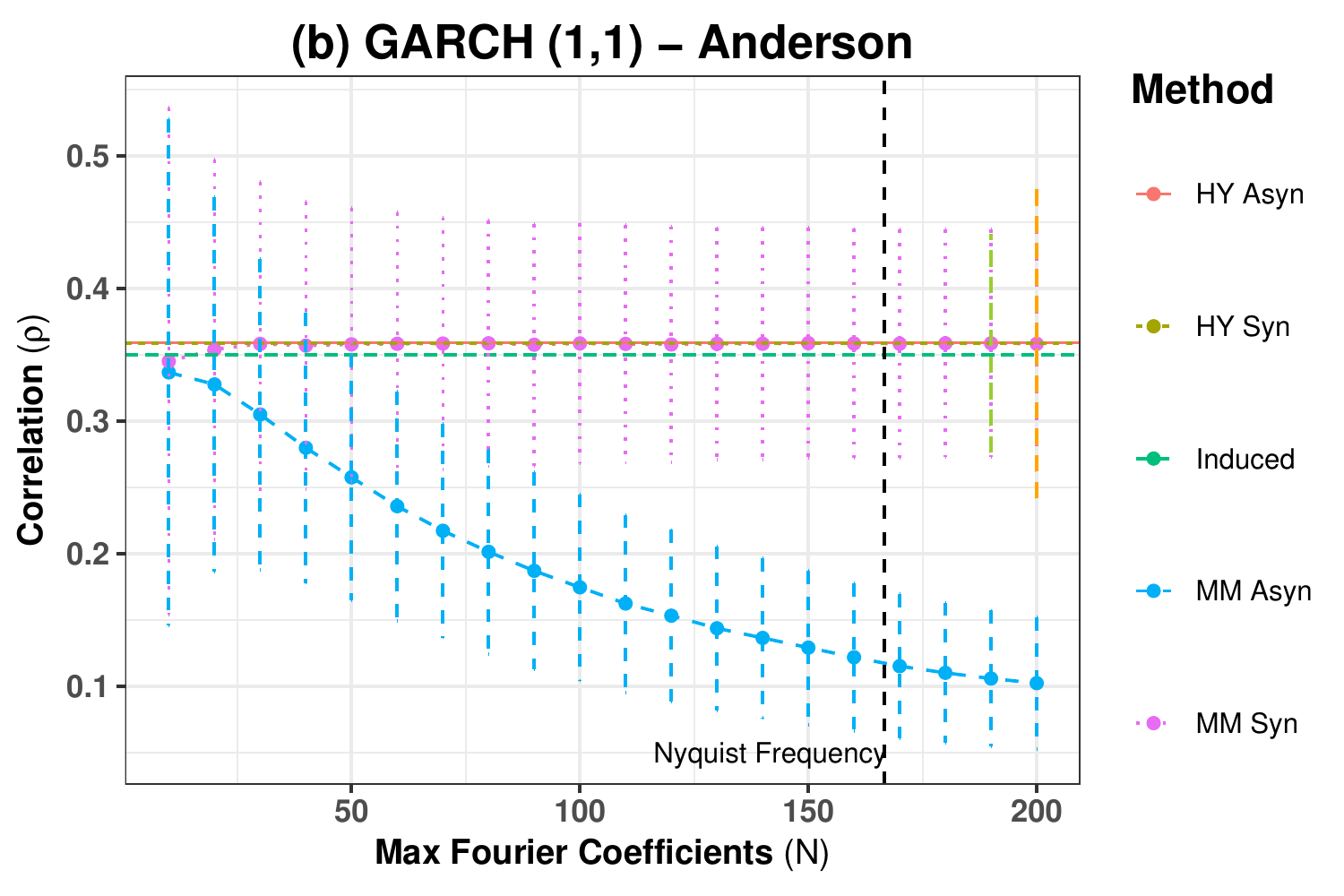}
\end{subfigure}
\begin{subfigure}{0.5\textwidth}
\includegraphics[width=\textwidth]{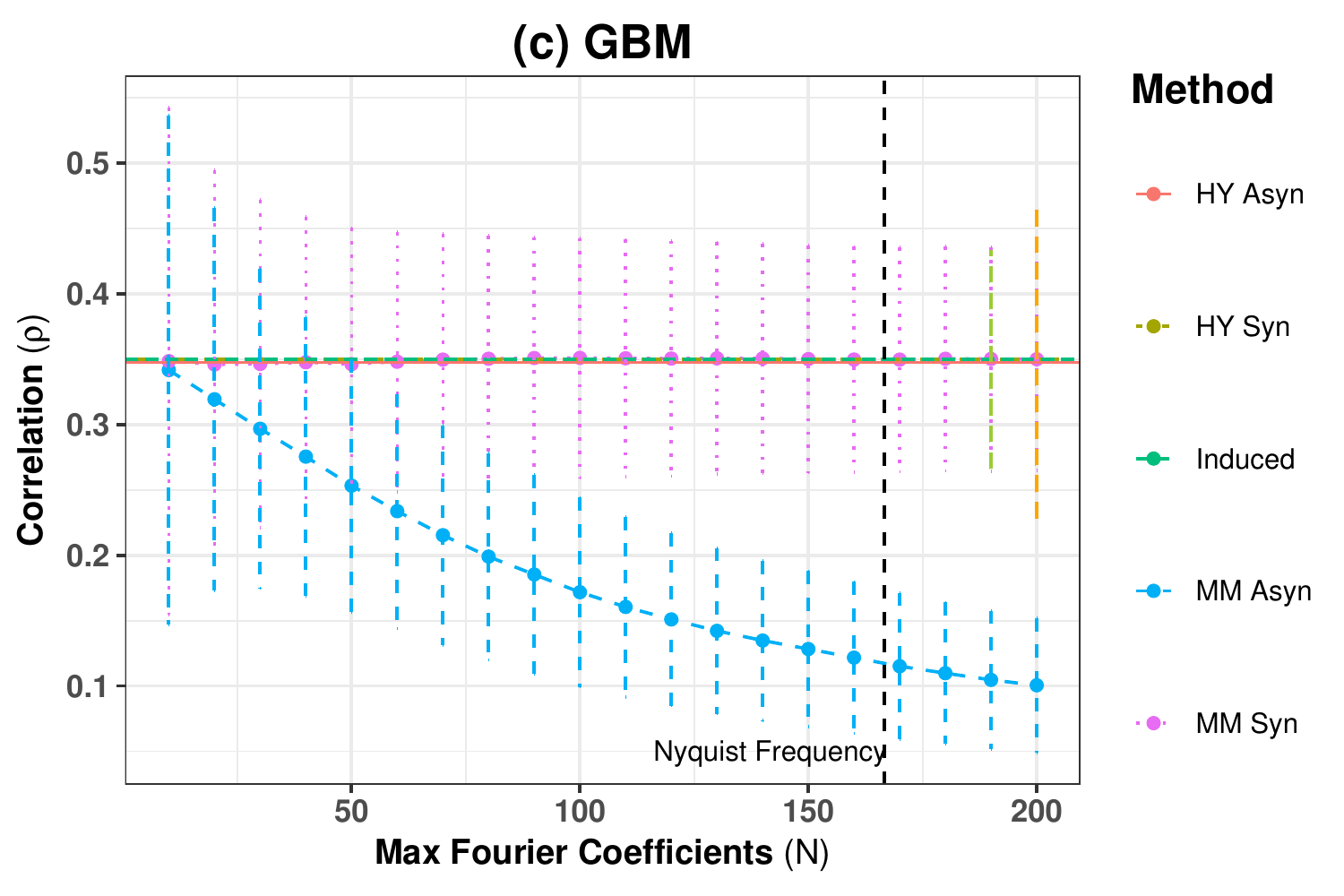}
\end{subfigure}
\hfill
\begin{subfigure}{0.5\textwidth}
\includegraphics[width=\textwidth]{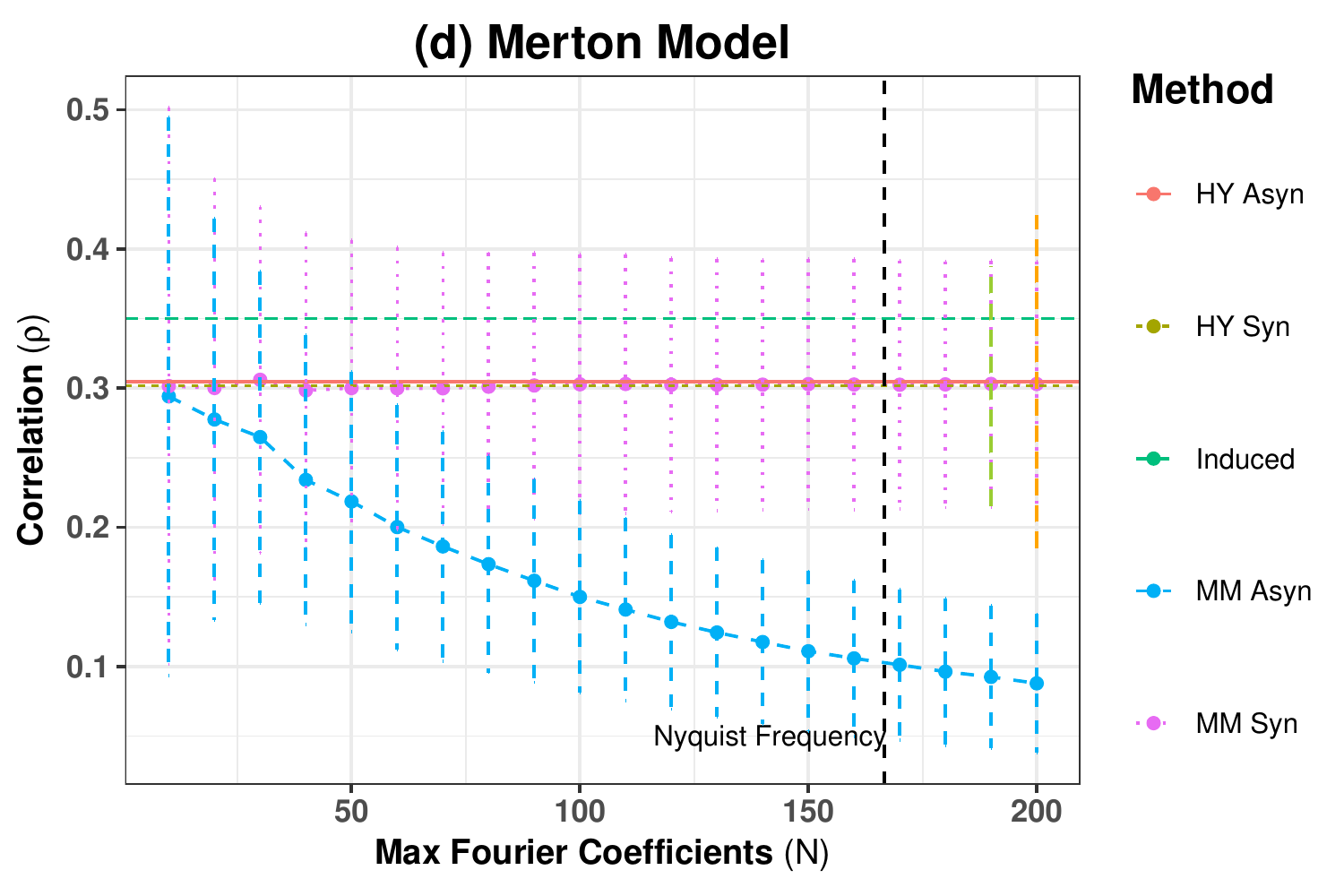}
\end{subfigure}
\begin{subfigure}{0.5\textwidth}
\includegraphics[width=\textwidth]{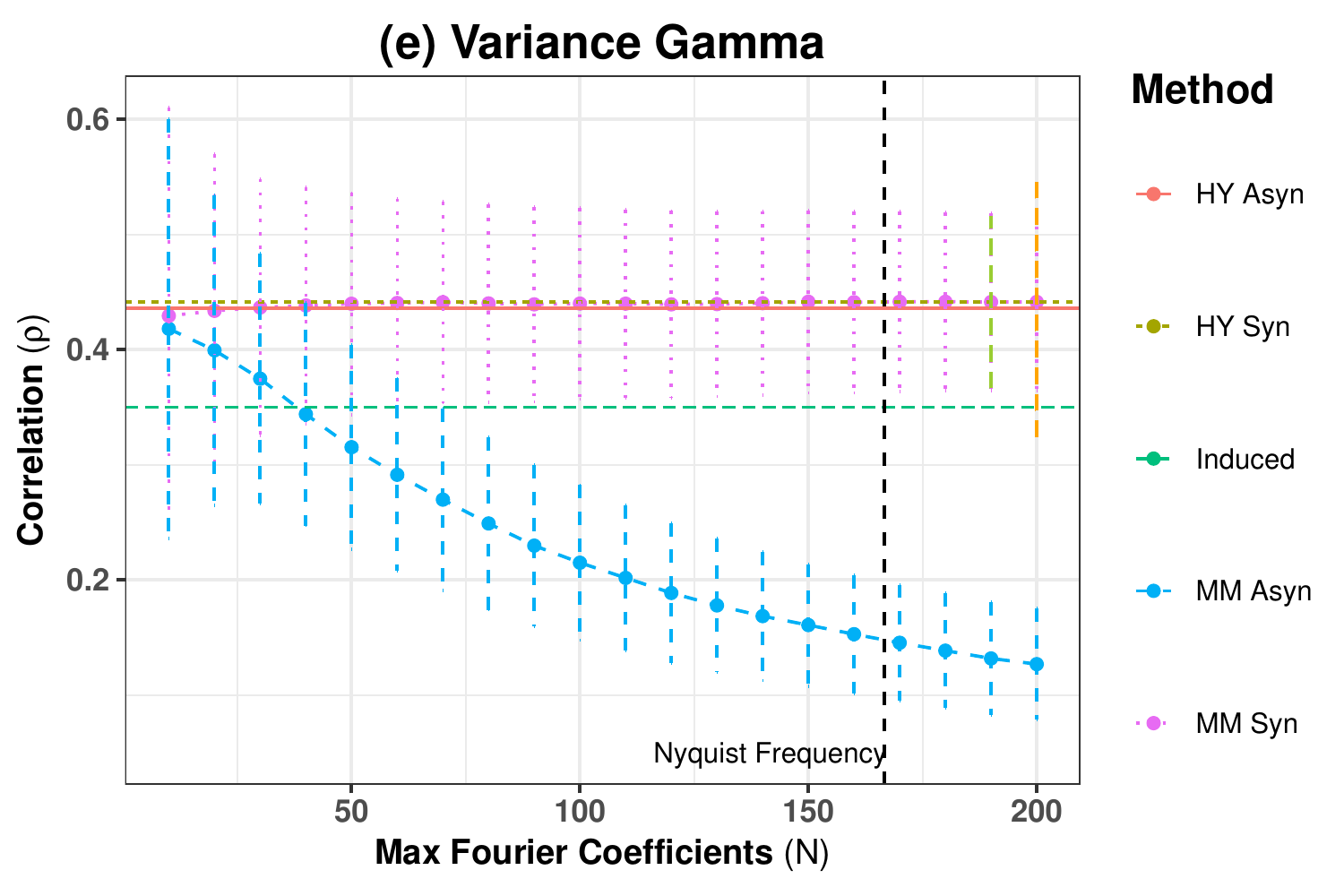}
\end{subfigure}
\hfill
\begin{subfigure}{0.5\textwidth}
\includegraphics[width=\textwidth]{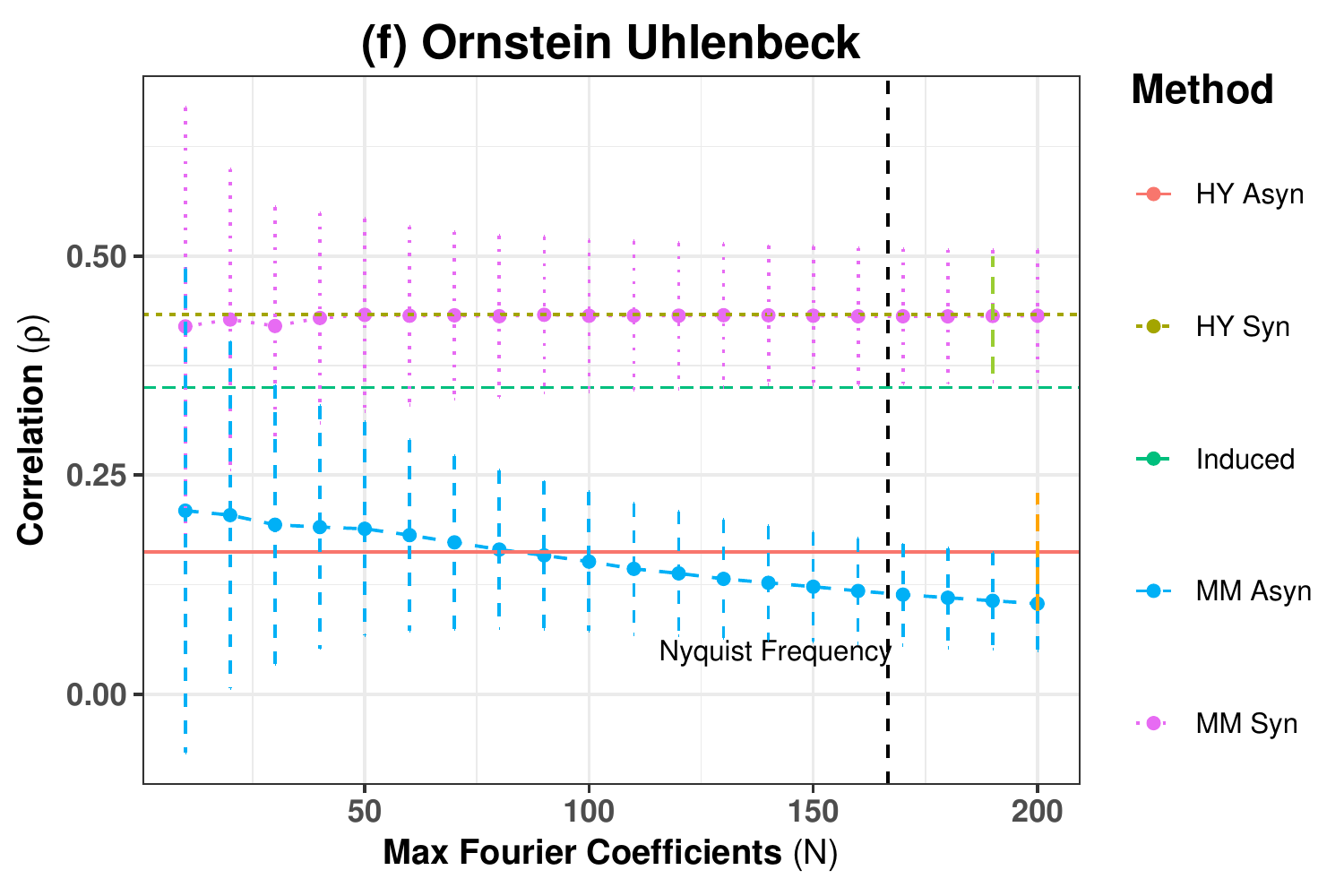}
\end{subfigure}
\caption[The effect of asynchrony.]{Comparing the two estimators by introducing asynchrony through sampling the time series with exponential inter-arrival times. The parameters used are the same as the previous experiments. We note that (a) uses the specification of GARCH (1,1) by \cite{RENO2001} while (b) uses the specification of \cite{AB1998}. As per the legend, the synchronous (pink dots) and asynchronous (blue dots) average MM estimates using Algorithm \ref{algo:ComplexFT} differ more as the number of Fourier coefficients (N) increase. The synchronous (dark green line) and the asynchronous (orange line) average HY estimates using Algorithm \ref{algo:HY} in general recover the same estimates except for (f) - the OU process. The error bars are plotted as the standard deviation of the estimates across the replications. The green line is the induced correlation for each of the processes set to be $\rho = 0.35$. Additionally, the Nyquist frequency is indicated by the black dashed line. The results are obtained through 1,000 replications. The figures can be recovered using the R script file \href{https://github.com/rogerbukuru/Exploring-The-Epps-Effect-R/blob/master/Monte\%20Carlo\%20Plots/Reno\%20Extended.R}{Reno Extended.R} on the GitHub resource \cite{PCRBTG2019}.}
\label{fig:RE}
\end{figure*}

Drawing attention back to comparing the two estimators: we modify the experiment slightly. Specifically, for Figure \ref{fig:RE}, the experiment is conducted by first simulating price paths of 10,000 seconds from the various stochastic processes from above, using their respective parameters as before \footnote{The Merton model has $\lambda_1 = \lambda_2 = 0.2$}. The asynchrony is induced by sampling the first asset with an exponential inter-arrival time with mean 30 seconds and the second asset with a mean of 45 seconds. The synchronous case here is achieved by forcing the first time series to be observed at the same time as the second time series. The rationale behind adjusting the experiment is so that the Nyquist frequency can be indicated. This gives us a view of the impact of aliasing. 

From Figure \ref{fig:RE}, the Nyquist frequency is calculated based on the \textit{average} sampling frequency, this is not the true cutoff required to avoid any aliasing. The true cutoff required is computed by first finding the highest sampling frequency present in the data, then computing the corresponding Nyquist frequency. Using the true cutoff is how we compute all the MM estimates in this paper except in Figure \ref{fig:RR} and \ref{fig:RE}. The rationale behind this is that we are trying to study the co-movement between high-frequency events, therefore picking a lower N to avoid market micro-structure noise will lead to the aliasing of the event data. Picking a lower N creates a smoothing effect due to aliasing of higher frequencies. This is useful to identify the true signal under the market micro-structure noise argument \cite{MS2008}. For identifying the co-movement of events picking a lower N is a bad choice.

Using the appropriate cutoff for the asynchronous cases in Figure \ref{fig:RE} results in all the correlations for the MM estimate diminishing to zero. This result, when combined with the drop in correlation as the \% of missing data increased in Figure \ref{fig:GBMcor}, indicates the important role played by the level of asynchrony in reducing correlation as a function time scale \cite{PI2007}.

From Figure \ref{fig:RE}, it is clear that for the MM estimates, the correlations decrease for the asynchronous case as the number of Fourier coefficients (N) increase; whereas for the synchronous case the correlations become closer to the synchronous HY estimates as N increases. The error bars are calculated to be the standard deviation from the estimates and decrease as N increases. This shows that the estimates become more accurate with more Fourier coefficients. The HY estimates are not a function of N, but rather provide a baseline to compare the MM estimate against. 

For Figure \ref{fig:RE} (a) through to (e), the asynchronous HY estimate recovers the synchronous HY estimate which is expected as Hayashi and Yoshida have claimed that the estimator is immune to the Epps effect - as represented by asynchrony. However, the HY estimator demonstrates an Epps effect when using the Ornstein Uhlenbeck (OU) process. This was not picked up by the first missing data experiments because of the nature of decimation. In the previous experiments asynchrony was induced through a missing data manner decimating uniformly sampled data, while in this experiment, the asynchrony is induced through exponential inter-arrival times to sample the price paths non-uniformly. Combined with the mean-reversion from the OU process, the sampling can pick up different co-movements between the price paths as a function of sampling frequency. An important point here is that the combination of sampling method and mean-reversion can lead to spurious lead-lag relations due to the high levels of asynchrony. The lead-lag representation of the Epps effect was investigated by \cite{RENO2001, MMZ2011}. This again highlights the compromised nature of the HY estimator for high-frequency finance applications. 

Although the HY estimate may be immune to the Epps effect in some missing data representations, it is not immune to the Epps effect arising from either the lead-lag representation \cite{GO2008} nor from smaller synchronous sampling intervals \cite{EPPS1979}. This later perspective will be further demonstrated in real financial data analysis using temporal averaging in later sections. When levels of asynchrony are high - common for high-frequency data, the HY estimator has the effect of deleting observations \cite{SFX2010}.

We have still presented results simulated from the classical continuous-time stochastic process perspective. This remains inconsistent with a truly the event-based view. It should be noted that one more realistic simulation approach could be to extend our results to study how the correlations behave when the process is simulated using a multivariate Hawkes point process \cite{HAWKES1971,BDHM2013,BMM2015}. Specifically \cite{BDHM2013} who considered lead-lags in the context of multiple scales and their impact on correlations to recover an Epps like effect. 

Researchers have also argued in the literature for the efficacy of the MM estimator from the market micro-structure noise perspective. An example is \cite{MS2008} who show that the MM estimator is unbiased for contaminated price processes with an appropriate choice of n and N, the given sample size and the number of Fourier coefficients respectively. Other authors have pointed out that the HY estimator is simply infeasible in the setting of market micro-structure noise \cite{SFX2010}. Once again, our critique of this relates to the underlying trading processes themselves and how this seems to be driven by order-flow. Some resolution of this could be achieve by comparing the relative impact of Hawkes simulated limit order flow with respect to combinations of limit order and market order-flows on various averaging scales, simulated with multivariate D-type Hawkes processes, perhaps using the approach of Large \cite{L2007}, but for multiple assets and trade types into a simulated order-book that implements price discover as a function of market structure and rules.

Finally, the SDE's chosen for this paper, by construction, have dimensionless correlations which does not depend on time. For this reason we could not directly study the Epps effect arising from smaller sampling intervals using our Monte Carlo experiments. However, from Figure \ref{fig:RR} and \ref{fig:RE} we use the Fourier methods to study smaller sampling intervals. It is known that correlations arising from asynchrony not only depends on the level of asynchrony but also on the sampling intervals chosen \cite{MMZ2011}. It then seems that the Epps effect arising from smaller sampling intervals and asynchrony are related. It then makes sense that there is a requirement to attempt to decompose these two factors and how they can contribute towards an Epps like effect. 

There is a potential problem. It has been analytically shown that an Epps effect arising from asynchrony, as a function of $\Delta t$, can be modeled by considering the correlation directly in terms of the sampling interval \cite{MMZ2011}:
\begin{equation} \label{eq:Mas}
	\tilde{\rho}_{\Delta t}^{12}=c\left(1+\frac{1}{\lambda \Delta t}\left(e^{-\lambda \Delta t}-1\right)\right).
\end{equation}
However the correlation only decreases and does not change signs. When studying real world data using different approaches to time averaging (See Section \ref{sec:averaging}) one can find examples of changes in the correlation at a particular scales and it is not always decreasing, but can also becoming positively correlated with a change in scale. This may be indicative of structural change in the underlying market as a function of averaging scales, above and below which the market dynamics changes, or merely random variations in the order flow itself?


\section{Data Description} \label{sec:description}

The data used in the analysis was obtained from Bloomberg \footnote{The TAQ data obtained from Bloomberg extracts time stamps up to the second and some trades, from either sampling or block trades effects, can have the same time stamp for which we aggregate using Algorithm \ref{algo:aggregation}.} and consists of Trade and Quote (TAQ) data from 10 equities listed on the Johannesburg Stock Exchange (JSE). The period considered is the week from 31/05/2019 to 07/06/2019. The equities considered are: FirstRand Limited (FSR), Shoprite Holdings Ltd (SHP), Absa Group Ltd (ABG), Nedbank Group Ltd (NED), Standard Bank Group Ltd (SBK), Sasol Ltd (SOL), Mondi Plc (MNP), Anglo American Plc (AGL), Naspers Ltd (NPN) and British American Tobacco Plc (BTI).
From the TAQ data, we only extract the Automated Trades (AT) that form the continuous trading process \cite{JSE}. We removed all overnight TAQ data and auction related data. We also made the pragmatic choice to remove the overnight returns across days.

\begin{table}[H]
\centering
\begin{tabular}{rrrrr}
  \hline
Tickers & ADV & $\hat{\sigma}_{D, \text{1HR}}^2$ & $\hat{\sigma}_{D, \text{10MIN}}^2$ & $\hat{\sigma}_{D, \text{1MIN}}^2$ \\ 
  \hline
BTI & 628652.6 & 0.0075 & 0.0135 & 0.0501 \\ 
  NPN & 558210.8 & 0.0017 & 0.0093 & 0.0494 \\ 
  AGL & 1150362.2 & 0.0040 & 0.0107 & 0.0341 \\ 
  (**)MNP & 340381.4 & 0.0059 & 0.0105 & 0.0457 \\ 
  SOL & 1209754.6 & 0.0057 & 0.0206 & 0.0868 \\ 
  SBK & 1885551.0 & 0.0037 & 0.0151 & 0.0664 \\ 
  NED & 903670.8 & 0.0048 & 0.0195 & 0.0821 \\ 
  ABG & 1321528.8 & 0.0080 & 0.0239 & 0.1080 \\ 
  SHP & 751731.0 & 0.0104 & 0.0200 & 0.0919 \\ 
  (*)FSR & 7698648.0 & 0.0055 & 0.0207 & 0.0715 \\ 
   \hline
\end{tabular}
\caption{The table indicates the Average Daily Volume (ADV) for the various tickers over the period of consideration where (**) and (*) indicate the least and most liquid ticker respectively. The table further shows the daily volatility for the various tickers by re-scaling the volatility computed using Algorithm \ref{algo:ComplexFT} from Figure \ref{fig:Closing} (a) to (c) with $\sqrt{24}, \sqrt{6*24}$ and $\sqrt{60*24}$ respectively. The fact that the re-scaled volatilities do not line up demonstrate that the price discovery in a continuous double auction market is not well represented as a Brownian motion.}
\label{tab:vol}
\end{table}


\section{Time Averaging} \label{sec:averaging}

In the analysis, we showcase the existence of the Epps effect in the JSE by plotting correlation heat-maps and indicating the average correlation magnitude plus minus one sample standard deviation of the correlation magnitude for various sampling intervals and various aggregation methods. We are able to demonstrate that lead-lag and asynchrony are insufficient in explaining the entirety of the Epps effect, but we also argue that there can be structural and hierarchical effects \cite{WG2014} where there are mixtures of correlations from different bottom-up and top-down effects that can create structural changes in correlations. If this is the case it we might require a multi-scale approach to covariance estimation, as suggested in \cite{Z2010}, to eliminate the notion of ``at the same time" \footnote{The multi-scale temporal perspective suggested in \cite{Z2010} remains viewed through the lens of the market micro-structure noise paradigm where the Epps effects is really biases due to asynchronicity and micro-structure noise; rather than an artifact of the world being fundamentally discrete and asynchronous with leads-lags in the averaging when event time is mapped into calendar time.}. Here we prefer to first alter the notion of time and link it to how we carrying event averaging for modeling. We would need to agree on what clock to using in a financial market analysis.  

We first consider the calendar time approach by averaging into calendar time bars. We then turn our attention to the 
intrinsic time paradigm; which presents significant statistical advantages. The benefits include the partial recovery of Normality and the IID assumptions, and it can deal with the issue of random and asynchronous nature of streaming trade data \cite{ELO2012A}. Furthermore, under this paradigm, the framework provided by \cite{DERMAN2002} can provide a link to recover the correlation in calendar time using the correlation computed from intrinsic time. However implicit in this perspective is an averaging operation that aggregates trade events and redefines the notion of both global and local times for the market being considered. Finally, we consider an approach to volume time averaging.

We refer the reader to our supporting materials and code on GitHub \cite{PCRGTG2019REPORT} for further details on the construction of the various data sets and data aggregation methods.

\begin{figure*}[hbt!]
\centering
\begin{subfigure}{0.245\textwidth}
\includegraphics[width=\textwidth]{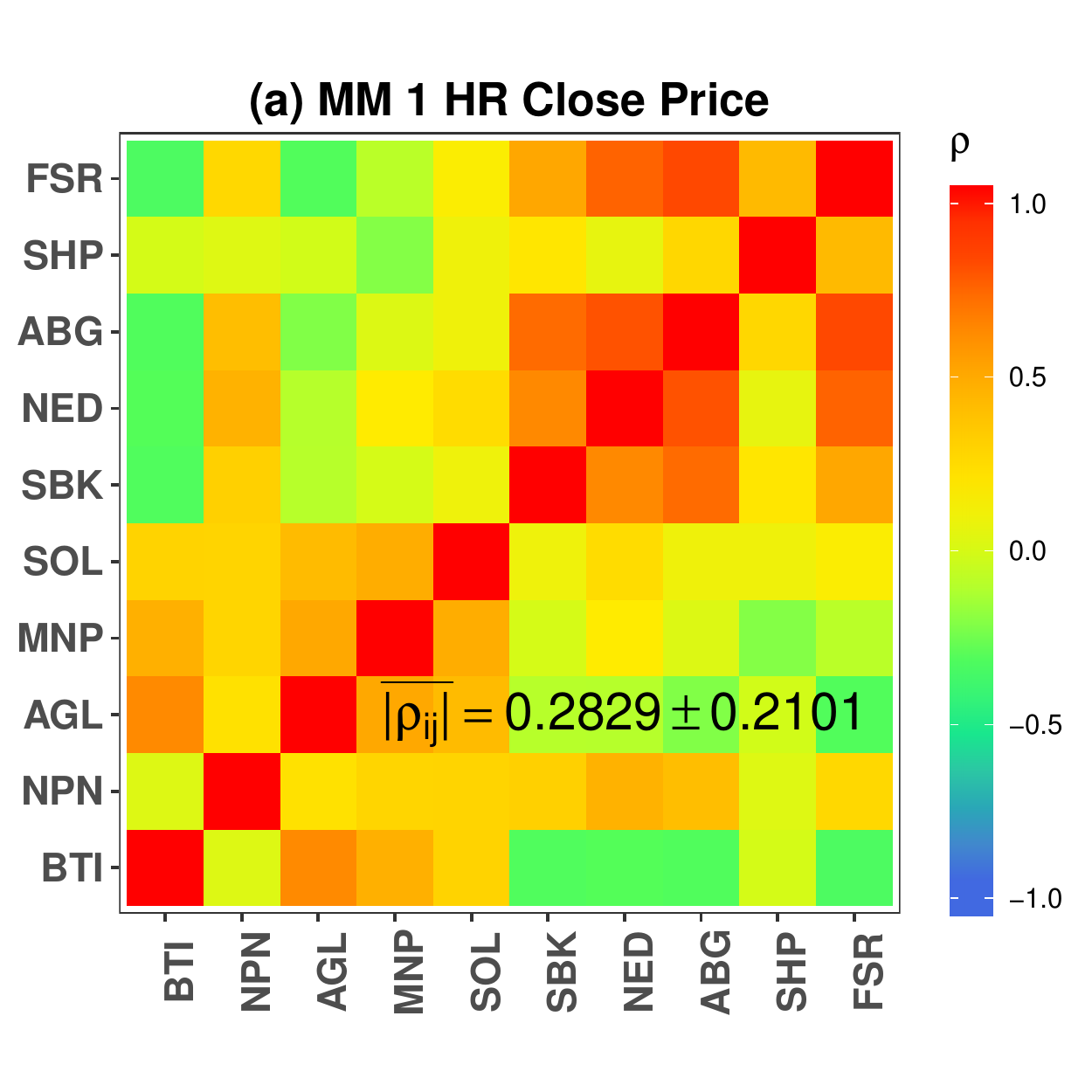}
\end{subfigure}
\begin{subfigure}{0.245\textwidth}
\includegraphics[width=\textwidth]{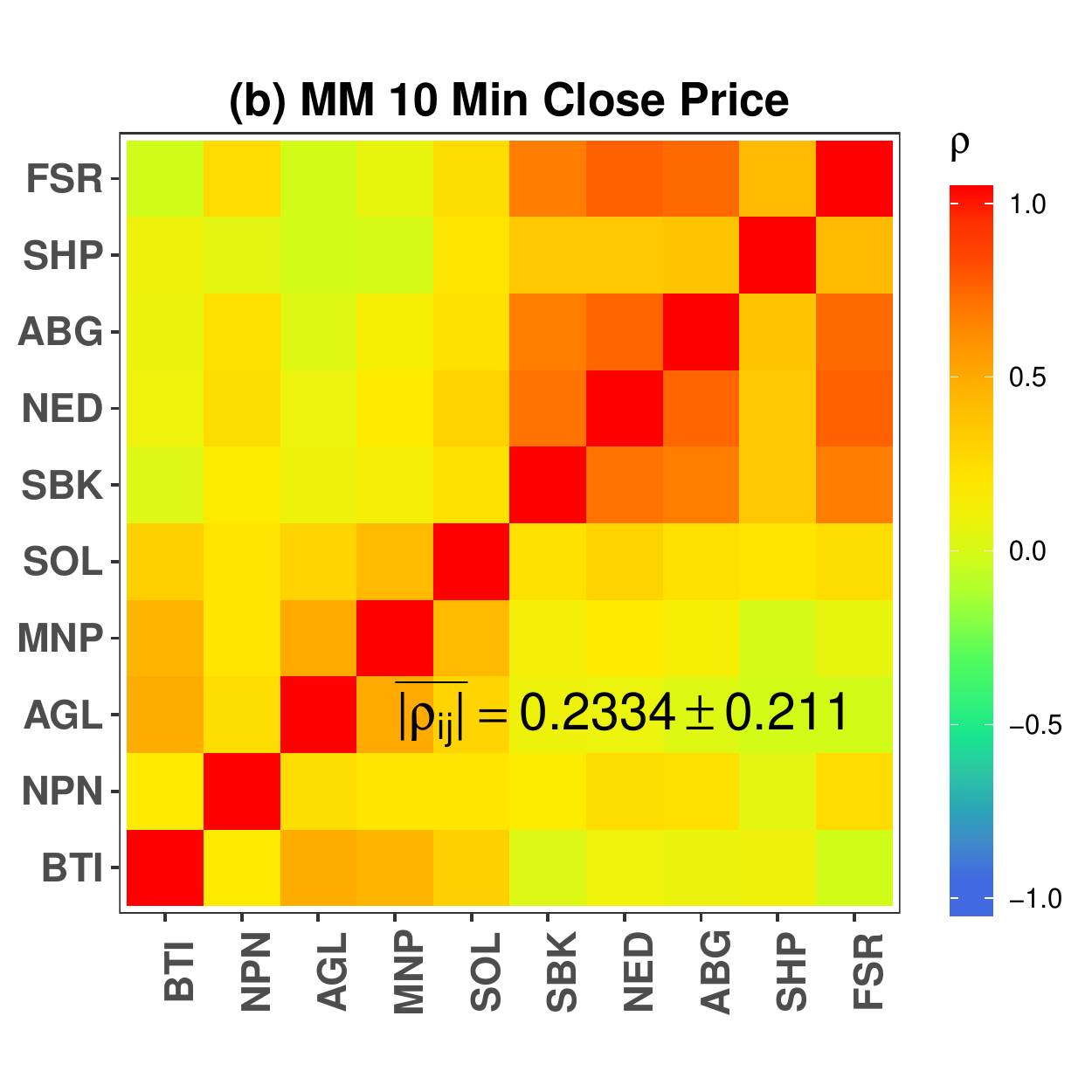}
\end{subfigure}
\begin{subfigure}{0.245\textwidth}
\includegraphics[width=\textwidth]{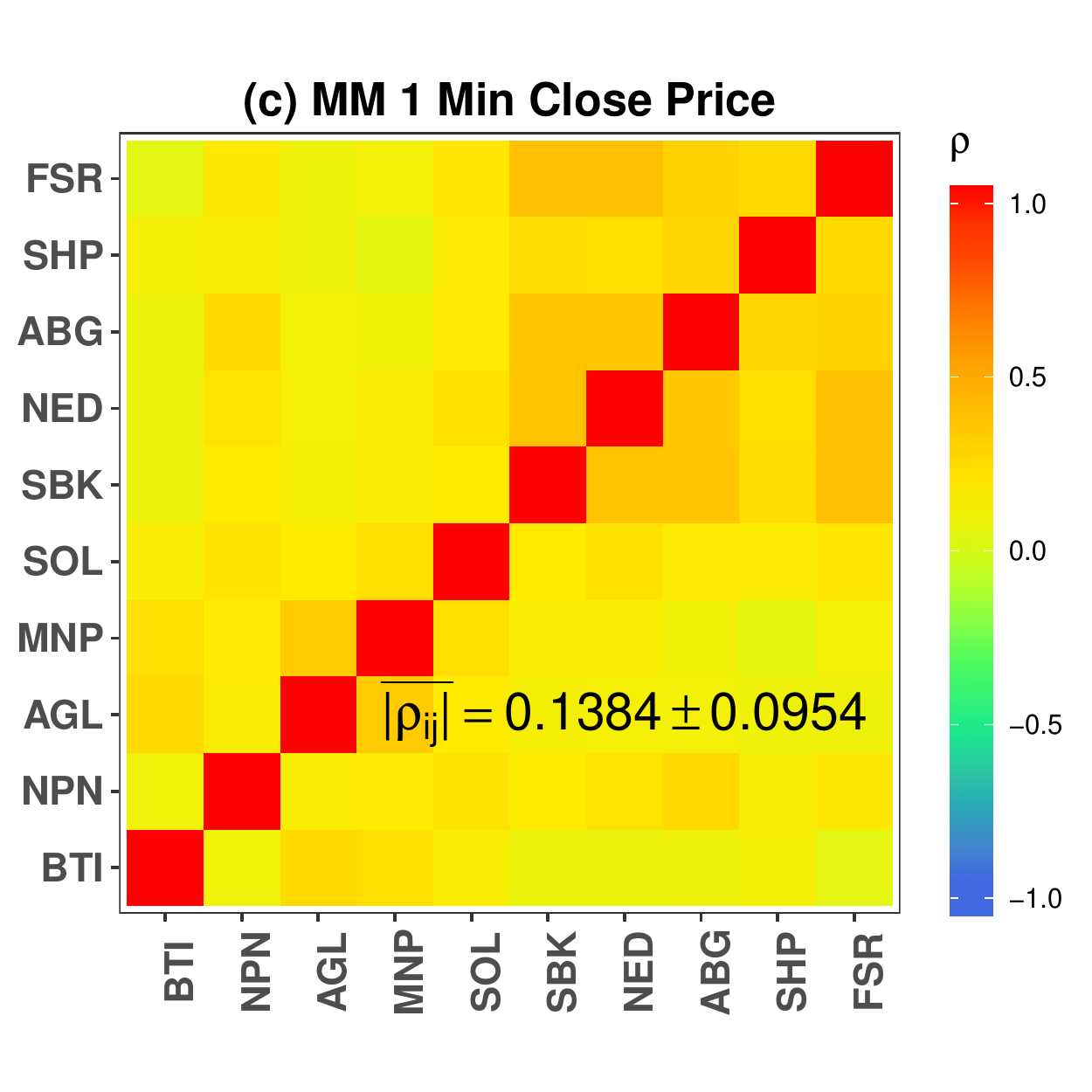}
\end{subfigure}
\begin{subfigure}{0.245\textwidth}
\includegraphics[width=\textwidth]{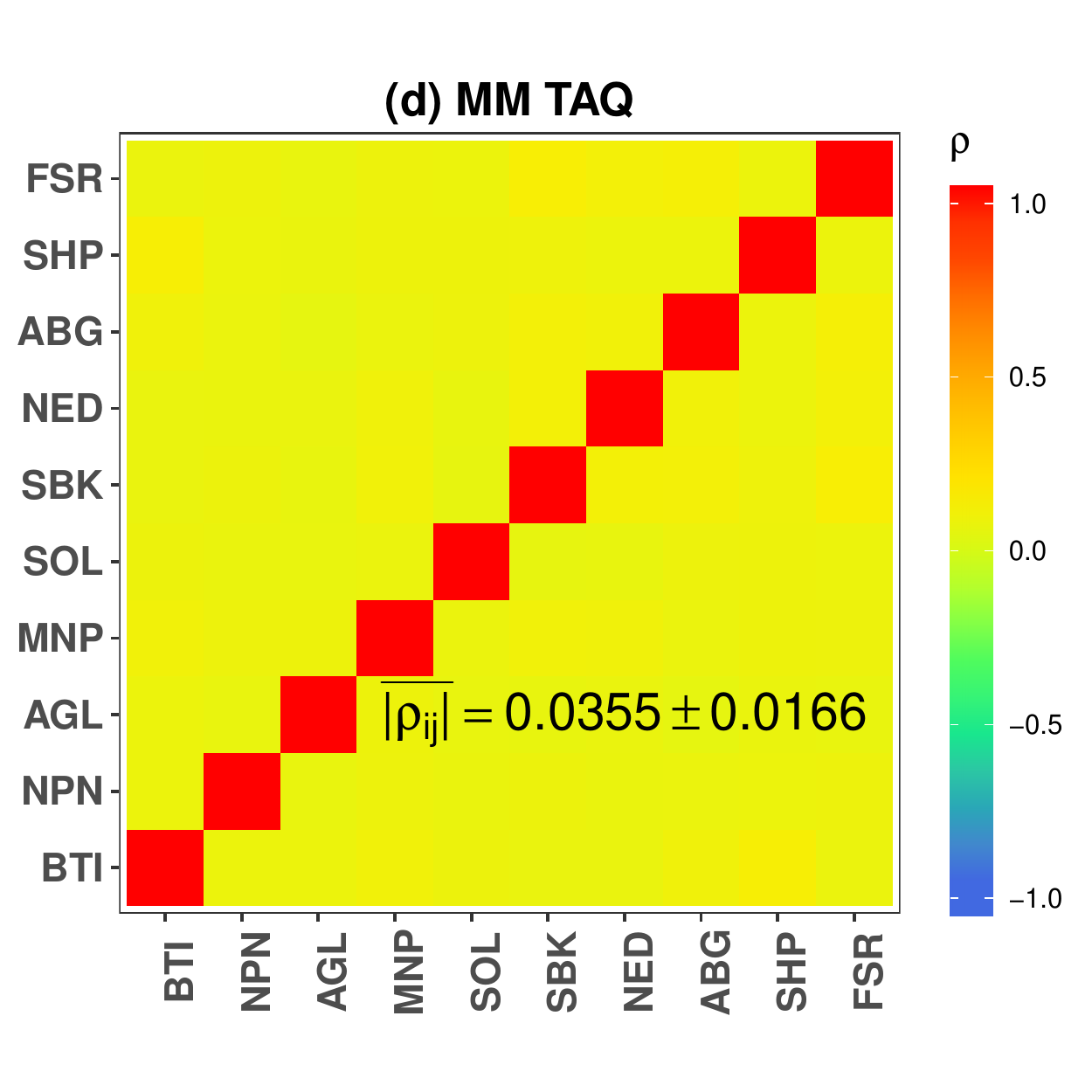}
\end{subfigure} \\
\begin{subfigure}{0.245\textwidth}
\includegraphics[width=\textwidth]{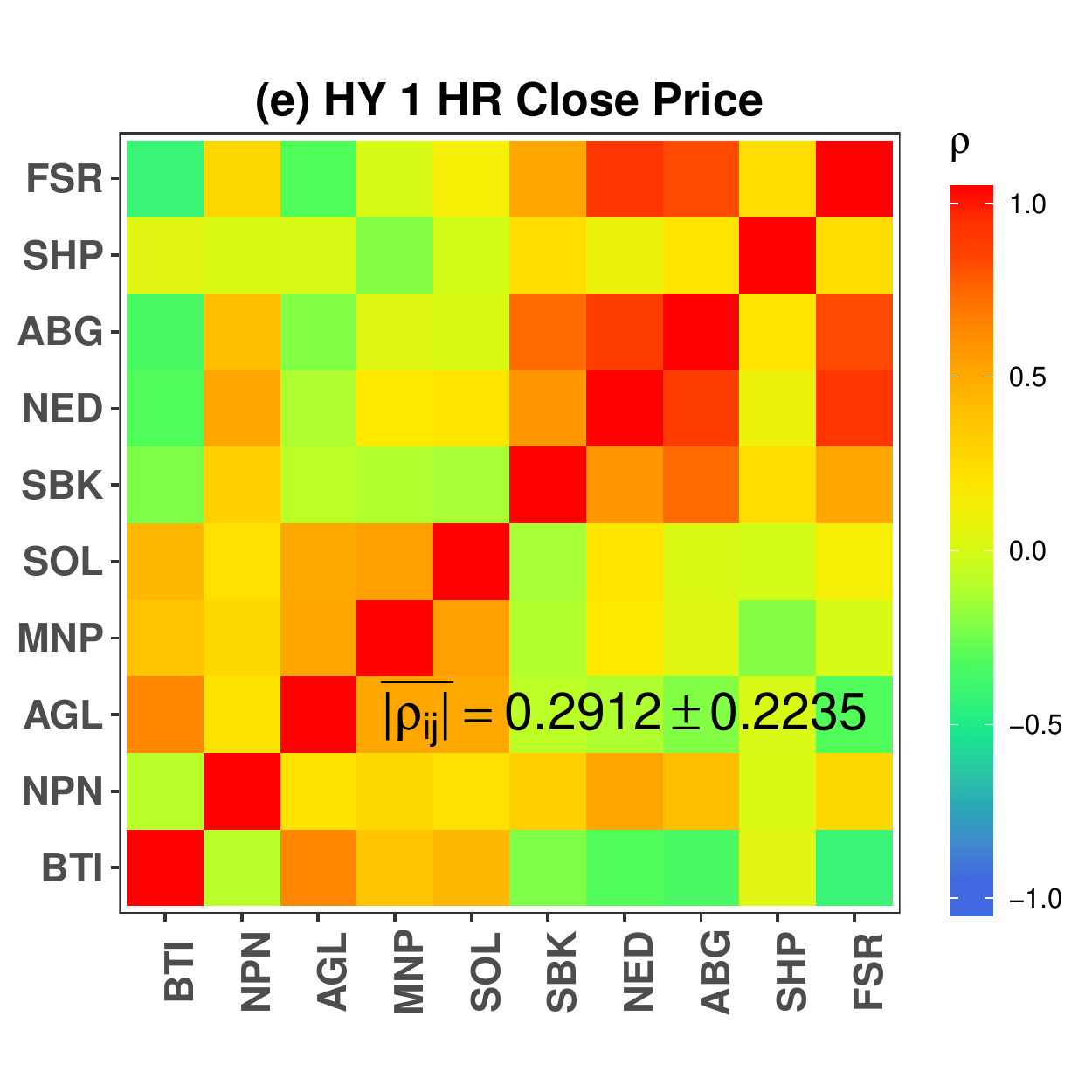}
\end{subfigure}
\begin{subfigure}{0.245\textwidth}
\includegraphics[width=\textwidth]{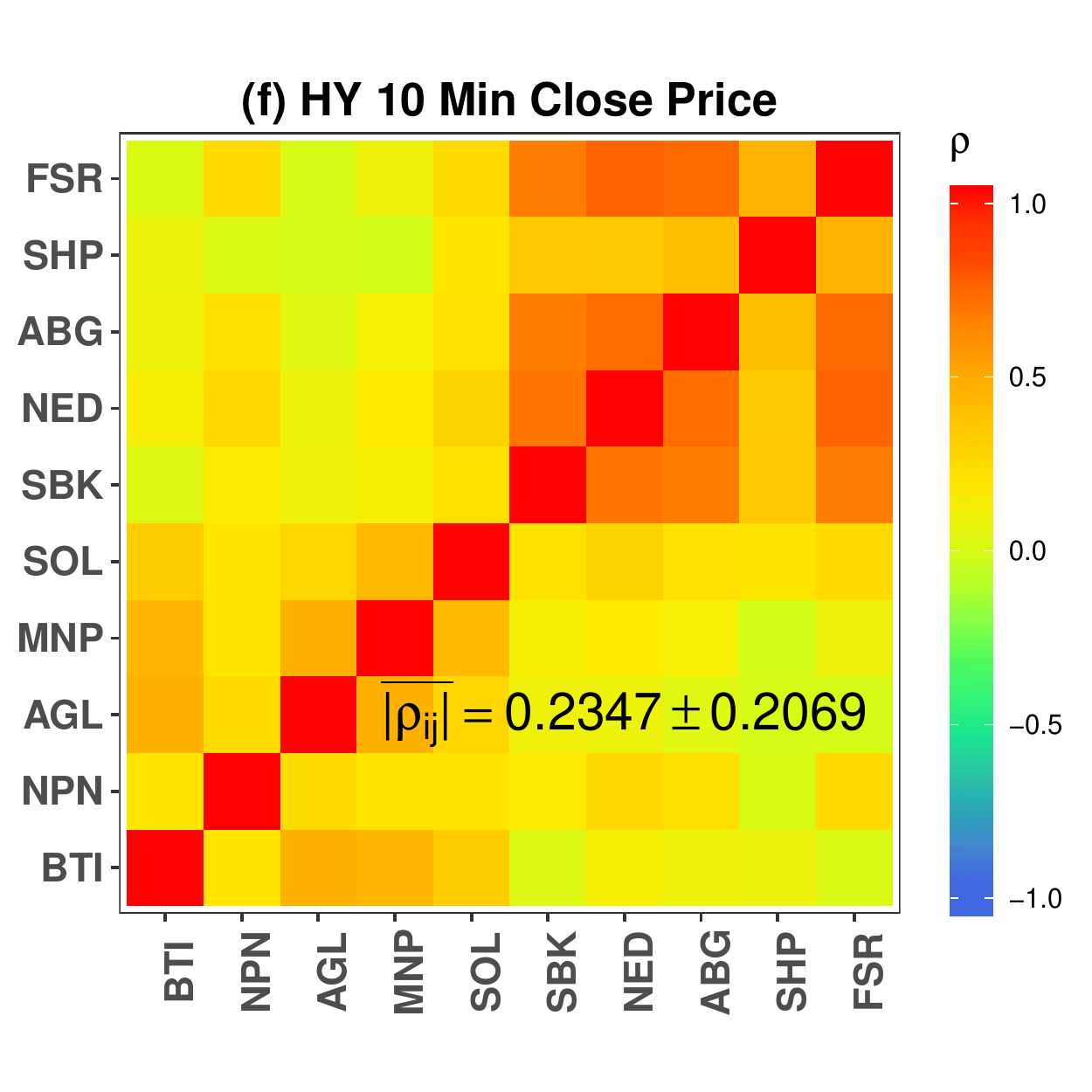}
\end{subfigure}
\begin{subfigure}{0.245\textwidth}
\includegraphics[width=\textwidth]{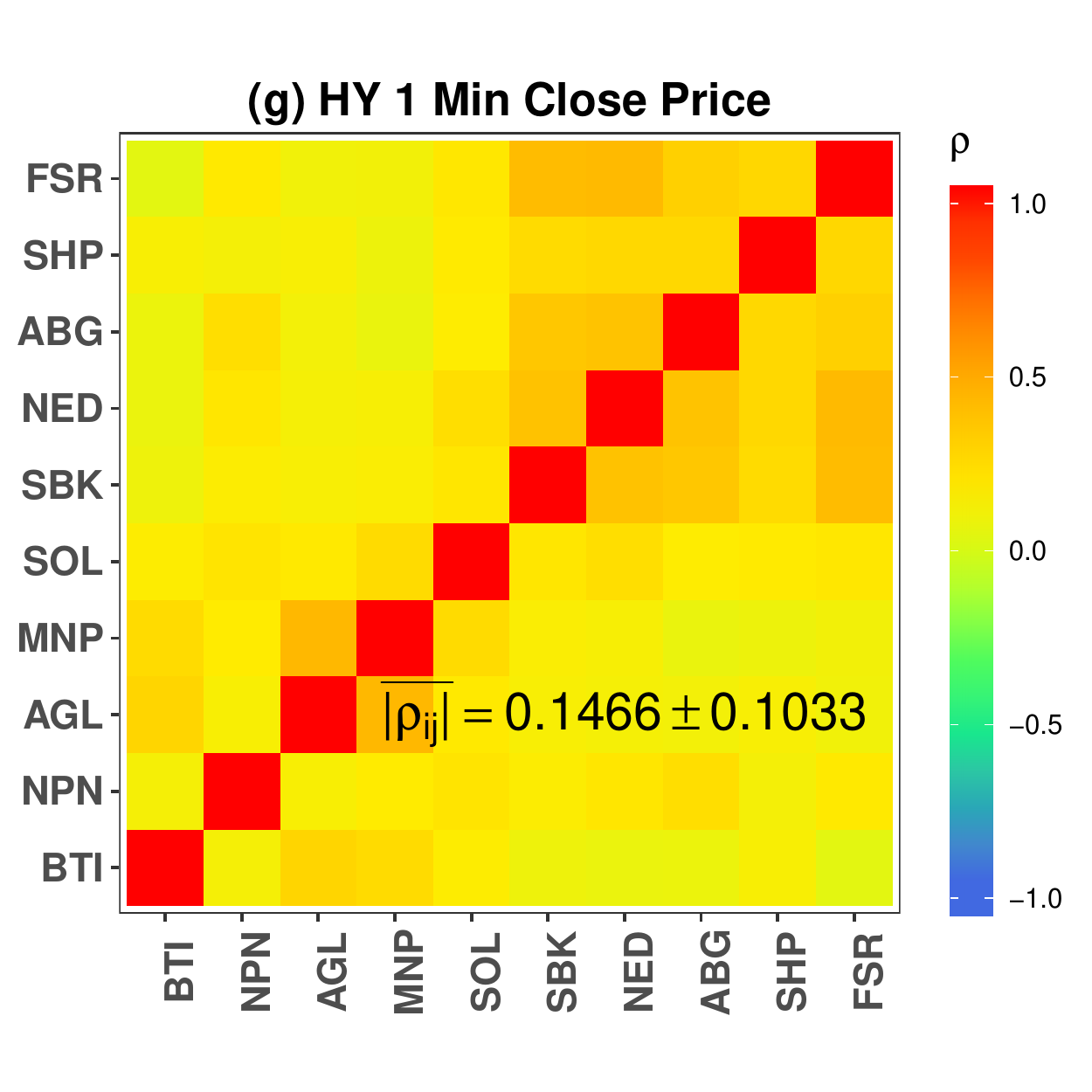}
\end{subfigure}
\begin{subfigure}{0.245\textwidth}
\includegraphics[width=\textwidth]{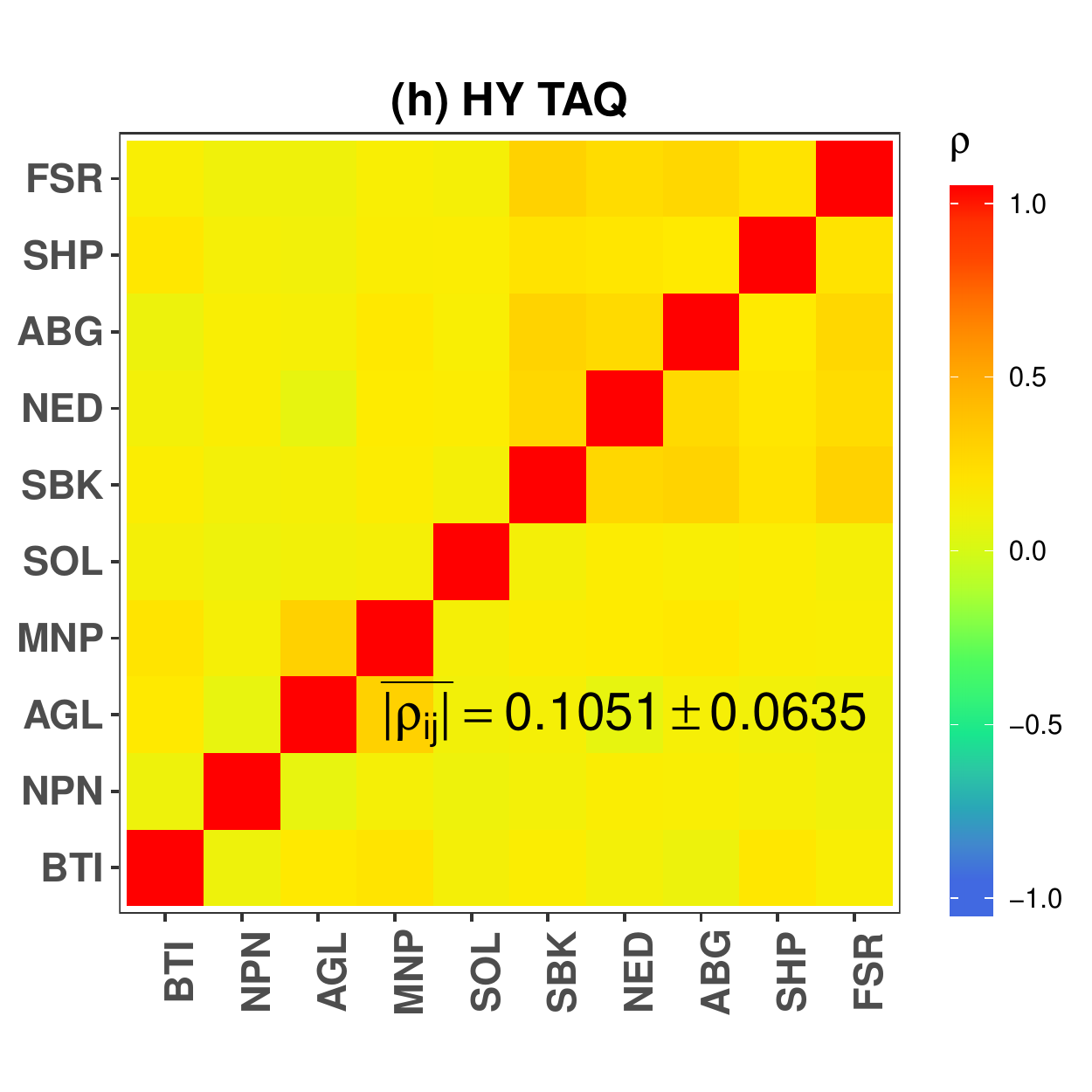}
\end{subfigure}
\caption[Comparing the two estimators with closing price aggregation.]{The Epps effect is visualised by aggregating TAQ data into closing bar prices. The closing aggregation decimates the TAQ sample by down-sampling and shifting the last trade in each bar to the end. From (a) to (d), we have the MM estimator applied to 1 hour, 10 minute, 1 minute closing bar data and TAQ data respectively using Algorithm \ref{algo:ComplexFT}. The insets provide the average correlations and their sample standard deviations. From (e) to (h), we have the HY estimator applied to 1 hour, 10 minute, 1 minute closing bar data and TAQ data respectively using Algorithm \ref{algo:HY}. The Epps effect \cite{EPPS1979} is demonstrated with both estimators and it persists with the MM but only slightly for the HY estimator on the TAQ data. The figures can be recovered using the R script file \href{https://github.com/rogerbukuru/Exploring-The-Epps-Effect-R/blob/master/Trade\%20Data\%20Heatmaps/Closing.R}{Closing.R} and \href{https://github.com/rogerbukuru/Exploring-The-Epps-Effect-R/blob/master/Trade\%20Data\%20Heatmaps/TAQ.R}{TAQ.R} on the GitHub resource \cite{PCRBTG2019}.}
\label{fig:Closing}
\end{figure*}

\begin{figure*}[hbt!]
\centering
\begin{subfigure}{0.245\textwidth}
\includegraphics[width=\textwidth]{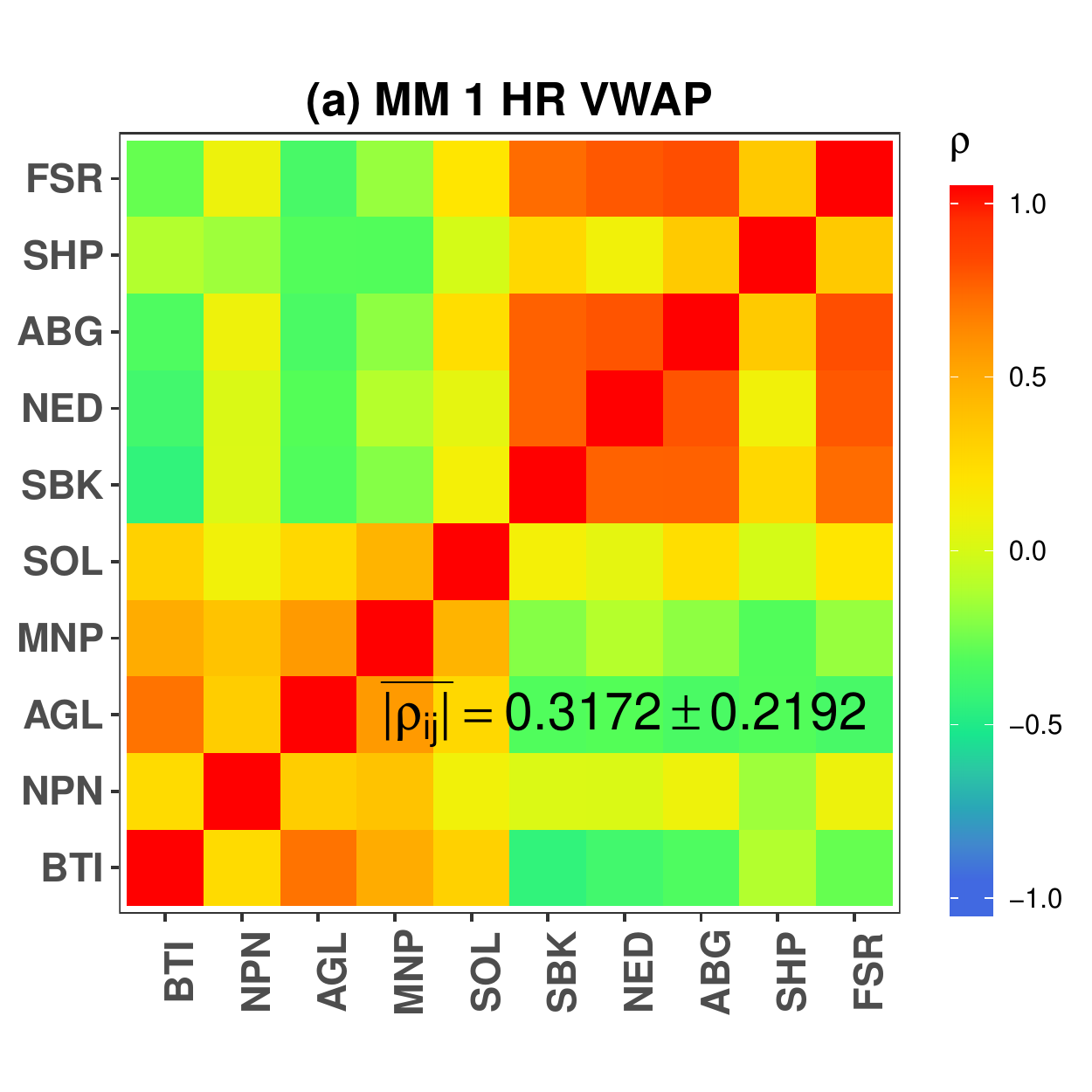}
\end{subfigure}
\begin{subfigure}{0.245\textwidth}
\includegraphics[width=\textwidth]{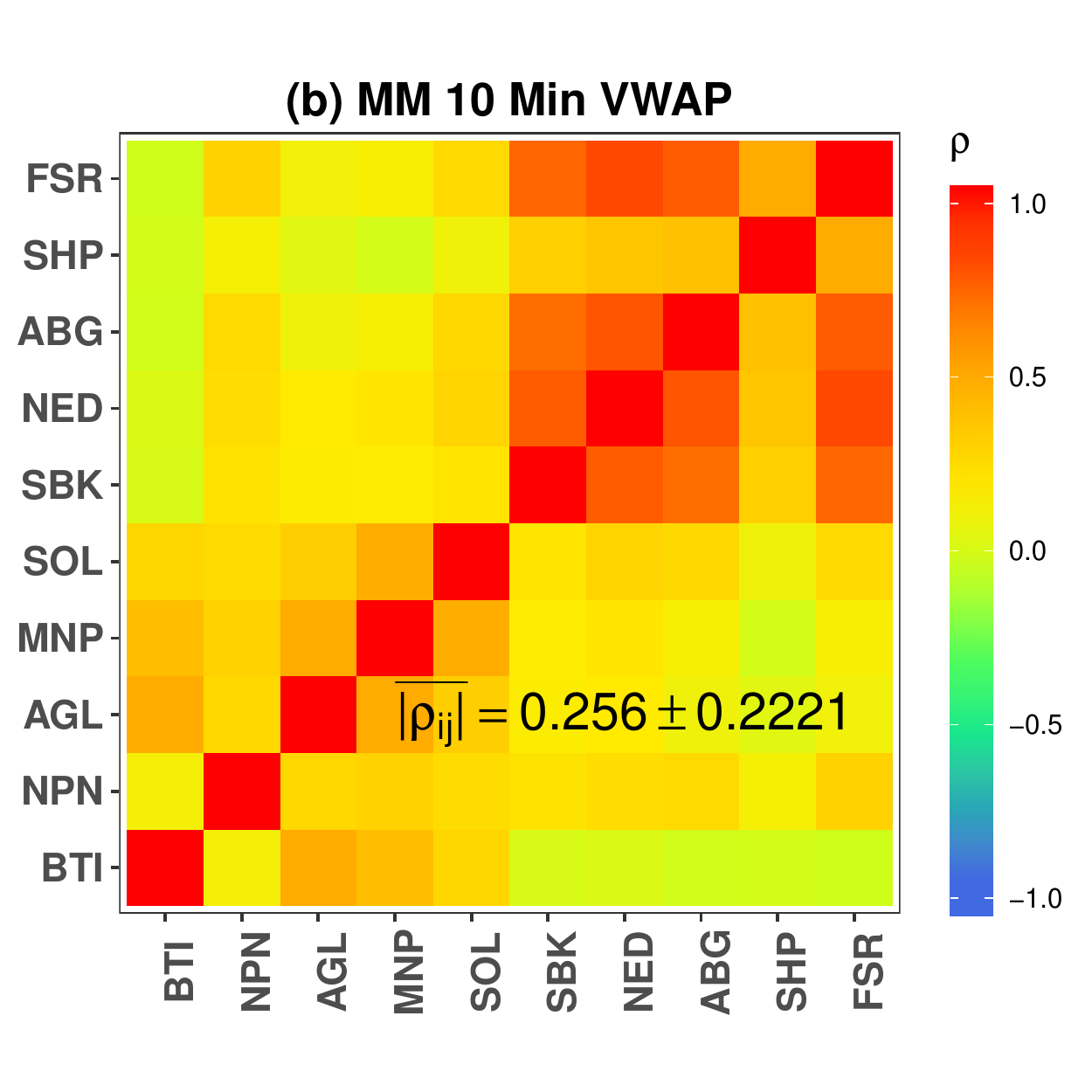}
\end{subfigure}
\begin{subfigure}{0.245\textwidth}
\includegraphics[width=\textwidth]{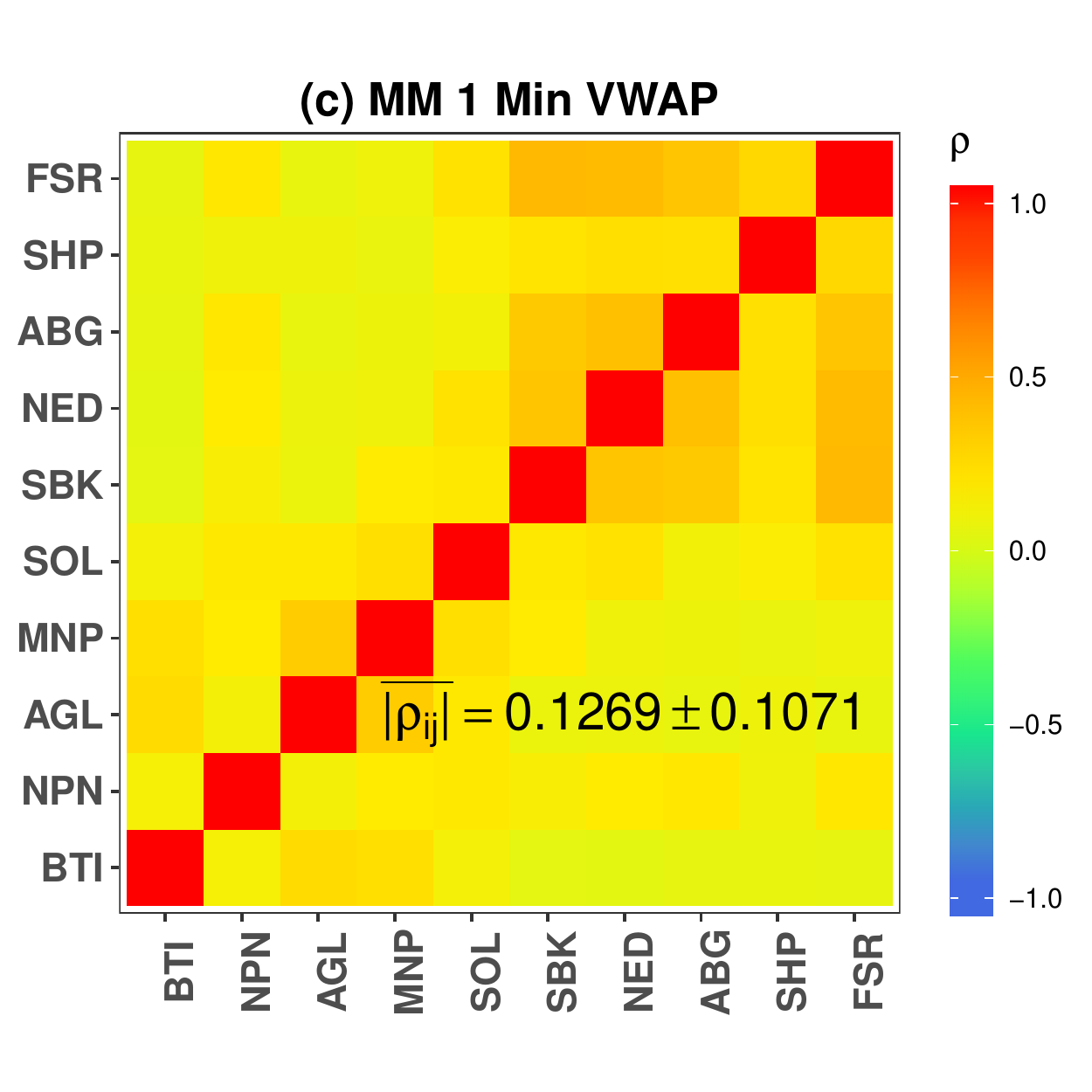}
\end{subfigure}
\begin{subfigure}{0.245\textwidth}
\includegraphics[width=\textwidth]{Figures/MM1WeekAsync.pdf}
\end{subfigure} \\
\begin{subfigure}{0.245\textwidth}
\includegraphics[width=\textwidth]{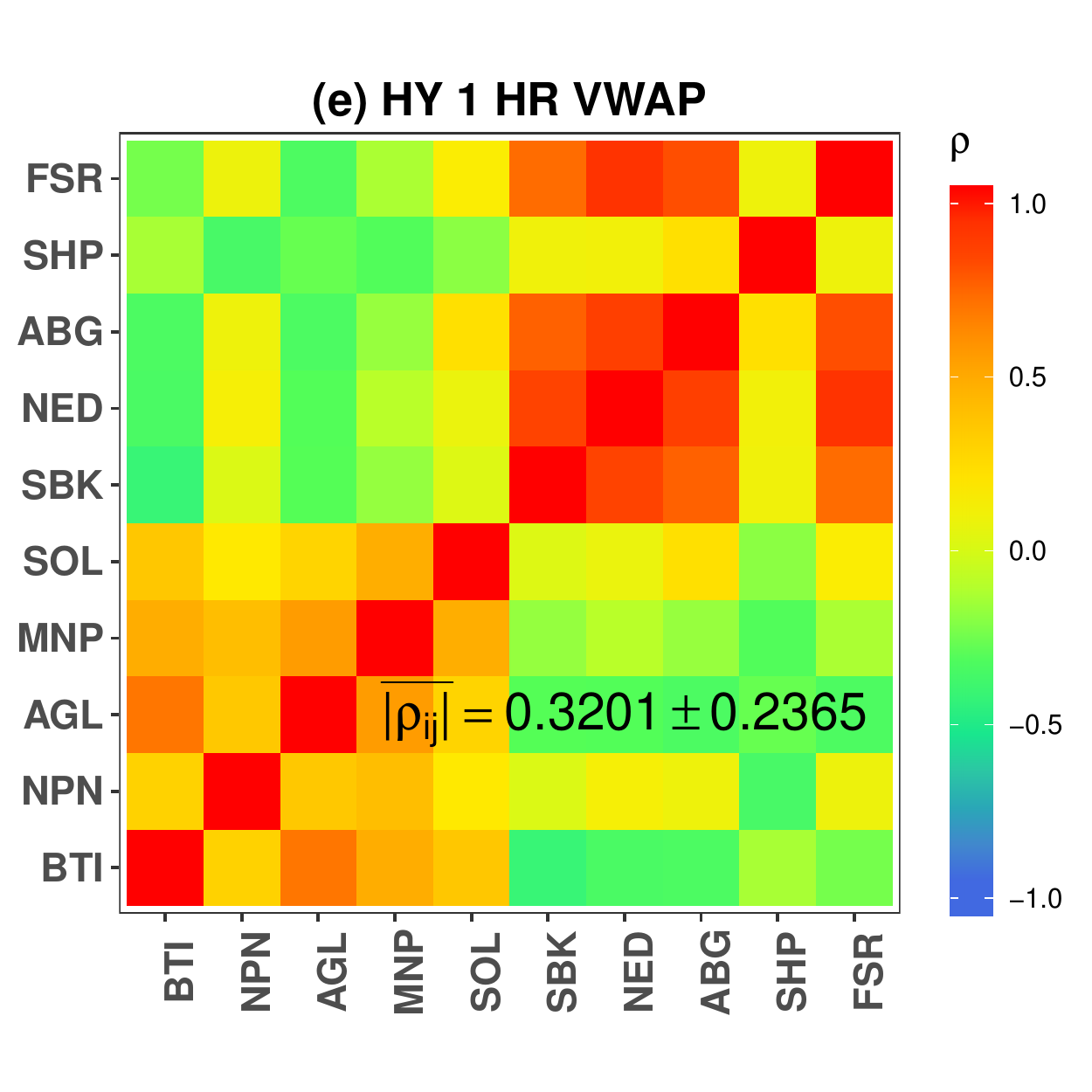}
\end{subfigure}
\begin{subfigure}{0.245\textwidth}
\includegraphics[width=\textwidth]{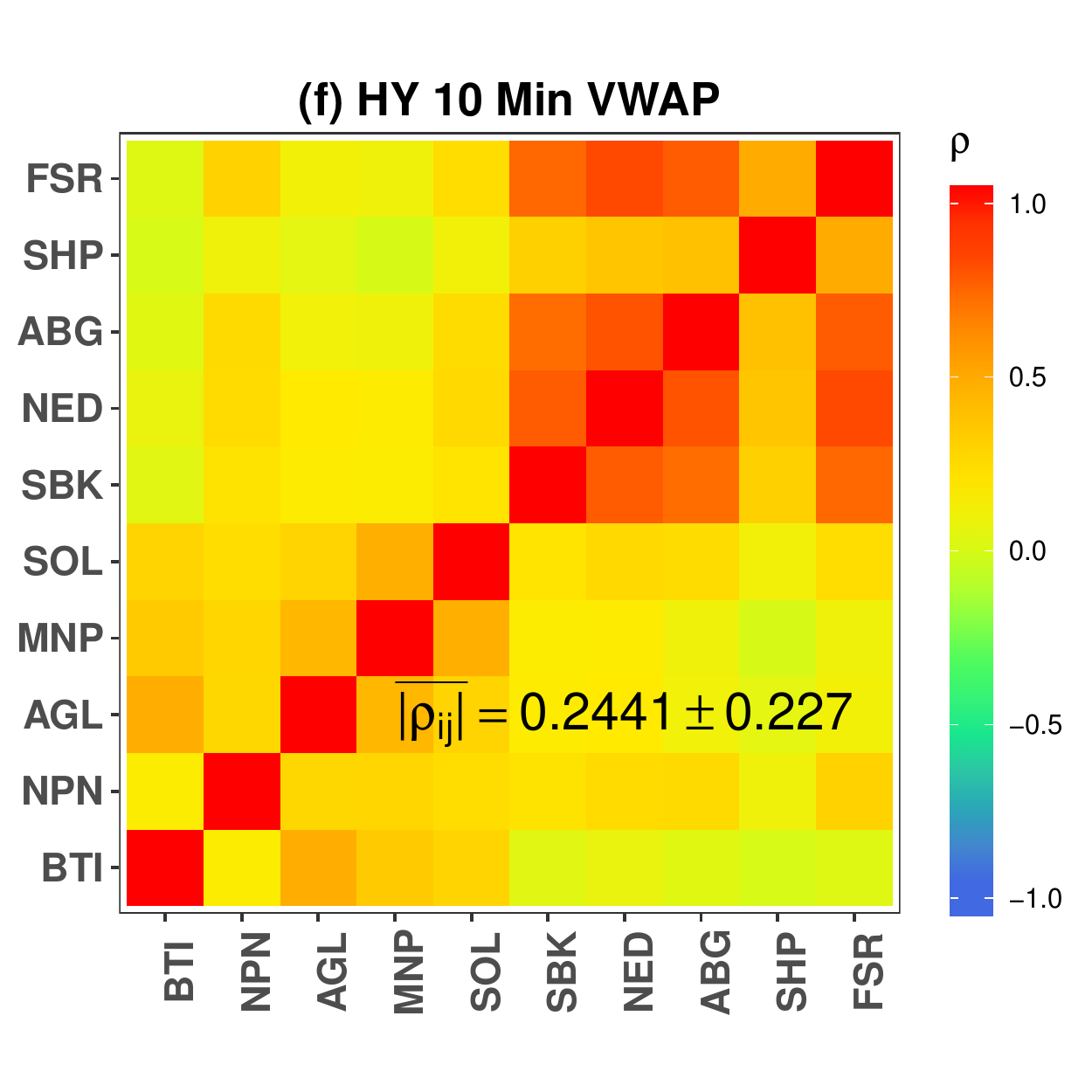}
\end{subfigure}
\begin{subfigure}{0.245\textwidth}
\includegraphics[width=\textwidth]{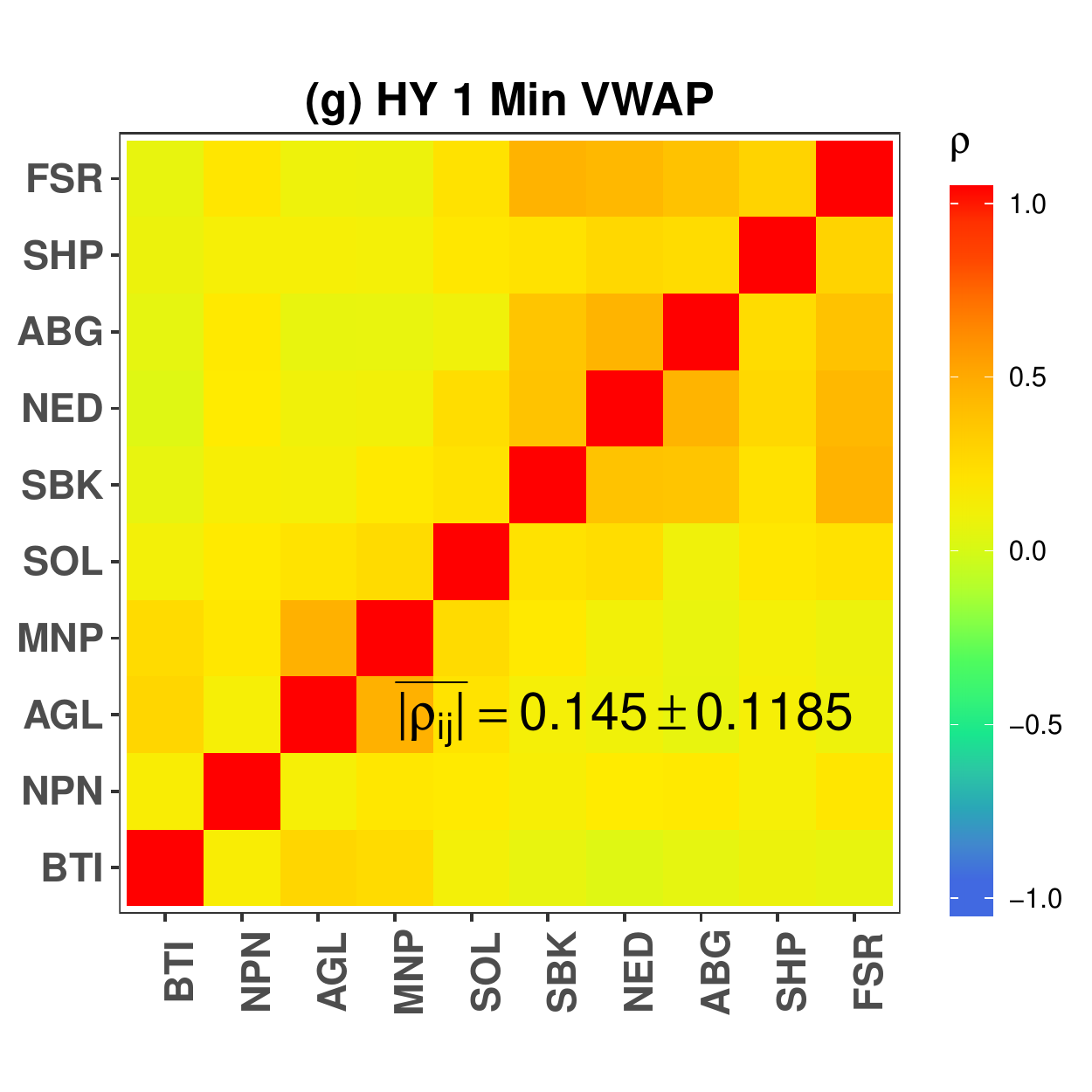}
\end{subfigure}
\begin{subfigure}{0.245\textwidth}
\includegraphics[width=\textwidth]{Figures/HY1WeekAsync.pdf}
\end{subfigure}
\caption[Comparing the two estimators with VWAP aggregation.]{The Epps effect is visualised by aggregating TAQ data into VWAP bar prices. The VWAP aggregation performs a volume average and shifts all the trades in each bar to the end. From (a) to (d), we have the MM estimator applied to 1 hour, 10 minute, 1 minute VWAP bar data and TAQ data respectively using Algorithm \ref{algo:ComplexFT}. From (e) to (h), we have the HY estimator applied to 1 hour, 10 minute, 1 minute VWAP bar data and TAQ data respectively using Algorithm \ref{algo:HY}. The average absolute correlations and their sample standard deviations are provided as insets. The Epps effect \cite{EPPS1979} is demonstrated with both estimators and it persists with the MM but only slightly for the HY estimator on the TAQ data. The figures can be recovered using the R script file \href{https://github.com/rogerbukuru/Exploring-The-Epps-Effect-R/blob/master/Trade\%20Data\%20Heatmaps/VWAP.R}{VWAP.R} and \href{https://github.com/rogerbukuru/Exploring-The-Epps-Effect-R/blob/master/Trade\%20Data\%20Heatmaps/TAQ.R}{TAQ.R} on the GitHub resource \cite{PCRBTG2019}.}
\label{fig:VWAP}
\end{figure*}

\subsection{Calendar time averaging} \label{ssec:calendar}

We begin the analysis by focusing on calendar time. To this end, we investigate sampling intervals decreasing from 1 hour, 10 minute, 1 minute bars to asynchronous TAQ data. In Figure \ref{fig:Closing} and \ref{fig:VWAP} (a) through to (d), we have the MM estimator applied to 1 hour, 10 minute, 1 minute and asynchronous TAQ data for closing and VWAP bars respectively. In Figure \ref{fig:Closing} and \ref{fig:VWAP} (e) through to (h), we have the HY estimator applied to 1 hour, 10 minute, 1 minute and asynchronous TAQ data for closing and VWAP bars respectively. Here prices are sampled and aggregated in calendar time. Closing prices are the result of decimating prices to the last price at the end of each calendar time interval. The VWAP prices is the volume averaged prices across the same calendar time interval.

The first point to notice from Figures \ref{fig:Closing} and \ref{fig:VWAP} is that the Epps effect is present for both estimators and for both closing and VWAP aggregation methods. Therefore the Epps effect is present at much higher sampling frequencies than what \cite{EPPS1979} originally considered. What is more interesting, is that when the sampling frequency increases to the highest available frequency, the correlation for the MM estimates drops towards zero as expected from the Epps effect, but for the HY estimates, the correlation from the TAQ data does not decay completely and remains close to the correlation from the 1 minute bars. Indicating that the HY estimator is manufacturing correlation, and that Epps like effects remain present in both estimators. Again we note that the existence of an Epps-like effect in the HY estimator is counter to some of the claims in the literature, that the Epps effect is a bias in the estimator for which the HY estimator is immune; we have demonstrated otherwise.

The second point to notice from Figures \ref{fig:Closing} and \ref{fig:VWAP} is that not only do we see the existence of the Epps effect, but there are also possible structural changes in the correlations for varying time intervals. One such example can be seen in Table \ref{tab:Closing} which shows the correlation values between equity tickers BTI and NPN from Figure \ref{fig:Closing}. The table shows the decay in correlation from 10 minute to asynchronous TAQ, which is the Epps effect demonstrated in \cite{MMZ2011}, but moreover, the change in correlation from 1 hour to 10 minutes switching from negative to positive and increasing in magnitude. This cannot be easily reconciled with the argument that lead-lag and asynchrony are sufficient in explaining the entirety of the Epps effect.

\begin{table}[]
\centering
\begin{tabular}{p{4cm}p{2cm}p{2cm}}
\hline
\multirow{2}{*}{Tickers: BTI \& NPN}& \multicolumn{2}{c}{Correlation $\rho$} \\  \cline{2-3}&MM &HY\\
\hline
1 Hour       (*)    & -0.0265                    & -0.1450 \\
10 Minute    (*)    & 0.1070                     & 0.1445  \\
1 Minute         & 0.0514                     & 0.0701  \\
Asynchronous TAQ & 0.0321                     & 0.0409  \\ 
\hline
\end{tabular}
\caption{The correlation values between two assets, BTI and NPN (from Figure \ref{fig:Closing}), are given using the two different estimators: Malliavin-Mancino (MM) and Hayashi-Yoshida (HY) respectively. The table demonstrates the existence of the Epps effect and highlights an example of how structural change in the correlation can manifest when the averaging scale is changed from 1 hour to 10 minute (See *). This provides one of many examples of the change in sign of correlations on different averaging time-scales. The Epps effect can be clouded by other scale dependent structural effects.} \label{tab:Closing}
\end{table}

The third lesser point to notice from Figures \ref{fig:Closing} and \ref{fig:VWAP} is that the top right corners for the various sub-figures are highly correlated. This reflects asset class specific properties of the stock universe we considered, and is because these are assets from the banking sector. These are highly correlated, the correlations obtained in calendar time tend to align with our prior domain informed expectations. We also notice that the correlation structure for the two estimates in the synchronous bar data are similar but not exactly the same (this is seen more clearly in Table \ref{tab:Closing}). From our results in the Monte Carlo experiments, we expect the correlation estimates for the two estimators to be exactly the same for the synchronous bar data. This discrepancy is due to some of the less liquid stocks not having sufficient trades within a given bar and therefore the bar data we have constructed are not truly synchronous. There is a liquidity effect. 

Finally, the closing price and VWAP aggregations are achieving different things: the first being a sample from end of the bar, and the latter being a weighted average of trades within a bar - yet, the aggregation methods did not manifestly change the correlation structure. Rather it seems that the VWAP aggregation accentuates the existing correlation structure in the closing price aggregation. This can be considered surprising if one retains a market micro-structure perspective because the closing price aggregation decimates the true price path and shifts the last trade in each bar to the end, while the VWAP changes the price path due to the volume averaging and shifts all the average trades in each bar to the end - yet the two methods produced correlation structures that are very similar. This makes sense if order-flow preserves the correlation structure of averaged top-down components of the market rather than correlation being encoded in each and every event from some bottom-up stochastic process. However, it is currently unclear which aggregation method most faithful represents the correlation structure at varying time scales in calendar time. For this reason we move on to an event time framework inspired by the intrinsic time approach provided by \cite{DERMAN2002}. This attempts to resolve 1.) asynchrony, 2.) the true correlation structure and 3.) the link between calendar time and intrinsic time through the choice of a volume time clock.

\subsection{Intrinsic volume time averaging} \label{ssec:intrinsic}

Here we considered an intrinsic time approach which considers time measured based off events unique to a particular instrument \cite{DERMAN2002}. We construct bar data with 8, 48 and 480 volume buckets equivalent on average to 1 hour, 10 minute and 1 minute bars in calendar time. For intrinsic time, we consider the framework provided by Derman \cite{DERMAN2002} which proposes an elegant link between intrinsic time and calendar time. Our implementation can be found in Algorithm \ref{algo:Derman}. Here prices are sampled uniformly in the average estimated intrinsic (volume) time unique to each stock and selected to provide the same amount of time intervals, on average, per trading day. 

\begin{figure*}[p]
\centering
\begin{subfigure}{0.245\textwidth}
\includegraphics[width=\textwidth]{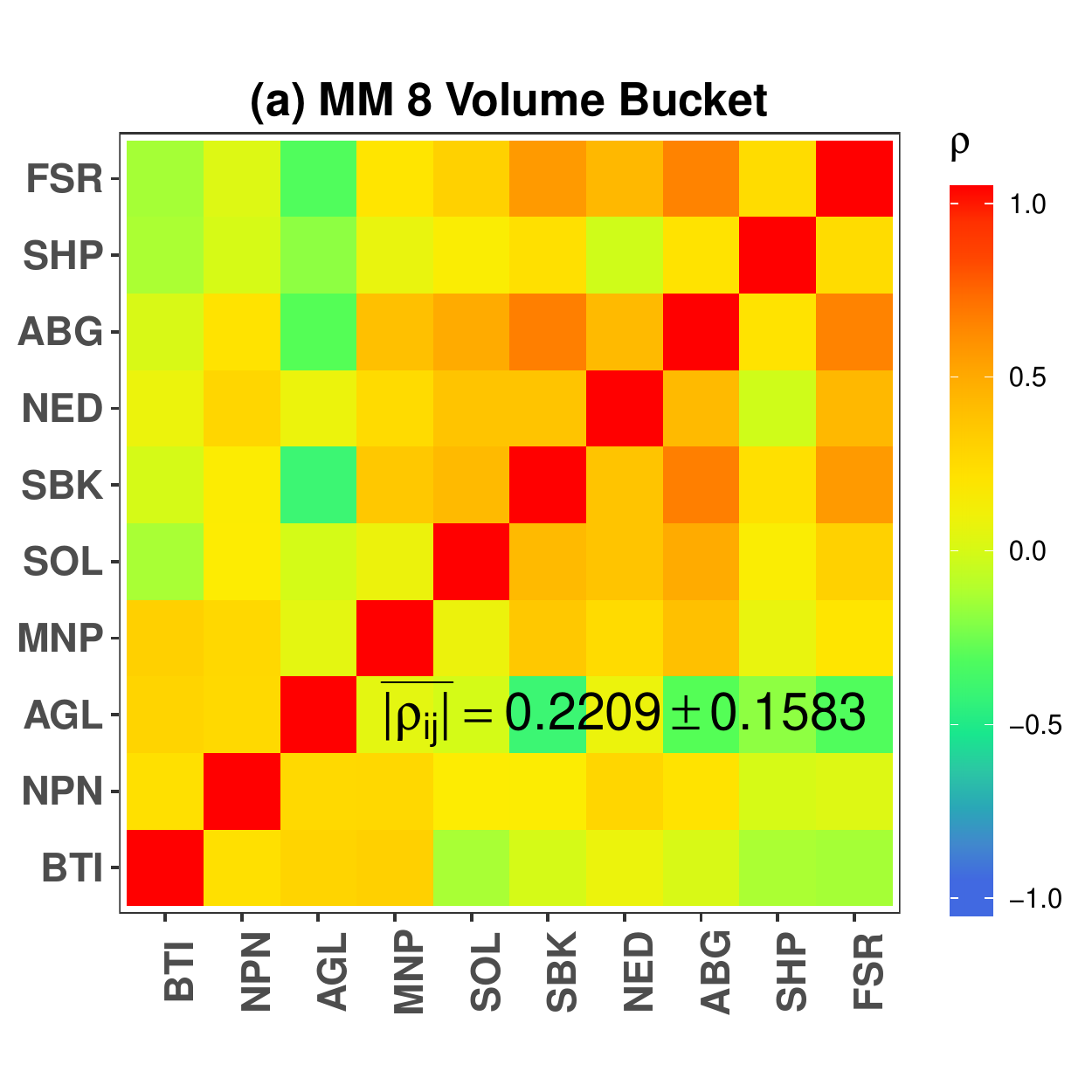}
\end{subfigure}
\begin{subfigure}{0.245\textwidth}
\includegraphics[width=\textwidth]{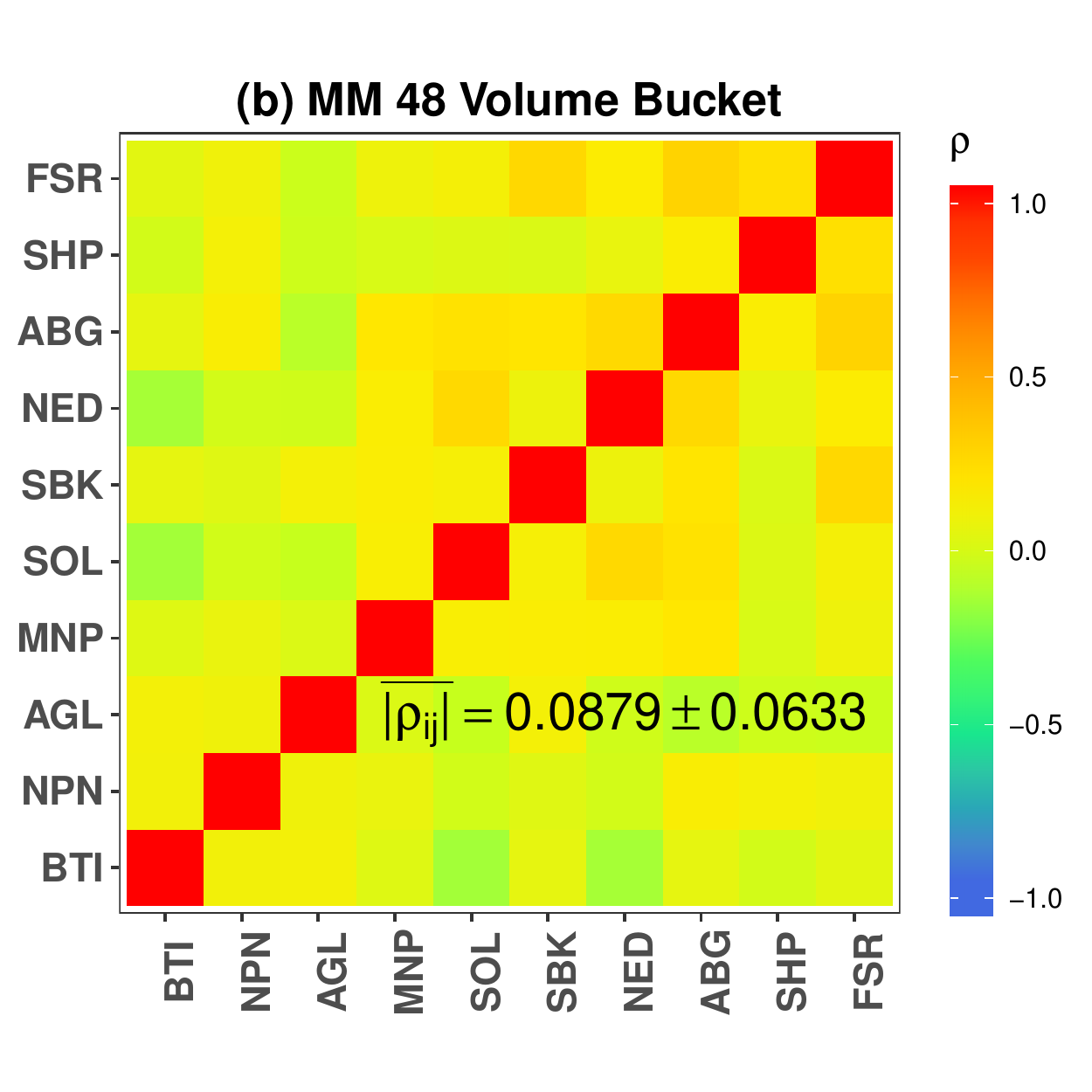}
\end{subfigure}
\begin{subfigure}{0.245\textwidth}
\includegraphics[width=\textwidth]{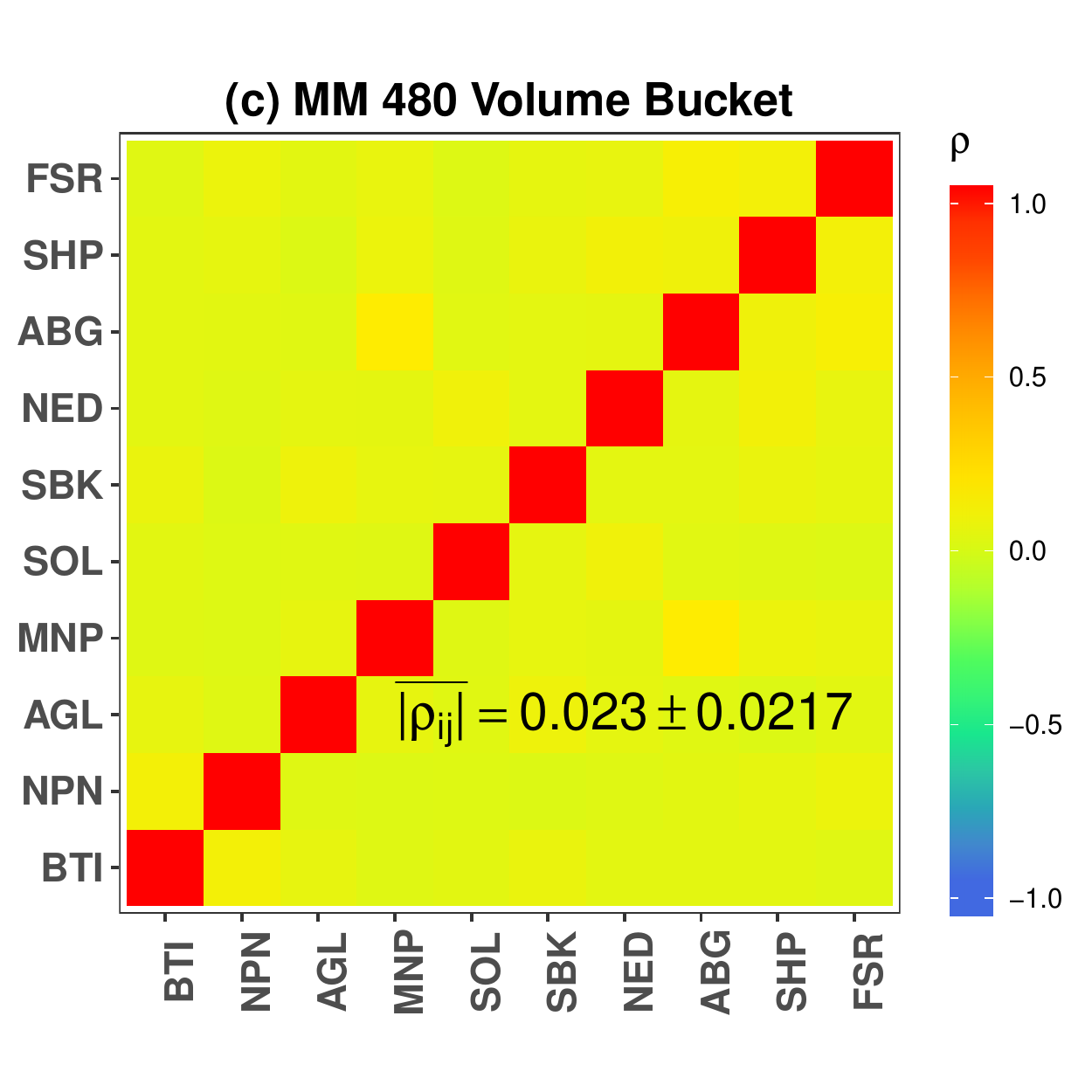}
\end{subfigure} \\
\begin{subfigure}{0.245\textwidth}
\includegraphics[width=\textwidth]{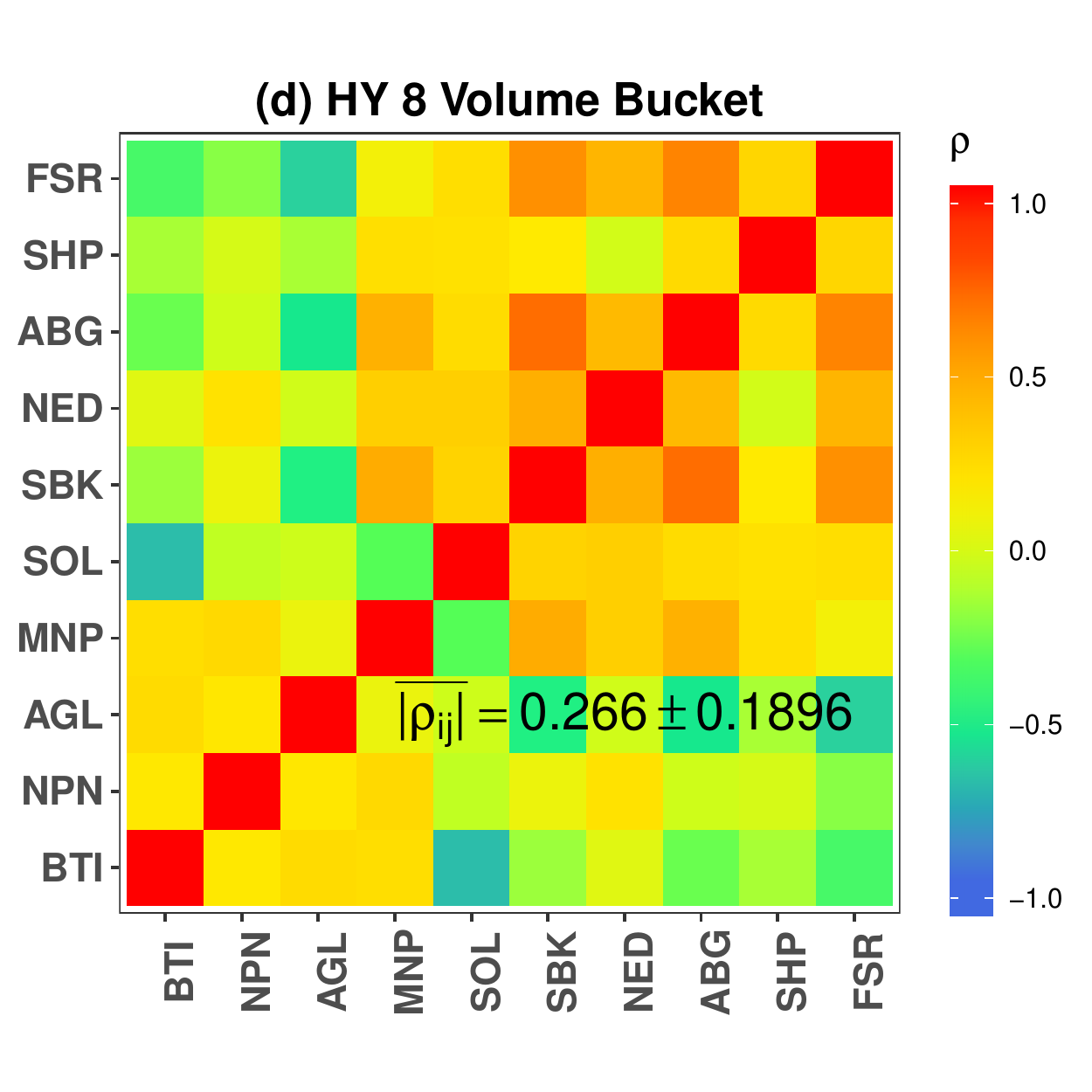}
\end{subfigure} 
\begin{subfigure}{0.245\textwidth}
\includegraphics[width=\textwidth]{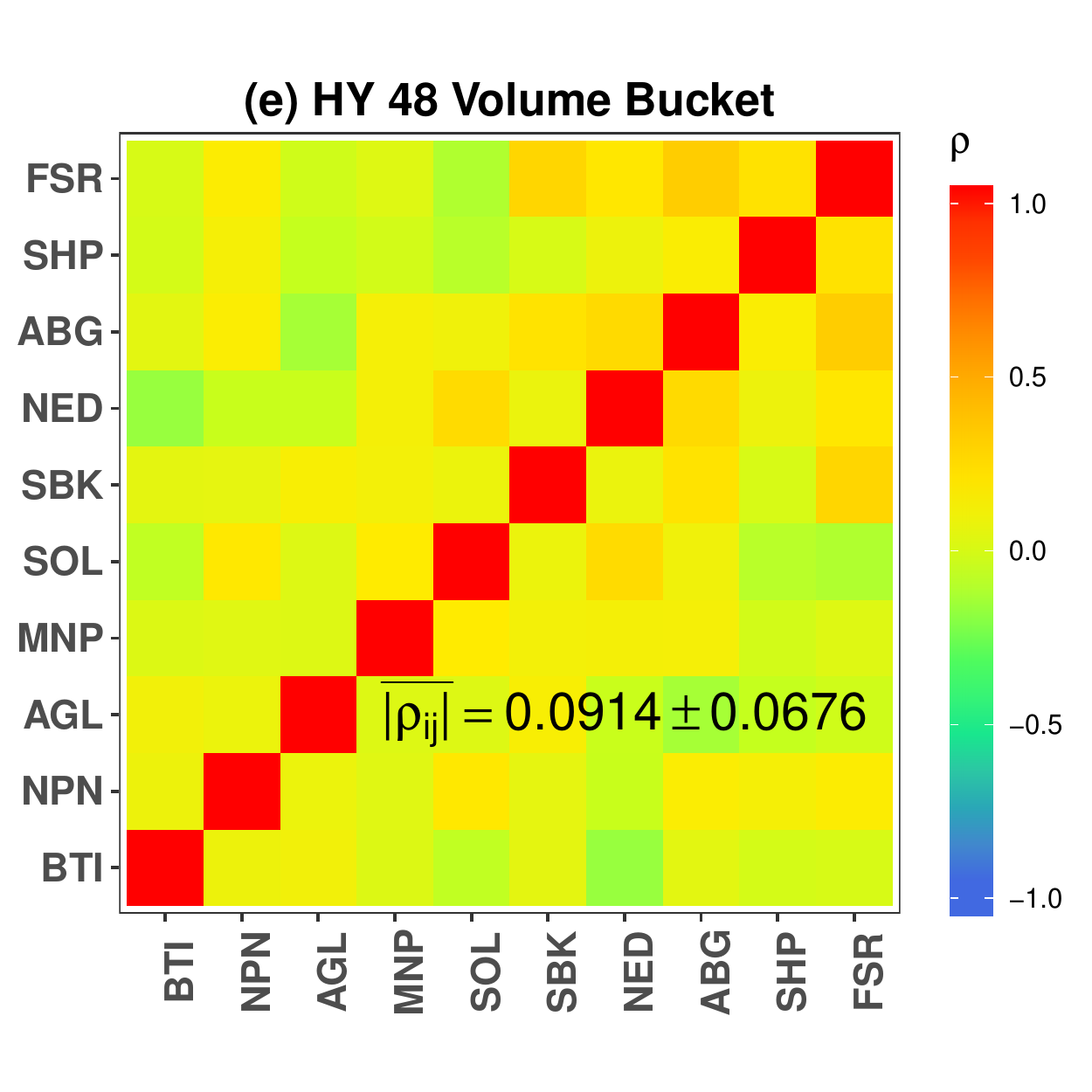}
\end{subfigure}
\begin{subfigure}{0.245\textwidth}
\includegraphics[width=\textwidth]{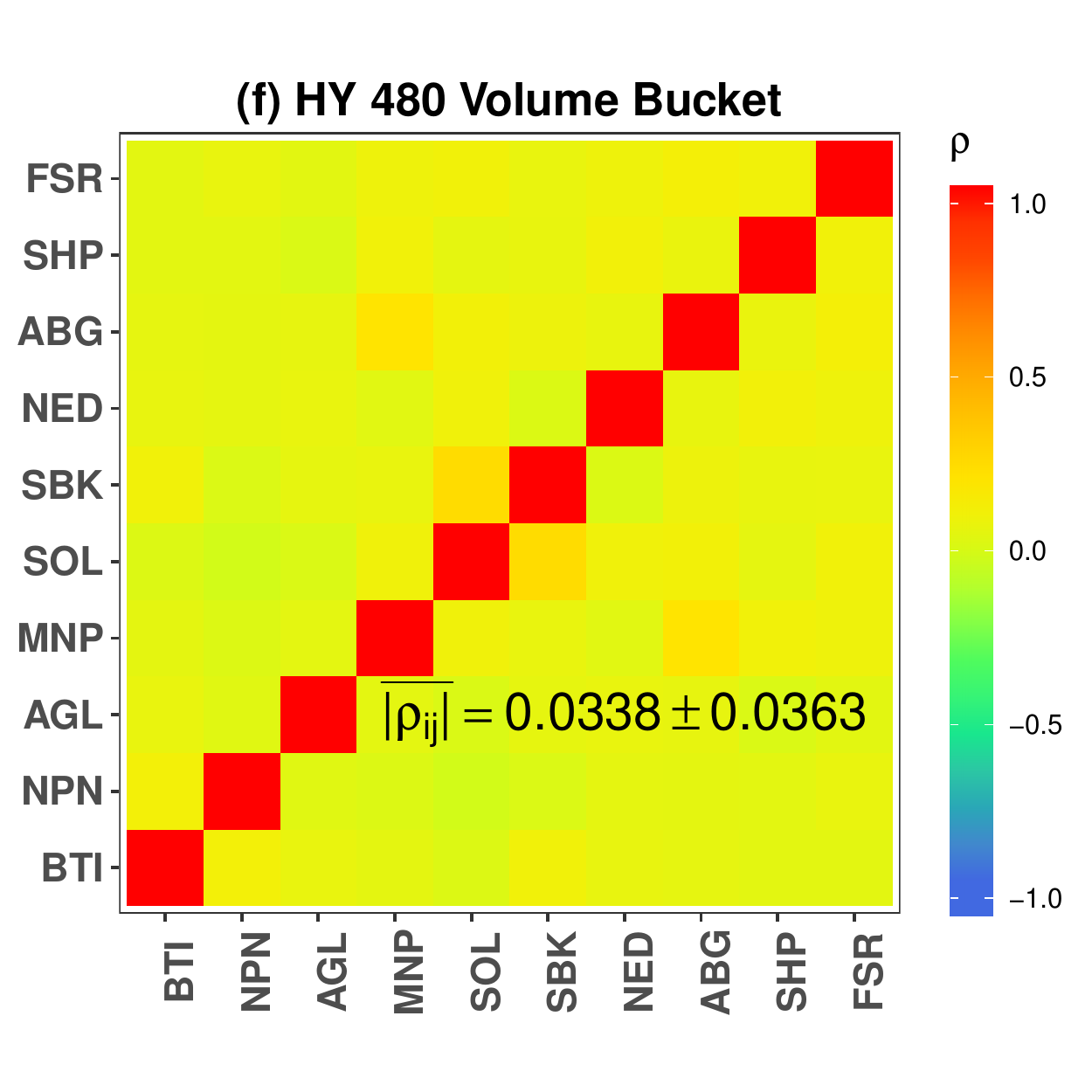}
\end{subfigure}
\caption[Comparing the two estimators using intrinsic time sampling \cite{DERMAN2002}.]{The Epps effect is visualised in intrinsic volume time using the stock specific volume time clocks \cite{DERMAN2002}. The aggregation is achieved by assuming each asset has their own intrinsic time scale given by \eqref{eq:Derman:1}, where $v_j$ is given by the average daily volume over the period of consideration divided by 8, 48 and 480 which corresponds to the calendar time equivalent of 1 hour, 10 minute and 1 minute bar data respectively. From (a) to (c) and (d) to (f), we have the MM and HY estimator applied to the calendar time equivalent of 1 hour, 10 minute and 1 minute bar using Algorithm \ref{algo:ComplexFT} and \ref{algo:HY} respectively. The average absolute correlations and their sample standard deviations are provided as insets. The Epps effect is clearly present for both estimators under the paradigm of Intrinsic time. The figures can be recovered using the R script file \href{https://github.com/rogerbukuru/Exploring-The-Epps-Effect-R/blob/master/Trade\%20Data\%20Heatmaps/IntrinsicTime.R}{IntrinsicTime.R} on the GitHub resource \cite{PCRBTG2019}.}
\label{fig:Derman}
\end{figure*}

\begin{figure*}[p]
\centering
\begin{subfigure}{0.245\textwidth}
\includegraphics[width=\textwidth]{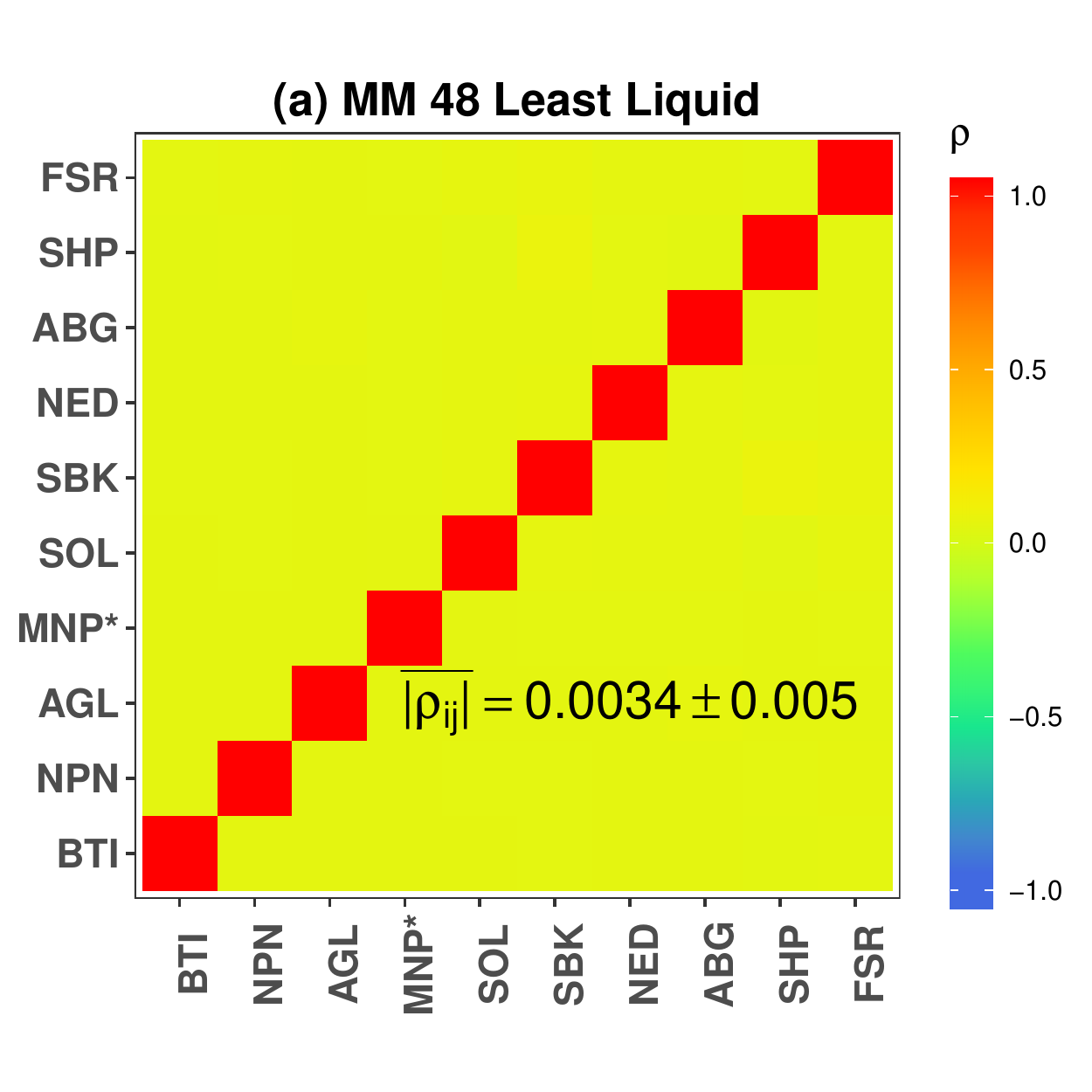}
\end{subfigure}
\begin{subfigure}{0.245\textwidth}
\includegraphics[width=\textwidth]{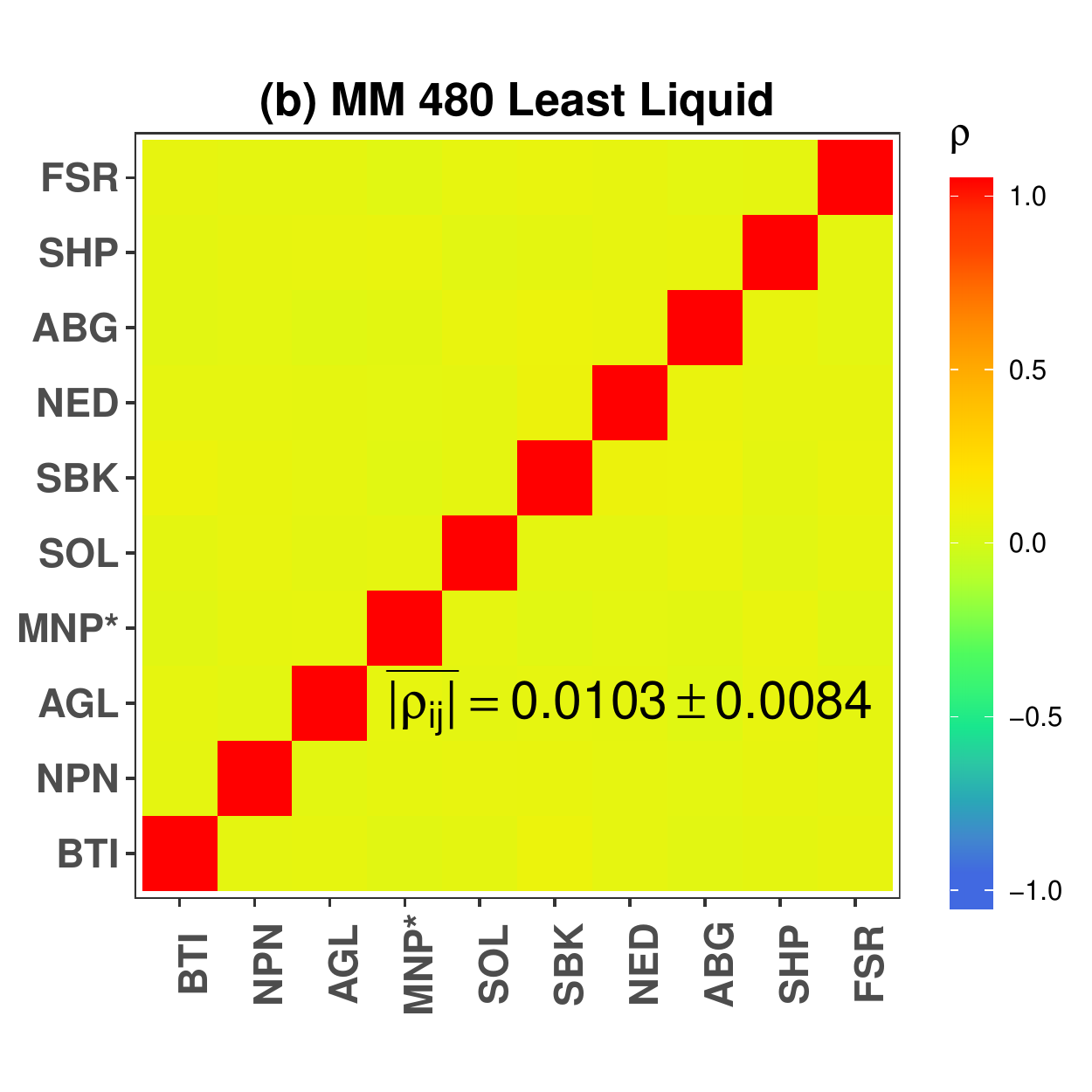}
\end{subfigure} 
\begin{subfigure}{0.245\textwidth}
\includegraphics[width=\textwidth]{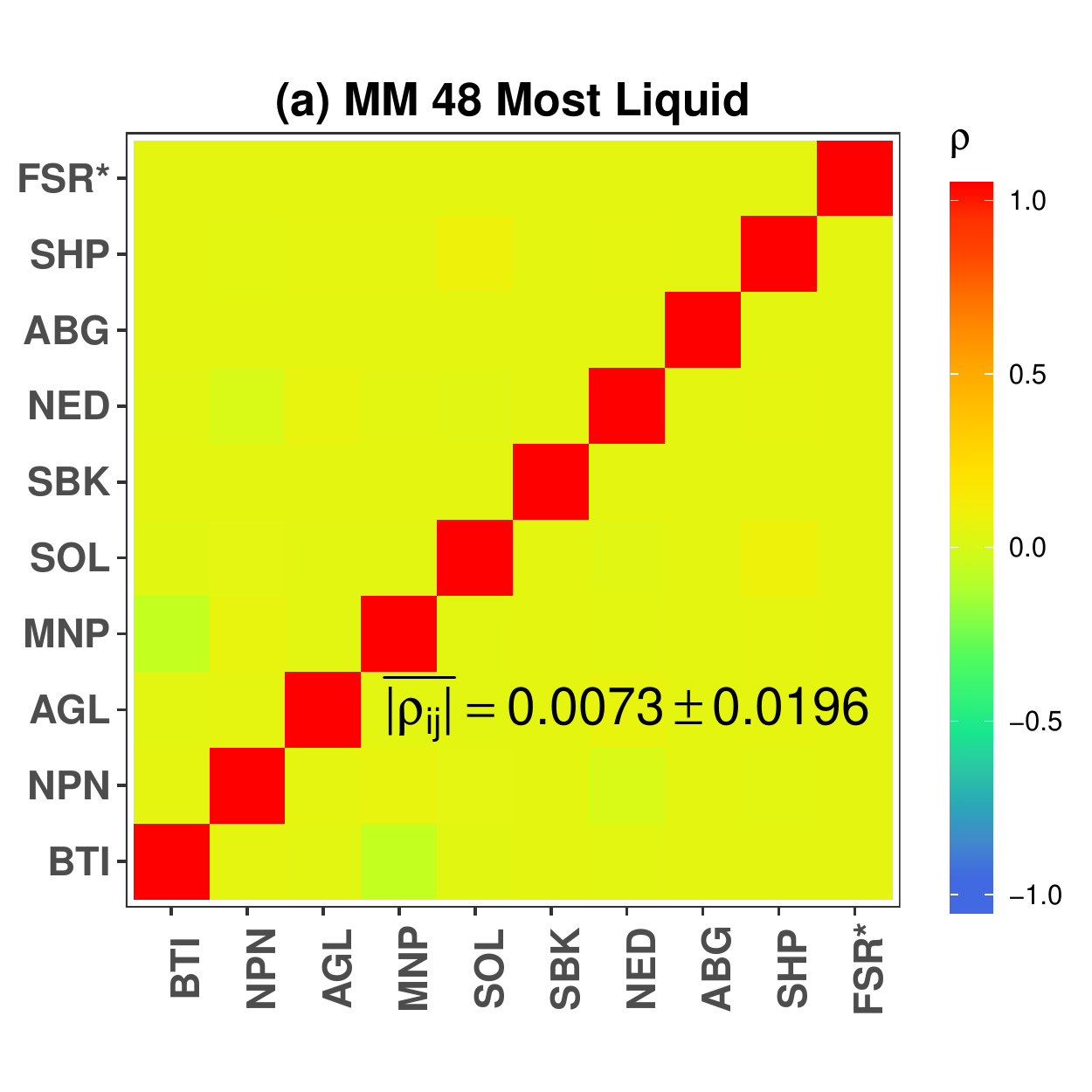}
\end{subfigure} 
\begin{subfigure}{0.245\textwidth}
\includegraphics[width=\textwidth]{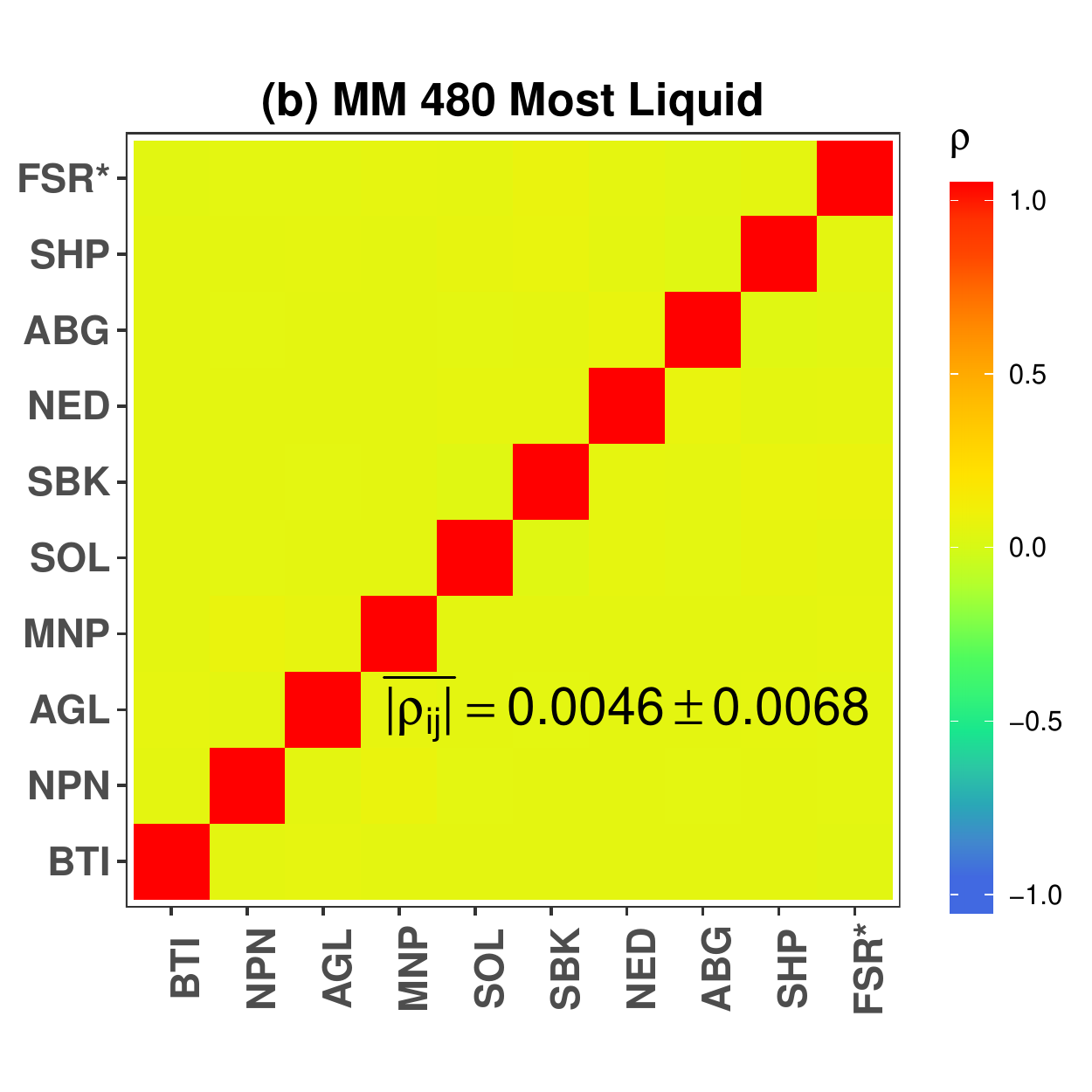}
\end{subfigure} \\
\begin{subfigure}{0.245\textwidth}
\includegraphics[width=\textwidth]{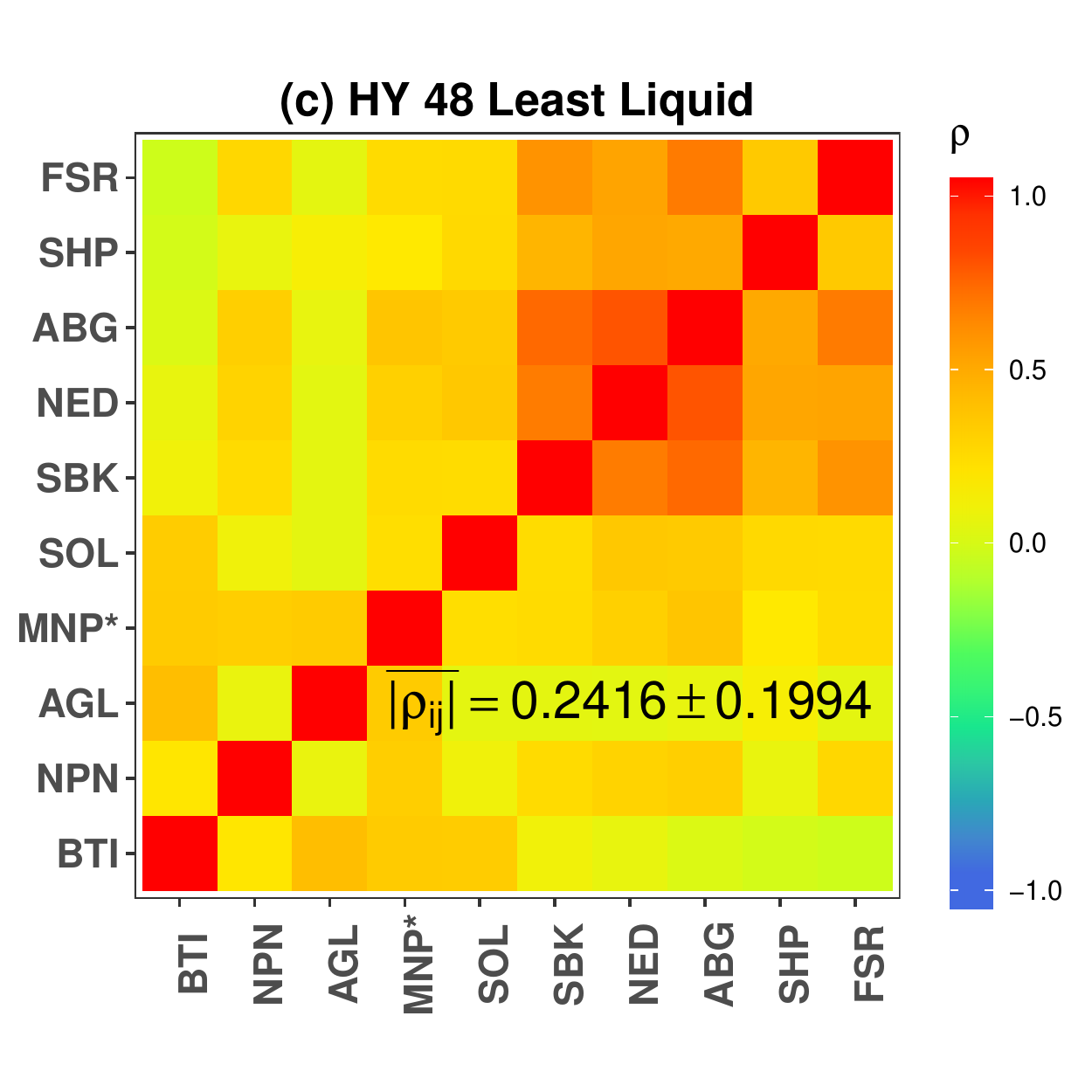}
\end{subfigure}
\begin{subfigure}{0.245\textwidth}
\includegraphics[width=\textwidth]{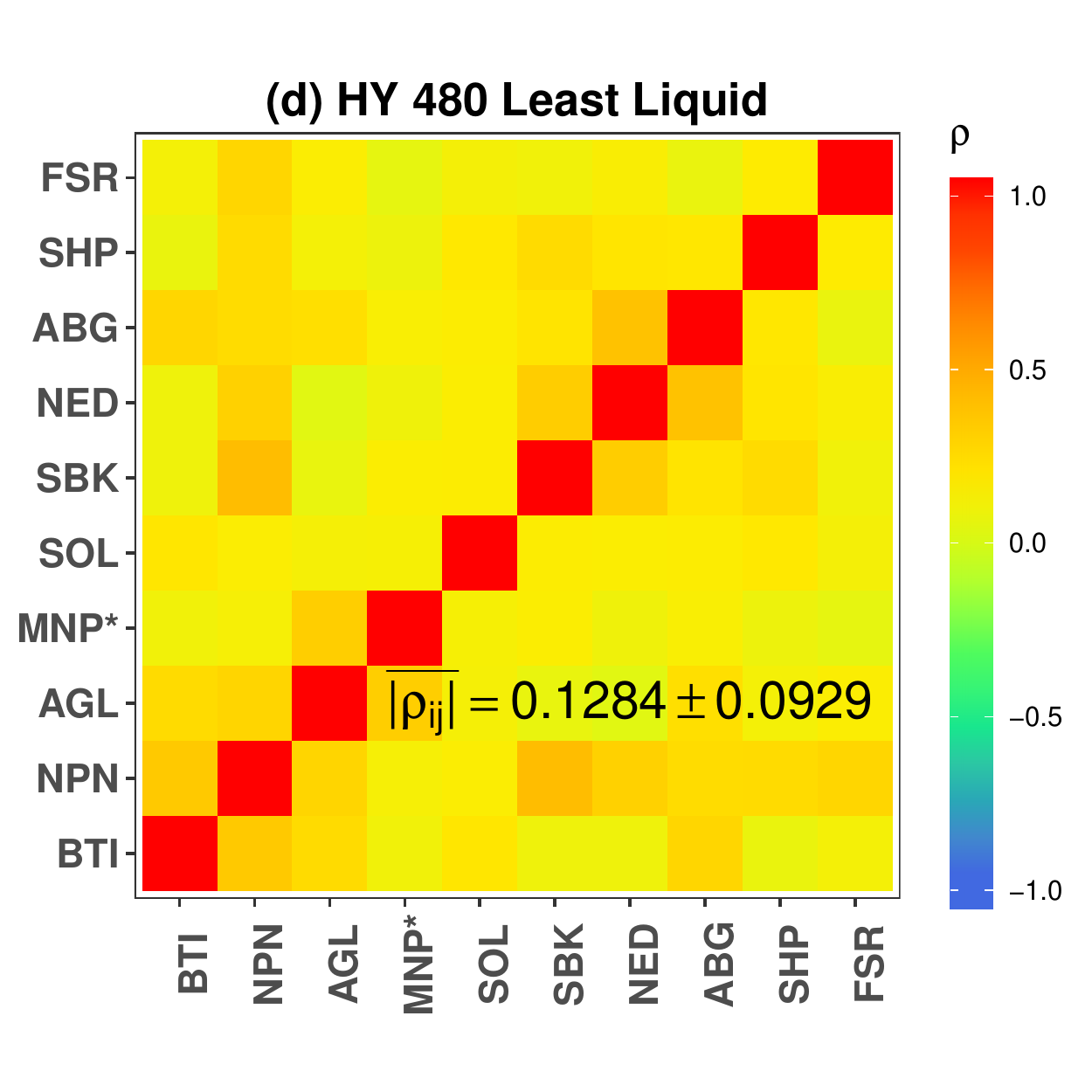}
\end{subfigure} 
\begin{subfigure}{0.245\textwidth}
\includegraphics[width=\textwidth]{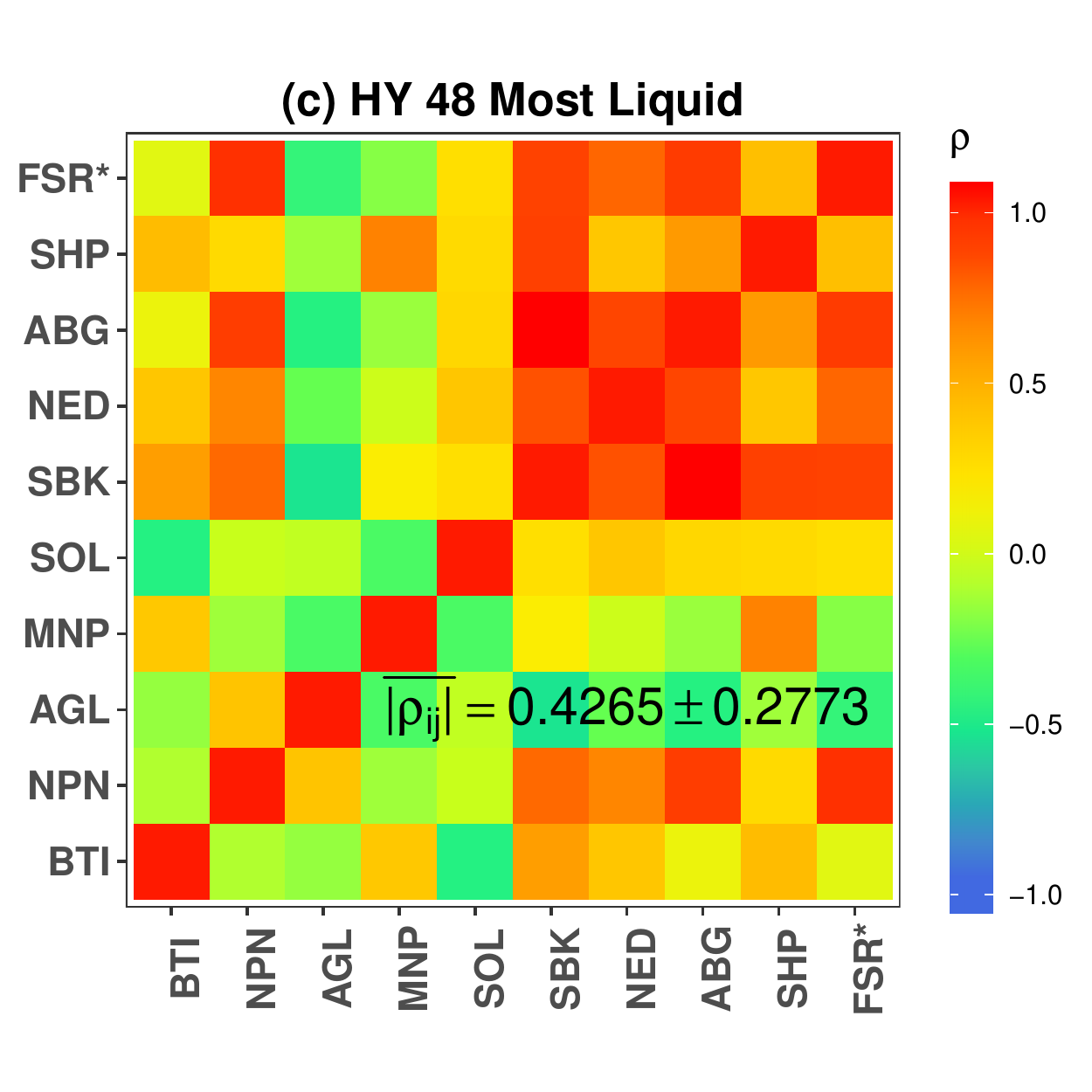}
\end{subfigure} 
\begin{subfigure}{0.245\textwidth}
\includegraphics[width=\textwidth]{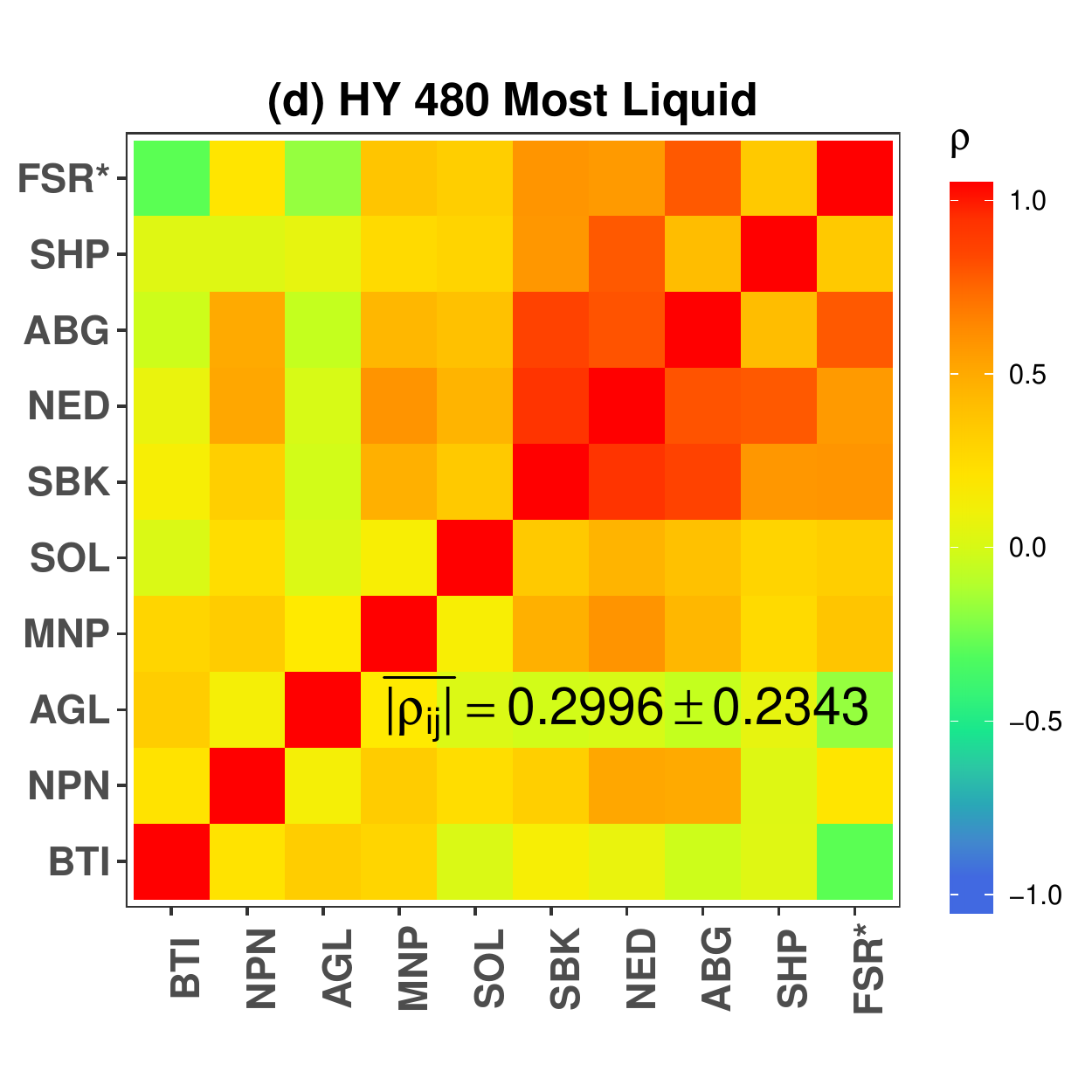}
\end{subfigure} 
\caption{We demonstrate the correlation structures along with the average absolute correlations and their sample standard deviations provided as insets using a synchronised volume time sampling scheme \cite{ELO2012A}. Here (a), (b) and (c), (d) is the MM and HY estimator applied to the volume equivalent of 10 minute and 1 minute calendar time averages with Algorithm \ref{algo:ComplexFT} and \ref{algo:HY} respectively. The baseline ``clock'' is indicated with (*) and the figures on the left (right) are synchronised according to the least (most) liquid tickers (respectively). Here the Epps effect is present with the HY estimator which seems to be manufacturing correlation for synchronising with the most liquid ticker. Whereas the MM does not demonstrate the Epps effect due to the sampling scheme and our choice of N. The figures can be recovered using the R script file \href{https://github.com/rogerbukuru/Exploring-The-Epps-Effect-R/blob/master/Trade\%20Data\%20Heatmaps/EventTime.R}{EventTime.R} on the GitHub resource \cite{PCRBTG2019}.}
\label{fig:Tim}
\end{figure*}

Under the intrinsic volume time paradigm, we investigate a framework proposed by \cite{DERMAN2002}. This framework assumes that each stock has their own intrinsic time scale which is constant through time. The constancy of the time scale is a key feature of this framework. Using the definition that the stocks' trading frequency $v_j$ is the number of intrinsic time ticks that occur for one calendar second, \cite{DERMAN2002} gives the relationship between the flow of calendar time $t$ and the flow of intrinsic time $\tau_j$ as 
\begin{equation}\label{eq:Derman:1}
	d\tau_j = v_j dt.
\end{equation}
More importantly, using equation \ref{eq:Derman:1} \cite{DERMAN2002} shows that the correlation in intrinsic time $\pi_{ij}$ is the same as the correlation in calendar time $\rho_{ij}$. This framework provides a method to create synchronous price paths in intrinsic time which can allow the recovery of the correlation in calendar time. 

Here instead of assuming $v_j$ to be the number of trades per calendar time second, we take $v_j$ to be the number of trades per unit of sampling interval considered. The correlations remain dimensionless and independent of the various time measurements.

To apply this sampling scheme, a volume bucket sizes must be chosen for each stock. This was determined by the approximate equivalent bar length in calendar time \footnote{To create the equivalent of 10 min calendar time bars in intrinsic time, we divide the average volume per day by 48.}. Since the volume bucket sizes are computed from the average volume per day, the price paths will not be fully synchronous due to different volume amounts traded per day for different stocks. Some stocks will have more (less) volume buckets if the trades for the day are above (below) the average for the stock (respectively). This can de-synchronise key events in the sampled data leading to leads and lags. 

This highlights the first issue with the framework. The assumption that each stocks' intrinsic time scale is constant through time is a strong assumption. Realistically the trading frequency can change over time depending on factors such as time of day, relevant news flow, announcements and order-flow events {\it e.g.} if news comes out that a company is about to undergo liquidation, traders may try to square their positions and increase the trading frequency. The second issue is that \eqref{eq:Derman:1} links the correlation in intrinsic time to that in calendar time; this assumption induces a continuity assumption from the calendar time into the intrinsic time given by the linear relationship - which re-scales the time given by a standard continuous-time stochastic processes. 

It must therefore be noted that it does not fully achieve the effect of converting stocks into intrinsic time - where time ticks purely on the events and a unit interval in intrinsic time can be thought of as a stochastic interval in calendar time with some stopping rule determining the number of trades counted. Nevertheless, it provides an informative lens through which to view the data if these assumptions hold.  

From Figure \ref{fig:Derman} (a) through to (c), we have the MM estimator applied to the calendar time equivalent of 1 hour, 10 minute and 1 minute bar data respectively. From figure \ref{fig:Derman} (d) through to (f), we have the HY estimator applied to the calendar time equivalent of 1 hour, 10 minute and 1 minute bar data respectively. It is clear that the Epps effect still exists under this different paradigm of aggregating TAQ data. 

We think this is interesting because this shows that the Epps effect does not only exist in calendar time, it is also present under the intrinsic time paradigm. What is even more interesting is that the correlation structures change depending on the sampling interval used, indicating that correlations are not dimensionless as required by \cite{DERMAN2002}; that the Epps effect seems to be intrinsically linked to the sampling intervals chosen. This attests to the idea that the Epps effect cannot be fully explained with asynchrony or lead-lag \cite{MMZ2011} simply in calendar time.  

Finally, the correlation structure in Figure \ref{fig:Derman} is very different to that of Figure \ref{fig:Closing} and \ref{fig:VWAP}, indicating that the correlations are not preserved across these various measurements of time as \cite{DERMAN2002} proved. Although we have highlighted a few of the pitfalls regarding this framework, this is a seamless framework, provided in the literature, which ties together the ideas from intrinsic time to calendar time.

This does not answer the question as to which estimator is the more efficient of the two; simply because the aggregation method creates data which is very close to being synchronous, therefore the two estimators behave similarly; as seen in Figure \ref{fig:Derman}. To this end, by employing the ideas from the intrinsic time framework \cite{DERMAN2002} but informed by the event time paradigm \cite{ELO2012A, ELO2012B} we created our own method of aggregating TAQ data. A method specifically focused on determining how the two estimators differ in framing the dependence of correlation on averaging scales. 

\subsection{Synchronised volume time averaging} \label{ssec:event}

Here we propose a synchronised volume clock sampling scheme. We use volume time aggregation to preserve some of the relative asynchrony of events through sequential volume aggregation inspired by \cite{ELO2012A,ELO2012B} but synchronised by using the same choice of volume interval. This volume time aggregation is achieved by setting the most liquid (and then the least liquid) asset as the baseline ``clock'' and events are then counted in terms of calendar time and each asset accumulates trades in calendar time until their respective volume buckets reach a predesignated threshold, at which time aggregated prices and quotes are presented with the associated calendar time stamp. The implementation can be found in Algorithm \ref{algo:LiningUpEvents}. Prices are sampled in volume time with the size of the volume interval is synchronised across the assets, but the calendar time at which each volume time increment is realised is not.  

This volume time averaging scheme attempts to preserve the stochastic nature of intrinsic time in calendar time and retains the partial recovery of Gaussian price fluctuations. 

Figure \ref{fig:Tim} (a) and (b) \footnote{We excluded the equivalent of the 1 hour calendar time average due to liquidity issues when synchronised to the most liquid ticker. Whereas we note that when synchronised to the least liquid ticker, the correlation structure of the HY looks similar to that of the 1 hour VWAP correlation and the MM has correlation around zero.} is the MM estimator applied to the 10 minute and 1 minute calendar time equivalent sample and (c) and (d) is the HY estimator applied to the 10 minute and 1 minute calendar time equivalent sample. The sub-figures on the left of Figure \ref{fig:Tim} are synchronised according to the least liquid ticker (MNP See Table \ref{tab:vol}) and those on the right are synchronised according to the most liquid ticker (FSR See Table \ref{tab:vol}). 

The first thing to notice is that even under this asynchronous sampling scheme an Epps like effect is still present with the HY estimator. However, for the MM estimator, the Epps effect is not clear when synchronised according to the least liquid ticker. This is a result of the highly asynchronous nature of our sampling scheme and our choice of using the Nyquist cut-off (See Section \ref{ssec:asynchrony}). The second thing to notice is how different the two estimators are: the MM has correlation near zero while HY has much higher correlation - the 1 minute equivalent calendar time bar synchronised to the most liquid ticker in Figure \ref{fig:Tim} (d) has significantly higher correlation when compared to the other aggregation methods - indicating that HY is manufacturing correlation. 

Figure \ref{fig:Tim} demonstrates the potential bias in the HY estimator when computing the correlation of events because it compensates the missing observations through the multiple contributions to bring up the estimate and therefore manufactures correlations. The MM treats missing observations as truly missing observations, it does not compensate the correlation but rather uses a lossless interpolation between events, therefore it does not overestimate the correlation to the same extent. 

The HY estimator remains the better estimator of the two if the observed prices in the financial market are samples from an underlying continuous-time stochastic process; however the MM estimator is the better estimator of the two in a high-frequency finance world with truly discontinuous, discrete and asynchronous events.


\section{Concluding Remarks} \label{sec:conclude}

We have demonstrated that the MM and HY estimators differ under asynchronous conditions. Although we have argued for the efficacy of the MM estimator over the HY estimator from a data-informed view, we do not concretely show which estimator is the better of the two. 

The MM estimator seems more believable because it possesses lossless interpolation between event data and it correctly captures the existence of an Epps like effect. On the other hand, the HY presents the following issues: i.) deleting observations and ii.) having a bias under the presence of lead-lag. Furthermore, it is unclear what the multiple contributions are truly doing with regards to the various empirical inconsistencies in the Epps effect.

The two estimators only differ under tick-by-tick asynchronous trade data. This is most graphically seen in the relative difference using the volume time averaging in Figure \ref{fig:Tim} where the MM estimator has a clear loss in correlation and smaller scales while the HY estimator manufactures correlations. The sequence of visualisation experiments: the decimation of events in the closing price representing in calendar time (Figure \ref{fig:Closing}), volume averaging in calendar time (Figure \ref{fig:VWAP}), the intrinsic time experiments (Figure \ref{fig:Derman}), and then the synchronised volume time averaging (Figure \ref{fig:Tim}); all suggest that the loss in correlations as the averaging scale is reduced is a fundamental property. It remains plausible that the effect results from the need to map an averaged observable from multiple event time scales into a single continuous calendar time in a way the shifts and smears events creating leads and lags while mixing time-scales and various structural constraints.   

To definitively determine the stronger estimator we would need a framework that ties together the discrete asynchronous data generating processes, the continuous-time stochastic processes underlying the construction of the estimators and their asymptotics, a satisfactory method to synchronise the TAQ data, and appropriate averaging scales and clocks that map back to calendar time in a reasonable manner.

In closing, we are able to demonstrate an Epps like effect in the South African financial market data through various temporal averaging scales to argue that lead-lag and asynchrony are insufficient in explaining the entirety of the effect; rather it seems to be possible, if one can account for possible structural or hierarchical effects, that the Epps like effects may still be dominated by averaged leads and lags induced between the interplay of choosing averaging scales and mapping observables from event time into calendar time. This could be approached by simulating the order-flow directly with point processes from different order types and different correlated assets, and then estimating correlations on different averaging scales from the resulting mid-prices. Estimating the impact of such a simulated order-flow on the transaction prices themselves would require simulated price discovery, with a simulated order book and matching engine.   

\section{Acknowledgements}

The authors would like to thank Etienne Pienaar, Dieter Hendricks, Nic Murphy and Diane Wilcox for various conversations and insights relating to the methods investigated in the paper. TG would like to thank Diane Wilcox for introducing him to the topic of alternative approaches to covariance estimation, and Nic Murphy for shared work relating to volume time based trading algorithms. TG acknowledges the financial support from the University of Cape Town. PC and RB acknowledges the financial support of the National Research Foundation (NRF) for funding their study at the University of Cape Town.

\printbibliography

\newpage

\section{Algorithms} \label{app:algo}

\begin{table}[H]
\begin{algorithm}[H]
\begin{algorithmic}
\Require {\\ \begin{enumerate} 
\item P: (n x m) matrix of asynchronously sampled price.
\item T: (n x m) matrix of asynchronously sampled times.
\end{enumerate}}
\State{Re-scale the time $[t_{min}, t_{max}] \rightarrow [0, 2\pi]$.}
\State{I. Extract trading times and prices.}
\For {i = 1 to m} 
\State{I.1. Slice the non-uniformly re-scaled sampled times}
\State{for the $i^{th}$ object.}
\State $\tilde{\tau} \leftarrow \tau(i)$
\State{I.2. Slice the sampled data indexing the times}
\State{for the $i^{th}$ object.}
\State $\tilde{\varphi} \leftarrow \ln\left(p\left(\tilde{\tau}\right)\right)$
\State{I.3. Compute the returns.}
\State $\delta_j \leftarrow \tilde{\varphi}(\tilde{\tau}_{j+1}) - \tilde{\varphi}(\tilde{\tau}_{j})$
\State{II.1. Compute Fourier coefficients for all values of $k$.}
 \State $\tilde{c}^+_k \leftarrow \sum_{j=1}^{n-1} e^{ik\tilde{\tau}_j} \delta_j$
 \State $\tilde{c}^-_k \leftarrow \sum_{j=1}^{n-1} e^{-ik\tilde{\tau}_j} \delta_j$
\State{II.2. Gather Fourier coefficients for $i^{th}$ object}
\State{for all values values of $k$.}
\State $c^+(i) \leftarrow \tilde{c}^+$
\State $c^-(i) \leftarrow \tilde{c}^-$
\EndFor
\State{III. Compute the integrated volatility and co-volatility}
\State{over the time window for objects i and j.}
\State $\Sigma_{ii} \leftarrow \frac{1}{|K|} \sum_{k \in K} [c^+_k(i) c^-_k(i)]$
\State $\Sigma_{ij} \leftarrow \frac{1}{|K|} \sum_{k \in K} [c^+_k(i) c^-_k(j)]$
\State $\Sigma_{ji} \leftarrow \Sigma_{ij}$
\State $R_{ij} \leftarrow \frac{\Sigma_{ij}}{\sqrt{\Sigma_{ii}} \sqrt{\Sigma_{jj}}}$
\State \Return{($\Sigma$, $R$)}
\end{algorithmic} \caption{Malliavin-Mancino Estimator} \label{algo:ComplexFT}
\end{algorithm}
\caption{The Malliavin-Mancino Estimator (See Algorithm \ref{algo:ComplexFT}) computes the pairwise implementation of the Malliavin-Mancino estimator \cite{MM2002,MM2009} using a complex exponential formulation of the Fourier Transform. The R implementation can be found at \href{https://github.com/rogerbukuru/Exploring-The-Epps-Effect-R/blob/master/Estimators/ftcorr.R}{ftcorr.R} in the GitHub resource \cite{PCRBTG2019} and was provided by \cite{set2} and based on their MATLAB implementation.}
\end{table}

\begin{table}[H]
\begin{algorithm}[H]
\begin{algorithmic}
\Require {\\ \begin{enumerate} 
\item T: (n x m) matrix of asynchronous sampled times.
\end{enumerate}}
\State Set: {$t_{\text{min}} =$ minimum value of T}
\State Set: {$t_{\text{max}} =$ maximum value of T}
\For{j = 1 to m}
\For{i = 1 to n}
\State $\tau_{ij} = \frac{2 \pi (t_j - t_{min})}{t_{\text{max}} - t_{\text{min}}}$
\EndFor
\EndFor
\State \Return{($\tau$)}
\end{algorithmic} \caption{Time-Rescaling Algorithm} \label{algo:rescale}
\end{algorithm}
\caption{The Time-Rescaling Algorithm (See Algorithm \ref{algo:rescale}) rescales the trading times from $[0, T]$ to $[0, 2 \pi]$. The R implementation can be found at \href{https://github.com/rogerbukuru/Exploring-The-Epps-Effect-R/blob/master/Estimators/ftcorr.R}{ftcorr.R} in the GitHub resource \cite{PCRBTG2019} and is an auxiliary function and was based on the MATLAB implementation from \cite{set2}.}
\end{table}

\begin{table}[h]
\begin{algorithm}[H]
\begin{algorithmic}
\Require {\\ \begin{enumerate} 
\item P: (n x m) matrix of asynchronously sampled price.
\item T: (n x m) matrix of asynchronously sampled times.
\end{enumerate}}
\State{Loop through every element of $\Sigma$ [m x m].}
\For {i = 1 to m} 
\State{I.1. Compute the returns.}
\State $\delta_i \leftarrow \ln\left(p^i\left(t_k\right)\right) - \ln\left(p^i\left(t_{k-1}\right)\right)$
\For {j = 1 to m} 
\State $\delta_j \leftarrow \ln\left(p^j\left(t_k\right)\right) - \ln\left(p^j\left(t_{k-1}\right)\right)$
\State{I.2. Compute Kanatani's weight matrix for the}
\State{$i^{th}$ and $j^{th}$ stock.}
\State $W \leftarrow$ Kanatani weight for HY
\State{I.3. Compute $\Sigma_{ij}$.}
\State $\Sigma_{ij} \leftarrow \delta_i' W \delta_j$
\EndFor
\EndFor
\State{II. Compute the correlation matrix.}
\State $R_{ij} \leftarrow \frac{\Sigma_{ij}}{\sqrt{\Sigma_{ii}} \sqrt{\Sigma_{jj}}}$
\State \Return{($\Sigma$, $R$)}
\end{algorithmic} \caption{Hayashi-Yoshida Estimator} \label{algo:HY}
\end{algorithm}
\caption{The Hayashi-Yoshida Estimator (See Algorithm \ref{algo:HY}) computes the pairwise implementation of the Hayashi-Yoshida estimator. Its R implementation can be found at \href{https://github.com/rogerbukuru/Exploring-The-Epps-Effect-R/blob/master/Estimators/ftcorr.R}{ftcorr.R} in the GitHub resource \cite{PCRBTG2019} and was based on MATLAB implementation from \cite{set2}.}
\end{table}

\begin{table}[H]
\begin{algorithm}[H]
\begin{algorithmic}
\Require {\\ \begin{enumerate} 
\item $\tau$: (n x 2) matrix of re-scaled asynchronous sampled times.
\end{enumerate}}
\State{Initialize W matrix:}
\State Set: $W = 0$ ($N_i$ x $N_j$)  matrix of 0's
\State Populate the W matrix:
\For{j = 1 to $N_i$}
\For{i = 1 to $N_j$}
\If{$(t^i_{k-1}, t^i_k] \cap (t^j_{l-1}, t^j_l] \neq \emptyset$}
\State Set: $w_{kl} = 1$
\EndIf
\EndFor
\EndFor
\State \Return{(W)}
\end{algorithmic} \caption{Kanatani-Weight-Matrix} \label{algo:weights}
\end{algorithm}
\caption{The Kanatani-Weight-Matrix (See Algorithm \ref{algo:weights}) computes the Kanatani Weight matrix for the Hayashi-Yoshida estimator \cite{HY2005}. Our R implementation can be found at \href{https://github.com/rogerbukuru/Exploring-The-Epps-Effect-R/blob/master/Estimators/ftcorr.R}{ftcorr.R} in the GitHub resource \cite{PCRBTG2019}, and is an auxiliary function.}
\end{table}

\begin{table}[H]
\begin{algorithm}[H]
\begin{algorithmic}
\Require {\\ \begin{enumerate} 
\item n: number of price points to simulate.
\item $\mu$: (d x 1) vector of drift parameters.
\item $\Sigma$: (d x d) covariance matrix.
\item start price: (d x 1) vector of $S(0)$.
\end{enumerate}}
\State Procedure for the $i^{th}$ asset:
\begin{enumerate} 
\item Generate: Z $\sim N_d(0, I_{dxd})$
\small
\item Set: $S_i(t_{k+1}) = S_i(t_k) \exp\big[(\mu_i - \frac{1}{2} \sigma^2_i)(t_{k+1} - t_k) + \sqrt{t_{k+1} - t_k} \sum_{k=1}^d A_{ik} Z_k \big]$
\end{enumerate}
\State \Return{(S)}
\end{algorithmic} \caption{GBM Algorithm} \label{algo:GBM}
\end{algorithm}
\caption{The Geometric Brownian Motion (GBM) Algorithm (See Algorithm \ref{algo:GBM}) simulates a correlated GBM subject to the condition $S(0) =$ start price, where $A$ is the Cholesky decomposition of $\Sigma$. The R implementation can be found at \href{https://github.com/rogerbukuru/Exploring-The-Epps-Effect-R/blob/master/Monte\%20Carlo\%20Simulation\%20Algorithms/GBM.R}{GBM.R} in the GitHub resource \cite{PCRBTG2019} and was provided by \cite{GLASSERMAN2004}.}
\end{table}

\begin{table}[H]
\begin{algorithm}[H]
\begin{algorithmic}
\Require {\\ \begin{enumerate} 
\item n: number of price points to simulate.
\item $\mu$: (d x 1) vector of drift parameters.
\item $\Sigma$: (d x d) covariance matrix.
\item $\lambda$: (d x 1) vector of the Poisson process parameter.
\item a: (d x 1) vector of lognormal location parameter.
\item b: (d x 1) vector of lognormal standard deviation.
\item start price: (d x 1) vector of $S(0)$.
\end{enumerate}}
\State Procedure for the $i^{th}$ asset:
\begin{enumerate} 
\item Generate: Z $\sim N_d(0, I_{dxd})$
\item Generate: $N_i \sim$ Poisson $\left(\lambda_i \left(t_{k+1} - t_k\right)\right)$
\item Generate: $Z_2 \sim N_1(0, 1)$
\item Set: $M = a_i N_i + b_i \sqrt{N_i} Z_2$
\item Set: $X_i(t_{k+1}) = X_i(t_k) + (\mu_i - \frac{1}{2} \sigma^2_i)(t_{k+1} - t_k) + \sqrt{t_{k+1} - t_k} \sum_{k=1}^d A_{ik} Z_k  + M$ 
\item $S = \exp(X)$
\end{enumerate}
\State \Return{(S)}
\end{algorithmic} \caption{Merton-Model Algorithm} \label{algo:Mert}
\end{algorithm}
\caption{The Merton-Model Algorithm (See Algorithm \ref{algo:Mert}) simulates a correlated Merton Model. Subject to the condition $X(0) = \ln(\text{start price})$ and $A$ is the Cholesky decomposition of $\Sigma$. The R implementation can be found at \href{https://github.com/rogerbukuru/Exploring-The-Epps-Effect-R/blob/master/Monte\%20Carlo\%20Simulation\%20Algorithms/Merton\%20Model.R}{Merton Model.R} at the GitHub resource \cite{PCRBTG2019} and was based on the implementation by \cite{GLASSERMAN2004}.}
\end{table}

\begin{table}[H]
\begin{algorithm}[H]
\begin{algorithmic}
\Require {\\ \begin{enumerate} 
\item n: number of price points to simulate.
\item $\theta$: (2 x 1) mean reverting rate.
\item $\lambda$: (2 x 1).
\item $w$: (2 x 1) vector long term variance.
\item $\rho$: correlation.
\item starting variance: (2 x 1) vector of starting variance.
\item starting price: (2 x 1) vector of starting price.
\end{enumerate}}
\State Procedure for the $i^{th}$ asset:
\begin{enumerate} 
\item Generate: $Z \sim N(0, 1)$
\item Set: $\sigma_i^2(t_{k+1}) = \sigma_i^2(t_{k}) + \theta_i (w_1 - \sigma_i^2(t_{k})) (t_{k+1} - t_k) + \sqrt{2 \lambda_i \theta_i (t_{k+1} - t_k)} \sigma_i(t_{k}) Z$
\item Create: $\Sigma^*$ correlation matrix based on $\sigma^2(t_{k+1})$
\item Generate: $Z^* \sim N_d(0, I_{dxd})$
\item Set: $X_i(t_{k+1}) = X_i(t_k) + \sqrt{t_{k+1} - t_k} \sum_{k=1}^d A_{ik}^* Z_k^*$
\item $S = \exp(X)$
\end{enumerate}
\State \Return{(S)}
\end{algorithmic} \caption{GARCH Algorithm} \label{algo:Garch}
\end{algorithm}
\caption{The GARCH Algorithm simulates a correlated bivariate GARCH(1,1) process subject to the condition $X(0) = \ln(\text{start price})$, $\sigma(0) = \text{starting variance}$ where $A^*$ is the Cholesky decomposition of $\Sigma^*$. The implementation in R can be found at \href{https://github.com/rogerbukuru/Exploring-The-Epps-Effect-R/blob/master/Monte\%20Carlo\%20Simulation\%20Algorithms/GarchAndersen.R}{GarchAndersen.R}, while the specification from \cite{RENO2001} can be found in \href{https://github.com/rogerbukuru/Exploring-The-Epps-Effect-R/blob/master/Monte\%20Carlo\%20Simulation\%20Algorithms/GarchReno.R}{GarchReno.R}. Both can be found in the GitHub resource \cite{PCRBTG2019}.}
\end{table}

\begin{table}[H]
\begin{algorithm}[H]
\begin{algorithmic}
\Require {\\ \begin{enumerate} 
\item n: number of price points to simulate.
\item $\mu$: (d x 1) vector of drift parameters.
\item $\Sigma$: (d x d) covariance matrix.
\item $\beta$: (d x 1) scale parameter of Gamma.
\item start price: (d x 1) vector of $S(0)$.
\end{enumerate}}
\State Procedure for the $i^{th}$ asset:
\begin{enumerate} 
\item Generate: $Y_i \sim$ Gamma$\left(\left(t_{k+1} - t_k\right)/\beta_i, \beta_i\right)$
\item Generate: Z $\sim N_d(0, I_{dxd})$
\item Set: $X_i(t_{k+1}) = X_i(t_k) + \mu Y_i + \sqrt{Y_i} \sum_{k=1}^d A_{ik} Z_k$
\end{enumerate}
\State \Return{(X)}
\end{algorithmic} \caption{Variance-Gamma Algorithm} \label{algo:VG}
\end{algorithm}
\caption{The Variance-Gamma Algorithm (See Algorithm \ref{algo:VG}) simulates a correlated Variance Gamma process. The process is subject to the condition $X(0) =$ start price, and $A$ is the Cholesky decomposition of $\Sigma$. The implementation in R can be found at \href{https://github.com/rogerbukuru/Exploring-The-Epps-Effect-R/blob/master/Monte\%20Carlo\%20Simulation\%20Algorithms/Variance\%20Gamma.R}{Variance Gamma.R} in the GitHub resource \cite{PCRBTG2019} and was provided by \cite{GLASSERMAN2004}.}
\end{table}

\begin{table}[H]
\begin{algorithm}[H]
\begin{algorithmic}
\Require {\\ \begin{enumerate} 
\item n: number of price points to simulate.
\item $\mu$: (d x 1) vector of long term prices.
\item $\Sigma$: (d x d) covariance matrix.
\item $\theta$: (d x 1) vector of mean reverting rate.
\item start price: (d x 1) vector of starting prices.
\end{enumerate}}
\State Procedure for the $i^{th}$ asset:
\begin{enumerate} 
\item Generate: Z $\sim N_d(0, I_{dxd})$
\item Set: $X_i(t_{k+1}) = X_i(t_k) + \theta_i (\ln(\mu_i) - X_i(t_k)) (t_{k+1} - t_k) + \sqrt{t_{k+1} - t_k} \sum_{k=1}^d A_{ik} Z_k$
\item $S = \exp(X)$
\end{enumerate}
\State \Return{(S)}
\end{algorithmic} \caption{Ornstein-Uhlenbeck Algorithm} \label{algo:OU}
\end{algorithm}
\caption{The Ornstein-Uhlenbeck Algorithm (See Algorithm \ref{algo:OU}) simulates a correlated mean-reverting Ornstein Uhlenbeck process. The process is subject to the condition $X(0) = \ln(\text{start price})$ where $A$ is the Cholesky decomposition of $\Sigma$. The R implementation can be found at \href{https://github.com/rogerbukuru/Exploring-The-Epps-Effect-R/blob/master/Monte\%20Carlo\%20Simulation\%20Algorithms/Ornstein\%20Uhlenbeck.R}{Ornstein Uhlenbeck.R} in the GitHub resource \cite{PCRBTG2019}.}
\end{table}

\begin{table}[H]
\begin{algorithm}[H]
\begin{algorithmic}
\Require {\\ \begin{enumerate} 
\item $T_i$: time of trade.
\item $S_i$: price at which assets were exchanged.
\item $V_i$: volume exchanged.
\end{enumerate}}
\State Identify the unique trading times $t_j^*$, $j = 1,... M$.
\State Gather transactions with the same trading times into a set $\{J_j\}_{j=1}^M$.
\State Procedure for the $j^{th}$ set:
\begin{enumerate} 
\item Set: $$s_j^* = \frac{\sum_{i \in J_j}{\mathrm{price}_i*\mathrm{volume}_i}}{\sum_i{\mathrm{volume}_i}}$$
\item Set: $V_j^* = \sum_{i \in J_j} V_i$
\end{enumerate}
\State \Return{( $T^* = \{ t_j^* \}_{j=1}^M, S^* = \{s_j^* \}_{j=1}^M, V^* = \{ V_j^*\}_{j=1}^M$ )}
\end{algorithmic} \caption{Repeated-Trade Aggregation Algorithm} \label{algo:aggregation}
\end{algorithm}
\caption{The Repeated-Trade Aggregation Algorithm (See Algorithm \ref{algo:aggregation}) aggregates trades with the same time stamp using a Volume Weighted Average Price. The R implementation can be found at \href{https://github.com/rogerbukuru/Exploring-The-Epps-Effect-R/blob/master/Data\%20Creation\%20Algorithms/AsynchronousData.R}{AsynchronousData.R} in the GitHub resource \cite{PCRBTG2019}.}
\end{table}

\begin{table}[H]
\begin{algorithm}[H]
\begin{algorithmic}
\Require {\\ \begin{enumerate} 
\item $T_i$: time of trade.
\item $S_i$: price at which assets were exchanged.
\item $V_i$: volume exchanged.
\item $\tau$: bar length (Units of time).
\end{enumerate}}
\State Gather trades into their respective bars $\{ J_j \}_{j=1}^M$ determined by $\tau$.
\For{j = 1, ..., M}
\If{j = 1} \State{Set: $O_1 = S_1$}
\Else \State{Set: $O_j = C_{j-1}$ }
\EndIf
\State Set: $H_j = \max \{ S_i \in J_j \} $
\State Set: $L_j = \min \{ S_i \in J_j \} $
\State Set: $C_j = \text{the last} \  S_i \in J_j$
\State Set: $V_j^* = \sum_{i \in J_j} V_i$
\State Set: $$\text{VWAP}_j = \frac{\sum_{i \in J_j}{S_i*V_i}}{\sum_i{V_i}}$$
\State Set: $t_j^* = T_1 + j \tau$
\EndFor
\State \Return{($t^* = \{ t_j^* \}_{j=1}^M, O = \{ O_j \}_{j=1}^M, H = \{H_j \}_{j=1}^M, L = \{ L_j \}_{j=1}^M, C = \{ C_j \}_{j=1}^M, V^* = \{ V_j^* \}_{j=1}^M, \text{VWAP} = \{ \text{VWAP}_j \}_{j=1}^M$)}
\end{algorithmic} \caption{Bar-Data Algorithm} \label{algo:Synchronous}
\end{algorithm}
\caption{The Bar-Data Algorithm (See Algorithm \ref{algo:Synchronous}) creates OHLCV and VWAP bar data. The implementation in R can be found at \href{https://github.com/rogerbukuru/Exploring-The-Epps-Effect-R/blob/master/Data\%20Creation\%20Algorithms/SynchronousData.R}{SynchronousData.R} in the GitHub resource \cite{PCRBTG2019}.}
\end{table}

\begin{table}[H]
\begin{algorithm}[H]
\begin{algorithmic}
\Require {\\ \begin{enumerate} 
\item $T_i$: time of trade.
\item $S_i$: price at which assets were exchanged.
\item $V_i$: volume exchanged.
\item $v_j = \frac{\text{ADV}_j}{N}$: the bucket size where $\text{ADV}_j$ is the Average Daily Volume for asset $j$ over the period of consideration and $N$ is the number of buckets per day.
\end{enumerate}}
\State Expand the number of observations by repeating each observation $S_i$ as many times as $V_i$, resulting in $I = \sum_i V_i$ observations of $S_i$. Expand such that the initial ordering of $S_i$ is not lost.
\State Set: $\tau = 0$
\While{$\tau v_j < I$}{
\State $\tau = \tau + 1$
\State $\forall i \in [ (\tau - 1) v_j + 1, \tau v_j ]$, compute $$P_{\tau} = \frac{\sum_{i}{S_i*V_i}}{\sum_i{V_i}}$$
\EndWhile}
\State \Return{$P = \{P_1, ..., P_M\}$}, where $M = \tau$ at end of
\State while loop.
\end{algorithmic} \caption{Intrinsic-Time Averaging Algorithm} \label{algo:Derman}
\end{algorithm}
\caption{The Intrinsic-Time Averaging Algorithm (See Algorithm \ref{algo:Derman}) creates asset specific intrinsic time bar data using the framework provided by \cite{DERMAN2002}. The R code implementation can be found at \href{https://github.com/rogerbukuru/Exploring-The-Epps-Effect-R/blob/master/Data\%20Creation\%20Algorithms/IntrinsicTimeVolumeBuckets.R}{IntrinsicTimeVolumeBuckets.R} in the GitHub resource \cite{PCRBTG2019} and is adapted from Algorithm A.1. of \cite{ELO2012B}.}
\end{table}

\begin{table}[H]
\begin{algorithm}[H]
\begin{algorithmic}
\Require {\\ \begin{enumerate} 
\item $T_{ij}$: time of trade for $j^{th}$ asset.
\item $S_{ij}$: price at which the $j^{th}$ asset was exchanged.
\item $V_{ij}$: volume exchanged for $j^{th}$ asset.
\item $v = \frac{\text{ADV}^*}{N}$: the bucket size where $\text{ADV}^*$ is the Average Daily Volume for the least liquid asset over the period of consideration and $N$ is the number of buckets per day.
\end{enumerate}}
\State Unique trading times $\{ T^* \}_{i^*=1}^{N^*}$ across all $j$ assets, note that $N^* \geq N_j \ \forall j$ where $N_j$ is the number of trade times for asset $j$.
\State Set: $\tau = 0$
\For{$i^*$ in $T^*$}{
    \State Expand $S_{i^*j}$ as many times as $V_{i^*j}$ and append to 
    \State the storage set $A_j = \{ S_{kj}^* \}_{k=1}^{K^j}$ for the $j^{th}$ asset. 
	\While{Any ${K_j} > {v}$}{ 
	\State $\tau = \tau + 1$
		\For{$j^{th}$ asset}{
			\If{$K^j > v$}
				\State $\forall k \in [ 1, v ]$, compute $$P_{\tau j} = \frac{\sum_{k}{S_{kj}^*}}{v}$$
				\State Remove the first $v$ $S_{kj}^*$ from $A_j$.
			\Else
				\State $P_{\tau j} = \text{NaN}$
			\EndIf
		}\EndFor
	}\EndWhile
}\EndFor
\State \Return{$P_j = \{P_{lj}\}_{l=1}^M $} where $M = \tau$ at end of the loops
\end{algorithmic} \caption{Event-Time Averaging Algorithm} \label{algo:LiningUpEvents}
\end{algorithm}
\caption{The Volume-Time Algorithm (See Algorithm \ref{algo:LiningUpEvents}) creates asynchronous synchronised volume time bar data. The algorithm R implementation can be found at \href{https://github.com/rogerbukuru/Exploring-The-Epps-Effect-R/blob/master/Data\%20Creation\%20Algorithms/EventTimeVolumeBuckets.R}{EventTimeVolumeBuckets.R} at the GitHub resource \cite{PCRBTG2019}.}
\end{table}

\end{document}